# Single Biological Neurons as Temporally Precise Spatio-Temporal Pattern Recognizers

Thesis for the degree of “Doctor of Philosophy”

By

Vladislav (David) Beniaguev

This work was carried out under the supervision of

**Michael London and Idan Segev**

# Acknowledgments

Pursuing a PhD in computational neuroscience has taught me a lot about the world, about science, about academia, about the brain, and about myself.

For these learnings, I first and foremost would like to thank all the “Edmond and Lily Safra Center for Brain Sciences” (ELSC) PhD program faculty and students without whom this would have been a completely different experience.
ELSC faculty is composed of many dedicated teachers and researchers and the ELSC students I studied with are among the most wonderful gifts I have ever received in life. Working alongside such curious and thoughtful individuals is infectious.

I would also like to thank my supervisors Idan Segev and Mickey London.
They provided for me both the freedom I desperately wanted, and the guard-rail support I never knew I needed.

But truthfully, I admire them most for their courage.
The courage to follow their scientific beliefs, pursue research in an unpopular field of study and carry the torch for decades, despite substantial resistance by peers coming from every direction imaginable (“resistance” is somewhat of an understatement, at least from my point of view, although I’m certain they will be much more restrained).
During this PhD, I have witnessed firsthand that the world of academia is full of talented individuals, many of whom are highly intelligent, creative and have the capacity to work very hard. I’ve also observed that courage, on the other hand, is of much more limited supply, and I’m grateful to have had the opportunity to be supervised by courageous scientists.

# Abstract

*“Everything should be as simple as it can be, but not simpler.”*
Albert Einstein

This thesis is focused on the central idea that single neurons in the brain should be regarded as temporally precise and highly complex spatio-temporal pattern recognizers. This is opposed to the prevalent view of biological neurons as simple and mainly spatial pattern recognizers by most neuroscientists today. In this thesis, I will attempt to demonstrate that this is an important distinction, predominantly because the above-mentioned computational properties of single neurons have far-reaching implications about the various brain circuits that neurons compose, and on how information is encoded by neuronal activity in the brain. Namely, that these particular “low-level” details at the single neuron level have substantial system-wide ramifications.

In the introduction we will highlight the main components that comprise a neural microcircuit that can perform useful computations and illustrate the inter-dependence of these components from a system perspective.

In **chapter 1** we will discuss the great complexity of the spatio-temporal input-output relationship of cortical neurons that are the result of morphological structure and biophysical properties of the various ion channels in the neuron, in particular the NMDA-based receptor conductance that is involved in the operation of cortical excitatory synapses. We show that the input/output mapping of a single Layer 5 (L5) cortical neuron can be accurately fitted by a deep temporally convolutional neural network (DNN) with 5-8 layers.
This study was recently published in Neuron under the title “Single cortical neurons as deep artificial neural networks” (Beniaguev et al., 2021)

In **chapter 2** we will demonstrate that single neurons can generate temporally precise output patterns in response to specific spatio-temporal input patterns with a very simple biologically plausible learning rule. This is possible due to the addition of two biological features into the simplified Integrate and Fire (I&F) neuron model – multiple synaptic contacts emerging from a single pre-synaptic axon and temporal filtering due to the cable properties of dendrites. This work was published in bioRxiv under the title “Multiple synaptic contacts combined with dendritic filtering enhance spatio-temporal pattern recognition of single neurons” (Beniaguev et al., 2022), and is presently being submitted to a peer-reviewed journal.

In **chapter 3,** using the tools we have developed in **chapters 1 and 2,** we highlight new avenues of future inquiry (with some beginnings along those lines). Specifically, we use the differentiable deep network analog of a realistic cortical neuron as a tool to approximate the gradient of the output of the neuron with respect to its input. With this capability we demonstrate how one can determine where to place synapses (onto which dendritic locations) in order for the neuron to achieve a particular task (e.g. output the XOR of two of its inputs). We discuss some technical challenges with this approach. Finally, we zoom out and ponder on how one could tackle the larger question of what kind of useful computations could be performed by a single neuron? We describe a plan of action in great detail with the hope that it will serve as a guide for future work.

In **chapter 4** (an appendix) we expand **chapter 3** to describe extension of our ideas to neuronal networks composed of many realistic biological spiking neurons that represent either small microcircuits or even entire brain regions. In this chapter we describe in detail a set of modeling experiments with increasing complexity that attempt to use networks of realistic neurons (and their DNN analogues) to perform a specific set of computational tasks with the goal of determining which biological features are important for which computational aspects of the nervous system. Additionally, in this chapter, we also suggest a surprisingly simple recipe about how one can, in principle, determine the precise overarching computation that is performed by any microcircuit or architecture by utilizing information about synaptic update rules from experimental work (In this context we refer to the word “computation” as in David Marr’s highest level of analysis – i.e., determine the computational objective of the circuit). Namely, we describe how one can get at the computational level of analysis (in Marr’s framework) of a neuronal circuit by incorporating knowledge from the implementational and algorithmic levels of analysis. Specifically, we describe in great detail a way how this can be done by utilizing information about the biological learning rules implemented by the brain at the single synapse level and utilizing our deep neural network (DNN) representation of realistic single neurons (that we formulized in **chapter 1**) in neuronal circuits composed of such DNNs-neurons.

In the Discussion, we elaborate on how the principles provided in this thesis could be applied to study a wide range of neuronal circuits and gradually build towards computational level of understanding of all neuronal microcircuits in the brain, and from there building toward determining the computational objective of entire brain regions which, eventually, encompass the computational objective of the whole brain.

# Table of Contents

# Letter of Contribution

David Beniaguev is the first author and the main researcher in all the chapters of this thesis. Listed below are the co-authors of the papers and their respective chapters, and the contribution of each of the authors involved:

1. **Single Cortical Neurons as Deep Artificial Neural Networks**
   David Beniaguev, Idan Segev, Michael London
   Neuron 109 (17), 2727-2739. e3
   doi: https://doi.org/10.1016/j.neuron.2021.07.002

   D.B., conceptualization, methodology, investigation, visualization, software, validation, data curation, writing – original draft; I.S. and M.L., conceptualization, methodology, writing – review & editing, supervision, resources, funding.

2. **Multiple Synaptic Contacts Combined with Dendritic Filtering Enhance Spatio-Temporal Pattern Recognition of Single Neurons**
   David Beniaguev, Sapir Shapira, Idan Segev, Michael London
   bioRxiv 2022.01.28.478132;
   doi: https://doi.org/10.1101/2022.01.28.478132

   D.B., conceptualization, methodology, investigation, visualization, software, validation, data curation, writing – original draft; S.S., visualization, software; I.S. and M.L., conceptualization, methodology, writing – review & editing, supervision, resources, funding.

3. **Mapping any spatio-temporal function onto a single biological neuron using the differentiability property of the DNN that replicated the I/O of a neuron**
   David Beniaguev

4. **Beyond single neurons: mapping any input/output task onto a network of realistic spiking neuron models**
   David Beniaguev

## Personal motivation for this thesis: biology as engineering

Note, this section is not the scientific introduction to the thesis but rather just personal motivation. Some of the things below contain loosely grounded intuitions and hunches. This is the kind of motivation that usually is never written down on paper by self-respecting scientists (and with good reason). But some scientists clearly do have overarching motivations in their heads (and some don't, of course). Nevertheless, although this section is not very polished, it really is the common thread of thought that guides my thinking and brings all the pieces of this thesis together.

A brain is a carefully engineered highly complicated biophysical device. It equips an organism with the ability to understand its surroundings, both near via sensory processing (is there a cliff nearby?) and remote via memory (Paris is a beautiful city and croissants are sold there every day). It enables the organism to conduct carefully coordinated action sequences, both quick via motor system (like shooting a basketball), and slow with a long-term planning ability (like completing a PhD), and it allows the organism to learn and adapt to a constantly changing world and invent new action sequences in order to achieve its short term and long-term goals.

This device is engineered not by a designer, but by the process of evolution. To this date, despite studying it for more than a century, humanity has not been able to fully understand the intricacies and inner workings of the brain. Because the brain is a carefully engineered device, regardless of whether it is being designed explicitly by an engineer or by the process of evolution, common engineering principles apply when attempting to understand its structure and function. In fact, any living organism is essentially a carefully engineered entity (or "robot") that is specifically designed to function well in a particular environment. For this reason, it is my belief that the field of biology is best thought of not as a type of a "natural science" as it is commonly regarded, but rather that biological creatures are the engineering products of some technological tree of progress that was created not by man but by evolution. Consequently, I believe that the process of understanding biology is in essence the process of reverse engineering a highly elaborate and capable technological tree of non-human origin (evolutional), and we should attempt to understand it in a similar fashion to how we would have attempted to understand an alien artifact if discovered – as if it was purposefully engineered rather than being simply a complex naturally occurring phenomenon (like the weather or an erupting volcano as examples).

It is especially interesting that many engineering principles that were invented by humans in their attempt to build large and complex systems are surprisingly applicable also when attempting to understand biologically engineered entities, suggesting that there are perhaps a set of universal engineering principles that potentially govern all engineering efforts of all kinds.

To make the description above more concrete, below is a short and non-comprehensive list of engineering principles that are particularly relevant to the work that will be discussed in this thesis:

1. Composition of previously designed components into a larger component.
2. Common design language to precisely define a component's structure and function.
3. Use of design language to reuse and replicate previously designed components.
4. Repurpose of previously generated components to serve new functionality in new context.

5. The inputs and outputs of many different components share a common interface such that they can be easily combined and interact one with another.

Since the above principles are a bit abstract and very general, to make them more concrete below are two lists of these principles as applied to **specific examples** in biology and neuroscience.

Engineering principles as applied to **specific examples** in **biology** are:

- Different sequences of a small number of amino acids give rise to all proteins in living organisms.
- DNA (built from 4 chemical bases) serves as a universal language that allows storage of protein designs and their precise replication.
- Multiple different proteins interact with ATP molecules as part of their normal operation as a form of a common interface.
- Co-expression of several different proteins which are specifically designed to work together is often controlled by an expression of some master gene, which can be thought of as an individual component on its own.
- Entire cells can display many differences of expression patterns of various proteins and groups of proteins, giving rise to cells with varied functionality although they are all built out of the same components.
- Protein expression patterns of different cell types have overlapping proteins. The same protein can perform a different role depending on the context of what other proteins are expressed alongside it.
- Cell membrane separation between intracellular and extracellular spaces and cross membrane proteins can be thought of as implementing a common interface that allows interactions between groups of nearby cells via altering the concentration of molecules in the common extracellular fluid by secreting and absorbing different molecules (e.g., neurotransmitters).
- Groups of cells that interact with each other via the common extracellular space form larger tissue structures. Different tissue structures will have different functionality.
- Multiple tissues combined in a particular way form organs.
- Multiple organs form an organism.
- Multiple organisms form a species.
- Different species sometimes share large fractions of their basic building blocks (DNA/proteins) with other species (e.g., homo sapiens and Neanderthals).

Engineering principles as applied to **specific examples** in **neuroscience** are:

- A handful of different ion channels and membrane receptors control the influx and outflux of a small group of ions between a cell and the extracellular space. i.e. a small set of ion channels (hundreds) is shared across all neurons (billions).
- Physically, a piece of short cylindrical membrane "cable" can be composed in multiple different ways into specific tree-like shapes (both dendrites and axons).
- A small number of neuron types (hundreds). Each neuron receives (is activated by) neurotransmitters, usually from the axons of other neurons, and secrete, usually by their own axonal boutons, same/other neurotransmitters to affect other neurons in the network.
- Small number of neurotransmitter types (dozens) and neuropeptides that are secreted and absorbed by cells in the nervous system, serve as communication messengers between all cells in the nervous system.
- Each neuronal type is basically a specific subset of membrane channels and receptors (out of a large pool of options), specific shapes of dendrites and axons (out of a very large pool of options), specific internal machinery to implement how the composition of the membrane ion channels changes with time (functional plasticity and homeostatic plasticity) and how the shape of the dendrites/axons changes with time (structural plasticity). This effectively describes the current state + how this state evolves in response to external stimuli.
- A stereotypical and carefully engineered spike mechanism that is essentially common to all excitable neurons in all organisms. It initiates the secretion of neurotransmitters from neurons and is carefully preserved while propagating throughout the axonal arbor. It therefore serves as a binary "0" or "1" interface of each neuron of both input and output. With this interface, neurons can be combined and connected with each other simply, via synapses.
- The synapses as near universal connecting interface between neurons. There is a small number of synapse types (e.g., excitatory and inhibitory) and a handful of plasticity protocols that govern their evolution with time.
- Stereotypical and repeating connectivity patterns at various brain regions. Cortical and cerebellar microcircuits are very specifically conserved and the wiring within each one, although can change e.g., via structural plasticity, is rather restricted.
- Hierarchy and Modularity of different brain regions to become specialized at performing some specific computational tasks. The composition of output coming from several specific brain regions into new brain regions allows the creation of a new emerging "higher level" function. Output from any specific brain regions is often sent to a rather discrete set of output brain regions. This means that impairing a specific high level brain region will often lead not to a complete system failure but a shutdown of several specific functions, but impairing a lower-level brain region might lead to more extreme system failures.

Proper separation in abstraction is immensely useful in engineering, particularly during periods of rapid progress at some specific technological aspect. For example, a hard separation between hardware and software and full agreement about a common interface allowed software and hardware engineers to advance in tandem without the need to worry about what the other was doing. This is of course only a single example, but the history of technology is full of these examples.

In the chip industry there is a hard separation also between logical design of circuits and the physical implementation of them. This allowed different rates of progress on computer architecture design and fabrication processes to add up seamlessly with each other for over half a century.

Imagine a situation where the fabrication process of very large scale integrated circuits (VLSI) is stalled for a period of 20 years. In this case the best way to reach maximal efficiency is to create very specific hardware designs that utilize the specific kinks of the current fabrication process of the current era. Alternatively, when faced with the knowledge that every 2 years we have a better process in which some fixed architecture with fixed functionality will simply perform better, what types of hardware designs should one use? It is always the case that by "opening the abstraction" and "optimizing jointly" it's possible to create a more efficient device, since there are less constraints. But the process of designing such an integrative device is so specific and lengthy that any performance gains that are due to "breaking the abstraction" today don't carry to the next stage of fabrication technology. Since plainly waiting usually led to large improvements in fabrication tech after the invention of the transistor over the course of decades, those who "broke" the abstraction at any given time couldn't utilize the benefits of the next stage of progress and simply fell short. It's not that there weren't those who tried building more efficient hardware by utilizing the specific kinks of the specific process of the era. They simply died away.

In a way, one might say that the independent nature of speed of progress on different subparts of a large system, selects for a separately optimizable and decomposable system design in the long run. Creating a hard abstraction between two parts of a system might be beneficial compared to jointly optimizing both parts together. This is counter intuitive at first, but it usually intuitive to any engineer that worked on a large project that was comprised of many different subparts. Arguably, any biological species is such a "project". Similarly, an identical mechanism of operation could also drive the optimization process of biological evolution as it drives human technological progress. It could be that there is strong positive evolutionary pressure towards modularity and creating hard abstractions between biological components, precisely because it allows to optimize separately each component and yields to quicker rates of progress overall. It is therefore perhaps not a huge surprise that biology is full of modular components and hierarchies upon hierarchies of compositions of components and common interfaces between the various components at various hierarchies, etc. All this could just be the result of some universal set of engineering principles that govern all possible engineering processes, be them designer guided or evolutionary guided. Think about DNA and basic machinery to translate DNA sequences into proteins. Once this process is set, all it takes to create different functionality is to find specific sequences of DNA. Now assume that an alternative primitive lifeform emerged without the highly composable structure of the DNA -> proteins mechanism, and that functionality is somehow created in an entangled soup of molecules that happened to create something that resembles a single celled organism with several useful functions that help it survive. Now try to imagine the continuation of the evolutionary process, and how quickly the organism that is built with the hard separation between DNA and proteins will simply overtake the ecological niche of the simpler organism. Being non optimal today, can create speed for tomorrow. If the correct set of abstractions is created.

It is the overwhelmingly prevailing belief that “biology is messy” and is full of “undocumented spaghetti code”. It is my personal belief that this is very fundamentally incorrect, and that biology is best thought of as mostly ordered and organized, just that we don’t precisely understand the separation points of the abstractions in it. It could be that reverse engineering a radio (Lazebnik, 2002) or a microprocessor (Jonas and Kording, 2017) might be difficult if we don’t correctly understand what a transistor is or what is the binary representation of floating point numbers. Moreover, if indeed biology is highly organized and ordered, then looking for hard abstractions in biology and the brain might prove immensely useful towards the goal of better understanding them.

In this thesis, we identify single neurons as an important class of components in the biological brain and we attempt to abstract away all their inner workings, remaining with its essential input/output capabilities. Our goal is to do this comprehensively enough so that a new abstract single neuron model can be used as solid ground to stand on when attempting to further understand the biological brain – which is a large and complex engineered system. This unfortunately is not the case today, as the conceptual model most neuroscientists today have about the function of a single neuron are still mostly that the output of a neuron is determined by the sum of its excitatory and inhibitory synaptic inputs with a threshold operation in the axon. Consequently, it is my belief that, when standing on a somewhat shaky foundation regarding the understanding of the operation of a single neuron, the attempt to understand the function of various brain circuits is in many ways doomed to fail.

There are several reasons why single neurons are the best “hard abstraction” option to tackle. First, they are ubiquitous across many animal species with relatively small number of variations in them between species and within species. Second, the spike is a logical and consistent input/output (I/O) interface that connects most central nervous systems neurons. Third, experimental techniques today allow highly detailed inspection of the morphology and biophysics of single neurons. And lastly, detailed mathematical models of the inner workings of neurons are currently very mature.

There are several potential benefits to finding this encapsulation for single neurons. The most important of which is that there will be no need to further understand any “downwards” mechanisms that give rise to the functions of the O(100) neuronal types when we wish to understand “higher” brain functions. An additional benefit is that it will allow building on this “rock solid” foundation to understand “higher” layers of abstraction in the brain. Neurons are composed one with another via specific microcircuits and microarchitectures (Notable important examples: cortical microcircuit, cerebellar microcircuit, hippocampal circuit, basal ganglia circuit), and thus it is possible to understand the functionality of these microarchitectures via the composition of the previously established clear functions of the O(100) neuronal types via the connectivity (the connectivity is directly measurable via efforts like electron microscopy). When extending this one additional step further "upwards", we can explain the function of brain regions that are composed of macro architectures that are, e.g., the combination of several different cortical circuits specifically connected to serve some specific function.

How should a proper encapsulation/abstraction of a single neuron look like?

1. Input output encapsulation. Simply describing the function $y = f(x, w)$, where x is the input in form of multiple spike-trains the neuron receives, y is the binary output spike train the neuron emits, and $w$ are the internal parameters that related to its morphology, synapses, ion channel identities, ion channel distribution, etc.
2. Since plasticity occurs inside the single neuron as a function of its own activity, we must also describe how all intrinsic parameters are changing as a function of the full history. The full history can be described by the history of all past inputs, all past outputs, and all past internal parameters $\{x_t\}, \{y_t\}, \{w_t\}$ for t < 0, where t=0 is current moment. For the sake of mathematical simplicity and under the assumption that the neuron has an underlying algorithmic goal with regards to the statistics of its input, we will be satisfied with summary statistics of the history of the inputs alone, $P(x_t)$, and the steady state solution if some function of those summary 14tatistics. Overall, $w_{steady_state} = h(P(x_t))$. To make it clear, the hypothesis here is that the neuron cannot change its input but directly controls the evolution of its parameters via plasticity mechanisms and therefore indirectly affects the history of the output.
3. The two points above fully define the function of the single neuron for any new $P(x)$ that is driven at its inputs: $y = f(x, w_{steady_state}) = f(x, h(P(x_t)))$
4. Any neuron type can have a unique $f()$ and $h()$ function. There are only a handful of neuron types in the brains across multiple species, on the order of O(100). Therefore, our goal is to find $f()$ and $h()$ functions for each of them.

In summary, I view biological species of any kind as the engineering products of evolution, and the brain as perhaps the crown jewel information processing device achieved by evolution thus far. I suggest that any engineered system likely abides by some universal engineering principles, such a modular component structure, repeating and reusing of components and multiple layers of hard abstractions. I speculate that when optimizing complex systems, there might be a strong evolutionary pressure towards hierarchical and modular separation of subcomponents as it is useful in order to achieve rapid progress over long time scales. I suggest that at the current state of research, the single neuron might be the best point of attack for encapsulation and could serve as a perfect separation point for a "hard abstraction" barrier to build upon in the attempts to tackle the problem of understanding the brain as a large and complex engineered system. All of these things serve as my personal motivation for this thesis.

# Introduction

In this introduction I will briefly overview issues that are directly related to this thesis. I will start by discussing various perspectives on neurons as I/O devices, focusing on the spatio-temporal filtering aspect. I will next briefly highlight some perspectives on synaptic plasticity, as both Chapters 2 & 3 in this thesis make use of functional and structural plasticity. I will then highlight network level aspects as Chapter 4 elaborates on this level of description.

## Single neurons as complex spatio-temporal input-output devices

Neurons are the computational building blocks of the brain. The transformation from input to output of a single neuron has therefore been a major source of inquiry in neuroscience. Several key questions are (a) what is the precise input to output function of a single neuron? (b) what are the mechanisms by which this function is realized? and (c) what are the computational implications for the organism? With the recent development of sophisticated optical and electrical experimental techniques, our knowledge about the input output transformation and mechanisms involved has increased substantially. It has also become clear that many key neuronal types (e.g., cortical L5PC, hippocampal CA3 & CA1, cerebellar Purkinje cells) have a highly complicated I/O relationship. These neurons typically receive a barrage of thousands of synaptic inputs via their elaborated dendritic trees; these inputs interact with a plethora of local nonlinear regenerative processes, including the sodium mediated back-propagating action potential (Stuart and Sakmann, 1994), local NMDA-dependent dendritic spikes (Branco et al., 2010; Kastellakis et al., 2015; Polsky et al., 2004; Schiller et al., 2000), and a large calcium spike at the apical dendrite of cortical layer five pyramidal neurons (Larkum et al., 1999; Schiller et al., 1997). Incoming input axons from pre-synaptic neurons connect to the dendrite via input synapses which interact with above mentioned local nonlinear dendritic properties to eventually generate a train of output spikes in the neuron's axon. The axon is communicating the information that is encoded in the spike train via output synapses to thousands of other post-synaptic neurons. Indeed, as a consequence of their inherent nonlinear mechanisms, neurons can implement highly complicated I/O functions (Bar-Ilan et al., 2012; Behabadi and Mel, 2013; Cazé et al., 2013; Doron et al., 2017; Häusser and Mel, 2003; Hawkins and Ahmad, 2016; Katz et al., 2009; Koch and Segev, 2014; Koch et al., 1982; London and Häusser, 2005; Mel, 1992; Moldwin and Segev, 2018; Poirazi et al., 2003b, 2003a; Shepherd et al., 1985; Tzilivaki et al., 2019; Zador et al., 1991), and see recent work on nonlinear dendritic computations in human cortical neurons in (Gidon et al., 2020). Although the I/O transformation and mechanism of operation are largely known, the computational ramifications are still a major source of inquiry.

A classical approach to study the I/O relationship of neurons from a computational perspective is to construct simplified models that omits many of their detailed biological mechanisms. These models present highly reduced phenomenological abstraction of the neuron's I/O characteristics (Lapicque, 1907; McCulloch and Pitts, 1943). One such abstraction is the "perceptron" (Rosenblatt and F., 1958), which lies at the heart of some of the most advanced pattern recognition techniques to date (LeCun et al., 2015). However, the perceptron's basic function, a linear summation of its inputs and thresholding for output generation, ignores the nonlinear synaptic integration processes and the temporal characteristics of the output, which take place in real neurons. Some more recent modeling studies have addressed this gap (Gütig and Sompolinsky, 2006; Poirazi et al., 2003a; Polsky et al., 2004; Ujfalussy et al., 2018), but have either not considered fully diverse synaptic inputs distributed over the full nonlinear dendritic tree, nor did aim to capture the I/O transformation of neurons at a millisecond temporal precision of output spikes. Attempts to predict the spiking activity of neurons in response to somatic input current/conductance, rather than to dendritic synaptic input, could be found in (Jolivet et al., 2008; Naud et al., 2014) and in response to natural images in (Cadena et al., 2019; Keat et al., 2001).

Another common approach to study the I/O characteristics of neurons is to simulate, via a set of partial differential equations, the fine electrical and anatomical details of the neurons using the cable and compartmental modeling methods introduced by Rall (Rall, 1959, 1964; Segev and Rall, 1988). Using these models, it is possible to account for nearly all of the above mentioned experimental phenomena and explore conditions that are not accessible with current experimental techniques. While this is the only method to-date to account for the full I/O transformation in a neuron, this success comes at a price. Compartmental and cable models are composed of a high-dimensional system of coupled nonlinear differential equations, which is notoriously challenging to understand (Strogatz, 2001). Specifically, it is a daunting task to extract general principles that govern the transformation of thousands of synaptic inputs to a train of spike output at the millisecond precision from such detailed simulations, but see (Amsalem et al., 2020; Larkum et al., 2009; Magee and Johnston, 1995; Rapp et al., 1992; Schiller et al., 2000; Spruston et al., 1995; Stuart and Sakmann, 1994; Stuart et al., 1997; Wybo et al., 2021).

In **chapter 1** we propose a novel approach to study the neuron as a sophisticated I/O information processing unit by utilizing recent advances in the field of machine learning. Specifically, we exploited the capability of deep neural networks (DNNs) to learn very complex I/O mappings (Holden et al., 2019; Kasim et al., 2020; Senior et al., 2020), in this case, that of neurons. Towards this end, we trained Deep Neural Networks (DNNs) with rich spatial and temporal synaptic input patterns to mimic the I/O behavior of a layer 5 cortical pyramidal neuron model with its full complexity, including its elaborated dendritic morphology, the highly nonlinear local dendritic membrane properties, and a large number of excitatory and inhibitory inputs that bombard the neuron. Consequently, we obtained a highly computationally efficient DNN model that faithfully predicted this neuron's output at a millisecond (spiking) temporal resolution. We then analyzed the weight matrices of the DNN to gain insights Into the mechanisms that shape the I/O function of cortical neurons. By systematically varying the DNN size, this approach allowed us to: (i) characterize the functional complexity of a single biological neuron, (ii) pin down the ion channels-based and morphologically based origin of this complexity, and (iii) examine the generality of the resultant DNN to synaptic inputs that were outside of the training set distribution. We demonstrated that cortical pyramidal neurons, and the networks they form, are potentially computationally much more powerful and "deeper" than previously assumed.

## Single neurons as temporally precise computing devices

In recent decades it was shown in a set of experiments that when two neurons are connected they typically connect to each other via multiple synaptic contacts rather than a single contact (Holler et al. 2021; Silver et al. 2003; Feldmeyer, Lübke, and Sakmann 2006; Shepherd et al. 2005; Markram et al. 1997). Multiple synaptic contacts originating from a single pre-synaptic axon often impinge on different parts of the dendritic tree of the post-synaptic neuron (Feldmeyer et al., 2006; Holler et al., 2021; Silver et al., 2003). Interestingly, if the formation of synaptic contacts were just based on "Peters rule" (Peters and Feldman, 1976), namely based purely on the proximity of the axon-to-the-dendrite and independent of each other, then one would expect that the distribution of the number of such multiple contacts would be geometric, with one contact per axon being the most frequent case (Fares and Stepanyants 2009; Markram et al. 2015; Rees, Moradi, and Ascoli 2017). This is far from what was empirically observed where, e.g., around 3-8 synaptic contacts are formed between pre- and post- synaptic layer 5 cortical pyramidal neurons (Markram et al. 1997). This deviation from the distribution predicted by Peters' rule suggests that the number of synaptic contacts between two connected neurons might be tightly controlled by some developmental process and is thus likely to serve a functional purpose.

Several phenomenological models have attempted to explain how multiple synaptic contacts between the pre-synaptic and post-synaptic neurons are formed (Fares and Stepanyants, 2009), but very few studies have attempted to tackle the question of how might they be beneficial from a computational perspective. It is typically thought that this apparent redundancy overcomes the problem of probabilistic synaptic vesicle release, which results in unreliable signal transmission between the connected neurons (Rudolph et al., 2015). Several statistically independent unreliable contacts that sum together can reduce the variance of the post synaptic potentials (PSP). However, the same effect using a simpler mechanism could be achieved by multiple vesicles release (MVR) per synaptic activation (Holler et al., 2021; Rudolph et al., 2015) and does not require multiple synaptic contacts. Other studies addressed additional possible advantages for having multiple synaptic contacts between two neurons. Hiratani and Fukai (Hiratani and Fukai, 2018) demonstrated that multiple synaptic contacts might allow synapses to learn quicker. Note, however, that faster learning, although beneficial, fundamentally does not endow the neuron with the ability to perform new kinds of tasks. Zhang et al. (Zhang et al., 2020) model multiple contacts in the context of deep artificial neural networks but demonstrate no tangible computational benefit for it in that context. Jones et al. (Jones and Kording, 2021) also model dendrites in the context of artificial neural networks and demonstrate that when using a threshold linear dendrite model the classification performance of a single neuron with multiple contacts is improved compared to the single contact case. Several other studies use multiple synaptic contacts in the context of artificial neural networks, demonstrating some computational benefits (Acharya et al., 2021; Camp et al., 2020; Sezener et al., 2021).

In **chapter 2** we propose two key functional consequence of multiple contacts. Towards this end, we developed a simplified spiking neuron model that we termed the Filter and Fire (F&F) neuron model. This model is based on the Integrate and Fire (I&F) neuron model but incorporates two key additional features. (i) It takes into account the effect of the dendritic cable filtering on the time course of the somatic post-synaptic potential (PSP), whereby proximal dendritic synapses generate brief somatic PSPs whereas distal synapses give rise to broad PSPs (Rall 1964; Rall 1967). (ii) Each pre-synaptic axon makes multiple synaptic contacts on the F&F neuron model. Consequently, a single pre-synaptic spike gives rise to a PSP composed of multiple temporal profiles (we term this phenomenon "dendro-plexing" of the presynaptic spike). To analyze the memory capacity of this F&F model, we used the formulation of Memmesheimer et al. (Memmesheimer et al., 2014) developed for the I&F model. In this approach the model is trained to emit precisely timed output spikes for a specific random input pattern; the capacity is defined as the maximal number of precisely timed output spikes during some time period divided by the number of incoming input axons. We show that the memory capacity of the F&F model is three folds larger than that of the I&F model. We next show that the F&F neuron can solve real-world classification tasks where the I&F model completely fails. We further explored the effect of unreliable synapses on the memory capacity of the F&F model and demonstrate that multiple synaptic contacts dramatically reduce axonal wiring requirement in cortical circuits.

## Biophysical calcium-based plasticity - synaptic updates in the brain

Neurons are highly plastic devices, and they constantly change. It is believed that the most crucial change is related to the connection strengths between neurons, i.e. synaptic efficacies. One of the earliest descriptions of synaptic plasticity rules that might occur between neurons was provided by Donald Hebb (Hebb, 1949) stating that "When an axon of cell A is near enough to excite cell B and repeatedly or persistently takes part in firing it, some growth process or metabolic change takes place in one or both cells such that A's efficiency, as one of the cells firing B, is increased". And in fact, as a support for Hebb's hypothesis, long term potentiation (LTP) of synaptic connections (Bliss and Collingridge, 1993; Bliss and Lomo, 1973) and long term depression (LTD) of synaptic connections (Ito, 1989) were discovered. Over the years, many different protocols were uncovered that induce activity-dependent long lasting plasticity changes (Bliss and Collingridge, 1993; Shouval et al., 2010; Whitlock et al., 2006). Notable examples are high frequency stimulation of the presynaptic site (that cause post synaptic depolarization and perhaps spiking), pairing between induced spikes on the presynaptic site and voltage depolarization of the post synaptic neuron (using voltage clamp), and pairing between presynaptic spikes and postsynaptic spikes at millisecond precision (using current clamp), which is also known as spike timing dependent plasticity (STDP) (Bi and Poo, 1998; Markram et al., 1997b).

After the initial discovery, various additional STDP kernels were reported between different pairs of neuron types throughout the brain. Also, many studies started examining the theoretical implications of STDP on learning (Abbott and Nelson, 2000; Caporale and Dan, 2008; Feldman, 2012; Kappel et al., 2014; Nessler et al., 2013; Savin et al., 2010; Song et al., 2000). It was immediately clear that a mechanistic explanation of this phenomena will involve the back-propagating action potential from the soma to the dendrites because it was the only way for the soma/axon to communicate "backwards" with the synapse that were involved with generating an output spike. Additionally, Following the discovery of STDP many experimenters went exploring further this phenomenon. A notable example among which is a pairing of presynaptic spike with a post synaptic burst of spikes. Interestingly, the resulting kernel of this triplet pairing could not be explained (Shouval et al., 2010) by summating the pairwise STDP kernels. This was a first type of result among many plasticity results that STDP failed to account for.

In 2002, Golding et al.(Golding et al., 2002) reported that distal (apical) dendrites in hippocampal CA1 neurons underwent plasticity independently whether a back propagating action potential propagated backwards to the dendrites or not, whereas for proximal dendrites the back propagating AP was necessary for plasticity. They further dissected the circuit and concluded that blocking NMDA channels reduced the potentiation size of by approximately 50%, and that blocking Voltage Gated Calcium Channels (VGCC) also reduced the potentiation size by ~50%. Blocking both NMDA and VGCC abolished the potentiation almost completely. A summary of their results can be found in **Figure 1** below.

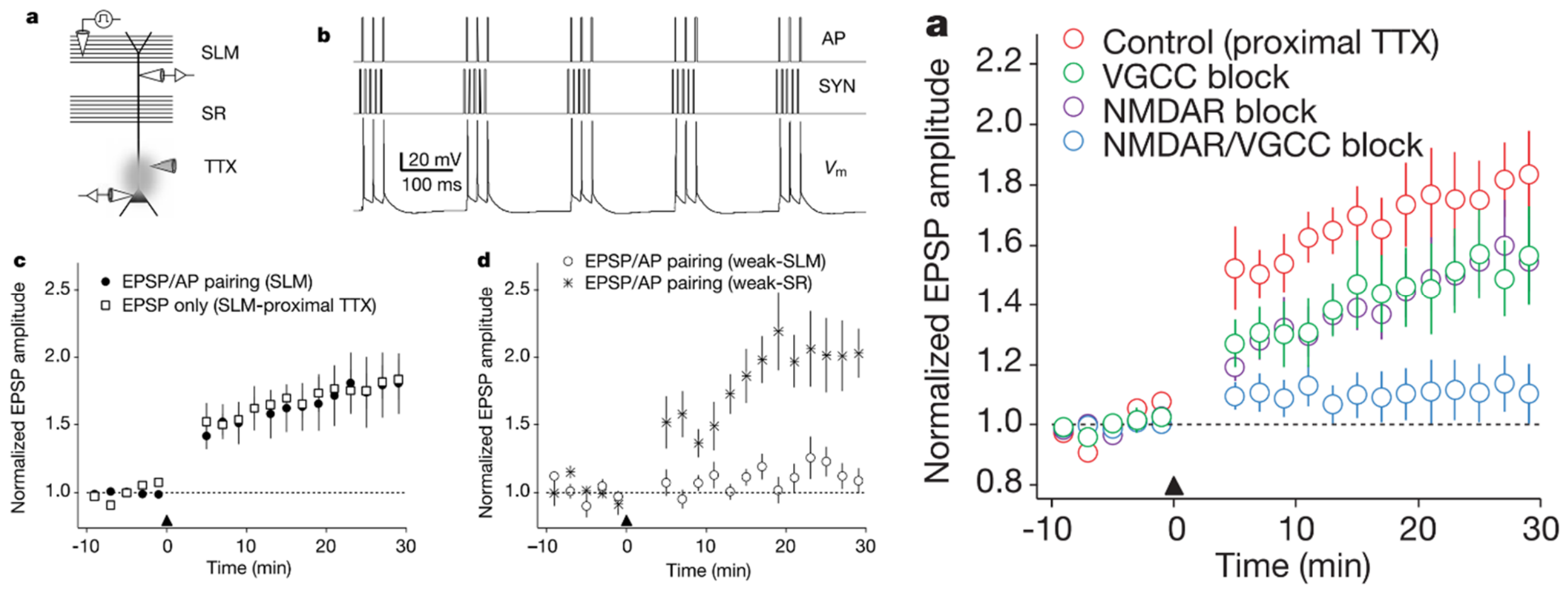

**Figure 1. Distal synapses undergo plasticity without BPAP.** Top left: experimental setup. SLM – distal dendrites. SR- proximal dendrites. Top center: pairing protocol of high frequency stimulation of synapses with a post synaptic burst. Bottom left: effect of proximal TTX (that blocks BPAP) on distal synapses (there is no effect). Bottom center: BPAP doesn't reach the distal dendrites but does reach proximal dendrites. Right: pharmacology suggests that NMDA and VGCC are together responsible for plasticity of distal dendrites. Adapted from Golding et al 2002

In 2006, Letzkus et al.(Letzkus et al., 2006) reported that the spike timing dependent plasticity depends on the location on the dendritic tree in cortex and that the kernels are shifted in time as a function of distance from soma. In 2016, Sandler et al.(Sandler et al., 2016) reported a novel form of synaptic plasticity that occurred at the distal dendrites of layer 5 pyramidal neurons of the cortex. This new form of plasticity did not include a pairing between presynaptic and post synaptic activation, but rather the EPSP that was recorded at the nexus increased substantially following repeated presynaptic stimulation alone.

These results required an explanation, and subsequently a calcium dependent plasticity rule was proposed to explain these results. A model proposed by Shouval (Shouval et al., 2002, 2010) was able to explain some of the triplet kernel and location dependence of synaptic plasticity. A similar yet slightly different model proposed by Graupner & Brunel (Graupner and Brunel, 2012) was able to explain all known plasticity phenomena (**Figure 2**). It was later demonstrated that this plasticity rule can be successfully applied to account for most known phenomena also in the context of a highly detailed simulation of a cortical circuit (Chindemi et al., 2022).

Local calcium signals are highly dependent on the distribution of various ligand gates and voltage gated ion channels on the dendritic tree, on the precise morphology of dendritic trees, and on the synaptic activation patterns of nearby synapses impinging on the dendritic tree. Therefore, if plasticity is the result of local calcium signals, then it might be best to study plasticity together with the effect of dendritic morphology and synaptic space-time activity to understand realistic synaptic plasticity. Note that that the proposed calcium-dependent plasticity mechanism could lead to initially unexpected results when considering together the complex dynamics that arises from the various dendritic components and potentially complicated synaptic activation patterns. For example, Bar Ilan at el. (Bar-Ilan et al., 2012) showed that inhibition can protect synapses from undergoing plasticity when placed in proximity in both time and space to an excitatory synapse. It is therefore evident that there is close entanglement between plasticity mechanisms, dendritic morphological and electrophysiological properties, and spatio-temporal synaptic activation patterns. It would thus be interesting to study this strongly coupled interaction further, and it perhaps even makes little sense studying synaptic plasticity without dendritic considerations due to this tight interdependency.

$$\tau\frac{d\rho}{dt} = -\rho(1-\rho)(\rho_\star-\rho) + \gamma_p(1-\rho)\Theta\left[c(t)-\theta_p\right] - \gamma_d\rho\Theta\left[c(t)-\theta_d\right] + Noise(t)$$

Synaptic Bi stability Term        Potentiation Term        Depression Term

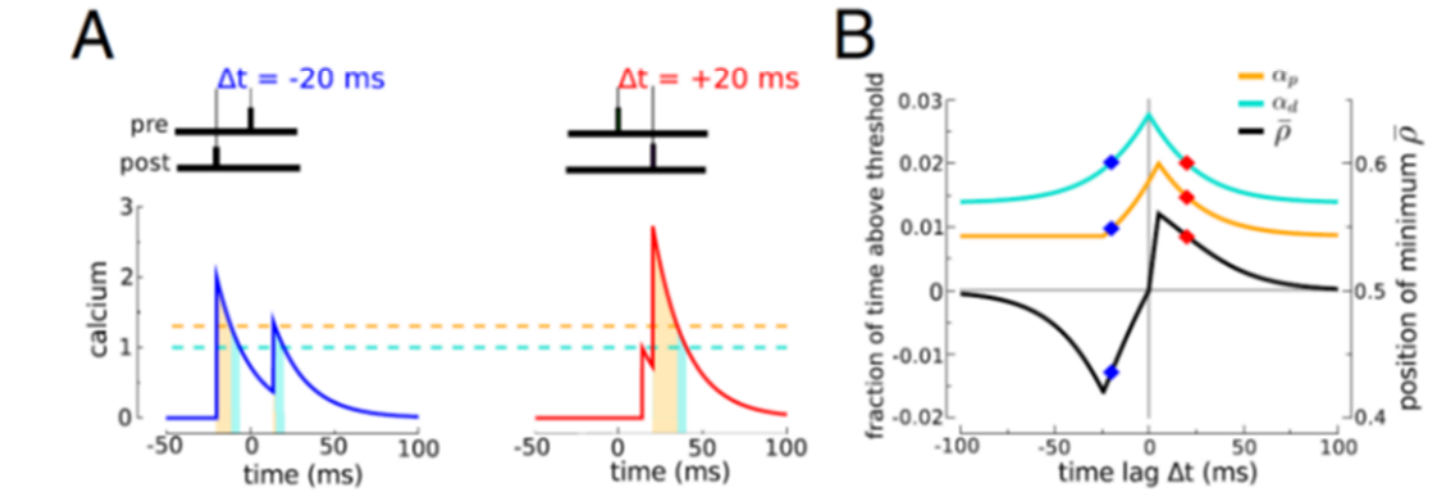

***Figure 2. Calcium Dependent Plasticity Model in Graupner & Brunel.*** *Top: the equation consists of 3 main terms. Synaptic bi-stability term that makes each synapse go towards 0 or 1 at steady state. Potentiation term that consists of a threshold of calcium concertation* $\theta_p$ *and potentiation rate* $\gamma_p$*. Same is applied also for depression.* ***A****. Top row: stimulation protocol of pre and post synaptic activation. Bottom row: calcium traces that result from stimulation and periods of potentiation and depression marked in color. When over the potentiation threshold (orange) the rate goes up, when above the depression threshold (blue) the rate goes down. In* ***B*** *The resulting potentiation and depression curves are depicted in color, and the combined standard STDP kernel in black. Note that for different parameter settings many different STDP kernels can be achieved. Adapted from Graupner & Brunel 2012*

It is important to keep in mind that calcium-based plasticity of AMPA receptors is not the only type of plasticity that underlies changes in the brain. Work by Losonczy et al.(Losonczy et al., 2008) showed that A type potassium channels in a dendritic branch (and not in a particular synapse) can also undergo plasticity in a fairly predictable manner following a specific pairing protocol of activating synapses at that branch region and inducing an AP at the soma. In fact, it is quite likely that all ion channels undergo local plasticity that is activity dependent according to some currently unknown conductance update rule.

It is also noteworthy that functional plasticity is not the only type of plasticity that occurs in neurons in the brain, and that structural plasticity also plays a major role in shaping neural circuitry. In fact, recent studies suggest that ~90% of excitatory synapses a neuron receives are somewhat negligible in the size in the EPSP amplitude they produce at the soma (Cossell et al., 2015). I.e., that only about 1 in 10 of the synapses are “substantial” synapses. This might hint at the fact that the main “goal” of functional plasticity could be to choose the subset of O(1,000) “best” synapses out of the O(10,000) total input synapses a neuron receives (where the term “best” is as yet unknown), and that those “silent” synapses are mainly waiting to undergo potentiation if the need arises for the neuron (or circuit) to change and adapt to new requirements. Furthermore, recent EM reconstructions (Kasthuri et al., 2015) suggested that in a radius of spine length around each dendritic segment that has a spine on it, there are about 10 very nearby axon crossings and presynaptic boutons that the spine could potentially connect to, but is not actually connected to. This means that while the neuron connects to O(10,000) nearby axons, the pool of potential axons to connect to is O(100,00). Together with studies about spine turnover rates in the neocortex, this might hint that the “goal” of structural plasticity is to find O(10K) “best” axons among the available O(100K) potential axons (again, “best” here is in some yet unknown sense). Together, these results suggest that any cortical neuron, by the combination of functional and structural plasticity, is in fact selecting O(1K) inputs to “listen” to out of O(100K) possible inputs nearby. This connectivity search process, along with a local calcium dependent plasticity rule, might lead with time to the accumulation of both anatomical and functional clusters on the dendritic tree (Kastellakis et al., 2015) that directly relate dendritic properties, synaptic plasticity mechanisms in the brain (Landau, 2022) and coordinated neuronal activity patterns.

To summarize, we have now reached a stage in the history of Neuroscience where the scientific community has a substantial degree of low-level understanding of various mechanisms at the neuronal and synaptic level (e.g., how various ion channels operate, how ionic currents flow along the dendritic tree, and how the ion channel distribution might change over time in an activity dependent manner). We can also envision how these low-level mechanisms are potentially inter-dependent one with another in a possibly complicated manner as we have suggested above. This motivates us to study further what higher-level single neuron phenomena might emerge from those intricately interacting low-level mechanisms.

## Neuronal types, micro-circuits, and macro architectures

There are multiple neuronal types in the brain (both multiple excitatory types and inhibitory types (Gupta et al., 2000)). Neuron types vary along the morphological shape and size, as well as their ion channel distribution across the membrane. Both the identities and density of ion channels vary across neuron types and also along different locations on the morphology within each neuron type. Internally, neurons might also vary along in the genetic programs they carry about how to adapt to changes (e.g., each neuron type might implement a slightly different synaptic plasticity algorithm). Together, these differences give rise to different input/output relationships for each neuron type. These neuron types also operate, inside the whole system, at specific input and output regimes (e.g. typical firing rates could be extremely low, such as for layer 2/3 excitatory neurons in the cortex, or very high, such as for medium spiny inhibitory neurons at the basal ganglia or the inhibitory Purkinje neurons in the cerebellum). Interestingly, although the number of neuron types could theoretically be infinite, it appears there are only several hundreds of neuron types throughout the brain of a particular animal species. This relatively small number makes them manageable to catalogue and fully describe in detail and various such attempts are currently undergoing (Cell Census Network, 2014; Druckmann et al., 2013; Gouwens et al., 2019; Gupta et al., 2000; Markram et al., 2015b)

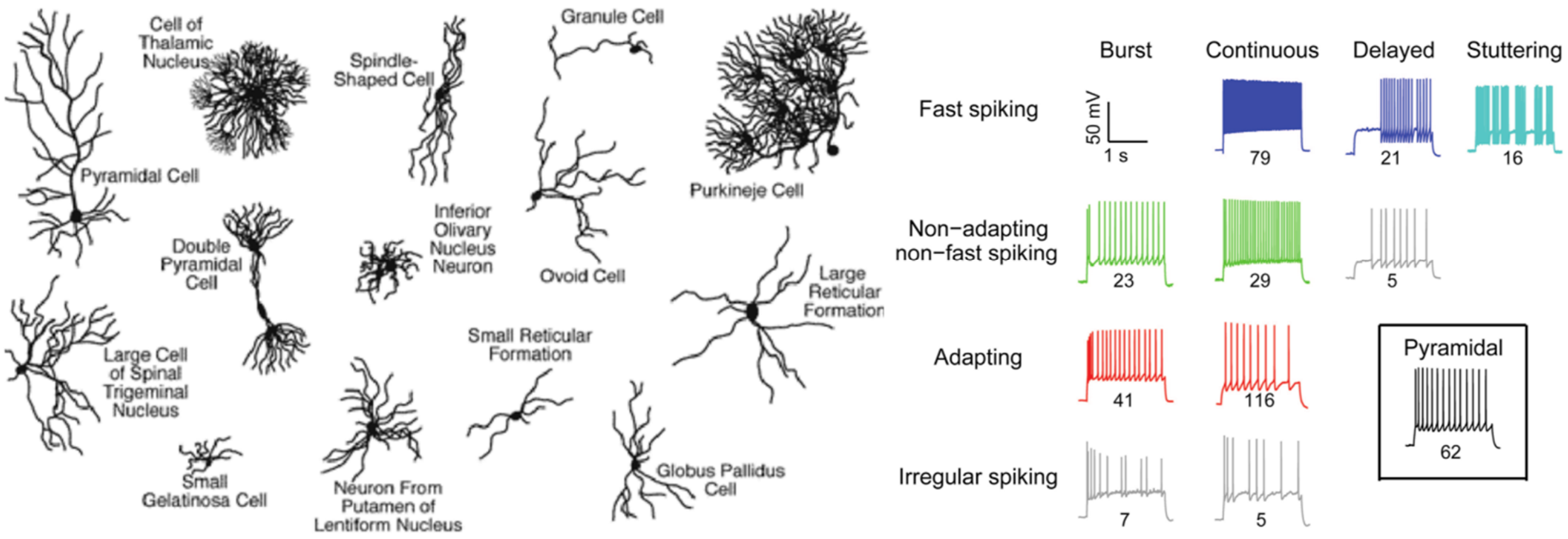

***Figure 3. Diverse Neuronal Types**, neuronal types are varied in both morphology and electrophysiology, properties that greatly affect their input/output function. Adapted from* (Druckmann et al., 2013; Stufflebeam, 2008)

The various neuronal types often interact with each other in various neuronal microcircuits throughout the brain in a stereotypical manner. A microcircuit could be defined as a small section of the brain that has a distinct connectivity pattern between a relatively small number of neuronal types among all possible types. Some notable examples might include the cortical microcircuit (Campagnola et al., 2022; Markram et al., 2015c), the cerebellar microcircuit (D'Angelo and Casali, 2012), and the hippocampal microcircuit (Aimone et al., 2014)). Of these, the cortical and cerebellar microcircuits are of particular interest since they appear to have a repeating structure that is replicated numerous times throughout the brain and comprise the majority of the brain volume (cortex) or of the number of neurons in the brain (cerebellum). Each instantiation of the same microcircuit serves a different purpose. E.g., the visual cortex vs the motor cortex.

The specific structure of a network microarchitecture greatly constrains the probability distribution of inputs that different neurons in the network see and therefore greatly shapes the function each neuron in a circuit will perform. Mathematically, for a simple conceptualization, one could think about a white noise random signal being filtered by a simple weighted moving average filter. The statistical properties of the output signal will greatly vary depending on the specific details of the filter. E.g. if it is a high pass filter, or a low pass filter, the frequency content or the autocorrelation of the resulting signal will vary significantly. Similarly, any complex nonlinear mathematical transformation applied to some input will shape and mold the input into having different properties.

We can view a network microarchitecture as a “complicated nonlinear filter” over the input stream and assert that it’s comprised of a different “filter” for each cell type in the network, resulting in a different probability distribution at this neuron’s input and output. Then, following the specific characteristics of the neuron I/O transformation and its plasticity rule, will therefore result in a different input-output relationship eventually learned by the cell. We later (in **chapter 4**) concretize these ideas in a simple yet highly illustrative Oja micro-circuit to perform PCA or ICA (Karhunen et al., 1997; Oja, 1982).

As we’ve previously stated, many brain regions exhibit highly constrained and specific connectivity patterns among themselves that appear like repeating structures throughout the brain. The most notable examples are the cerebellum the cortex, and the hippocampus as we previously stated. However, there is an additional layer of interestingness to note about these circuits. In particular, some similarities can be observed among these microcircuits, as if they are themselves belong to several “types” of neuronal microarchitecture with “different hyperparameters”. One such similarity is the sparse fan-out layer at the “entrance” to these circuits (Layer 4 (L4) in cortex, dentate gyrus (DG) in hippocampus and granule cell (GC) layer in the cerebellum). Another such similarity is the highly recurrent excitatory connections in L2/3 in cortex and CA3 in hippocampus. Another one is that the principal projection neuron type of these microcircuits receive two distinct types of inputs each. L5 cortical neuron’s basal dendrites receive mostly local feedforward inputs whereas apical dendrites receive input from distal cortical areas which are usually associated with feedback. CA1 hippocampal neurons receive input from local feedforward CA3 on their proximal dendrites and input from entorhinal cortex (EC) on the more distal dendrites that can be considered as feedback. Purkinje cerebellar neurons receive parallel fiber input (feedforward) and climbing fiber inputs (feedback) that are very different in the electrical way they activate the neuron. There are even similarities between hippocampus and cortex in the local innervation inhibitory neurons exhibit in the projection output layer (SOM cells innervate distal dendrites, PV cells innervate proximal dendrites and that both PV and SOM cells inhibit each other in both hippocampus and cortex). Along with what we’ve seen regarding the dependence of dendritic location on both the function of a neuron and the local plasticity rules it will apply, the highly specific connectivity patterns of different neuron types at relatively specific dendritic locations are most likely there to serve a predefined purpose to achieve some specific statistical calculation (and not another).

An additional similarity is that all projection neurons in these circuits appear to exhibit some form of “complex spike” that results due to feedback activation and specifically associated with large calcium influx into the cell (we’ve previously implicated calcium to be a critically important factor for synaptic plasticity). An interesting story might arise from these similarities. Consider for example a generic microcircuit-architecture of fan out input layer, then a recurrent layer, then a final “summarization of information” layer that can also be modulated by external feedback input. The cerebellum has an immense fanout input layer. The hippocampus has intermediate sized such layer, and cortex has a modest fanout input layer. One can consider the degree of fan out layer as a hyper-parameter. The same applies to the recurrent layer size, and degree of sparseness (CA3 in hippocampus appear to be much more densely recurrently connected compared to L23 PC layer in cortex) that can be an additional hyper-parameter. Another hyper-parameter could relate to the degree of influence of the feedback on the output projection layer (e.g., extremely strong feedback effect as in the climbing fibers input to the Purkinje neurons in the cerebellum, and less strong in cortical L5 and hippocampal CA1 neurons). The similarities and differences between these microcircuits are intriguing and deserve second, third and fourth looks.

The microcircuit contains repeating patterns of connectivity within the same brain. One step above that, all brains of same species exhibit highly specific long range connectivity patterns at the macro scale. This occurs at an even higher spatial scale than microcircuit connectivity – the connection between brain regions – and could be termed the marco-architecture of the brain. I.e., various cortical regions are connected one to another in a highly stereotypical way across different members of the same species. For example, the visual system in all primate brains exhibit the same macro-architecture of connections between various brain regions, as depicted in the famous Van Essen diagram (Felleman and Van Essen). We will touch upon micro-architecture and macro-architecture aspects of neuronal networks and how it's possible to study them further in **chapter 4**.

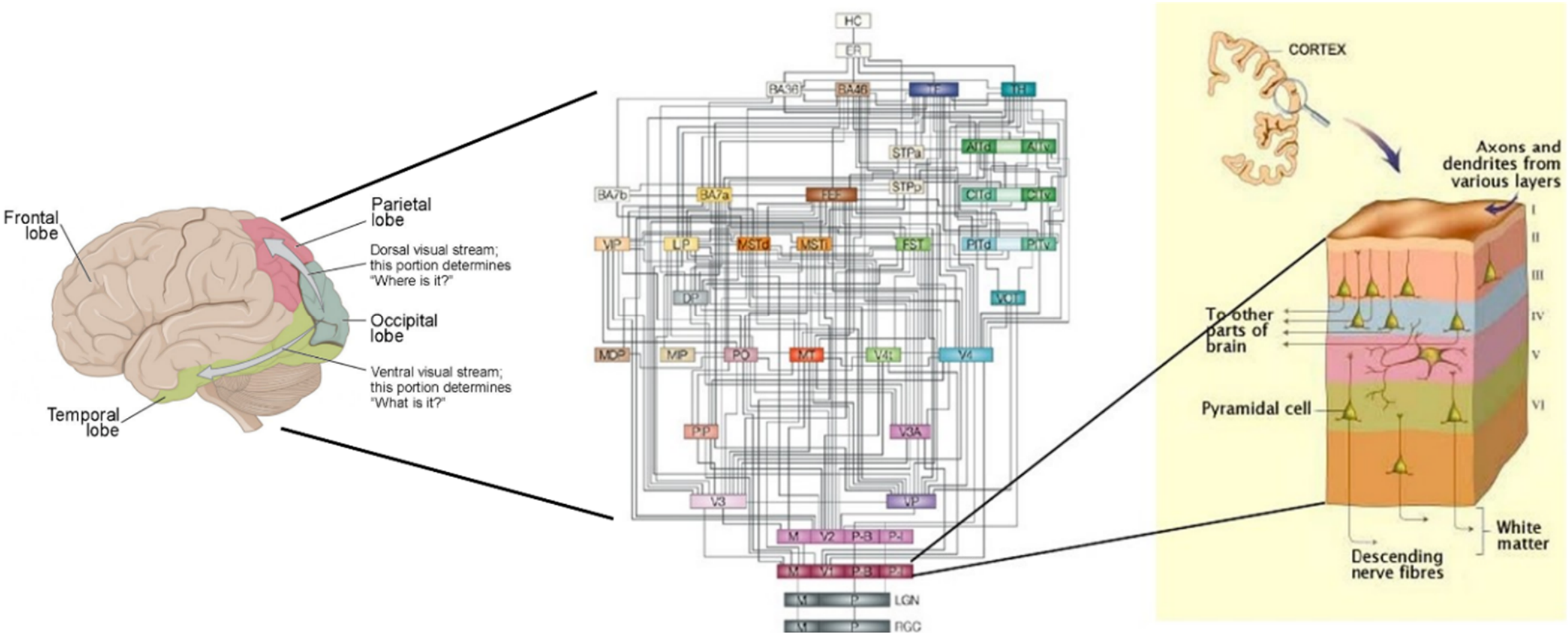

***Figure 4. From Macro-architecture to Micro-architecture to neuronal types (nano-architecture?)*** *Left: an illustration of the human brain with various brain regions illustrated. Middle: the famous Van Essen diagram depicting stereotypical connections inside the primate visual cortex. This can be termed macro architecture Right: An Illustration of a cortical column comprised of multiple cortical neuron types, a microarchitecture. adapted from* (Felleman and Van Essen)*.While searching for the statistical function of the cortical column, we must remember that a cortical column is only an important building block in a much larger system that has an objective at a higher level of abstraction. Here depicted is the famous Van Essen* (Felleman and Van Essen) *diagram for the visual system.*

# Chapter 1:

## Single Cortical Neurons as Deep Artificial Neural Networks

David Beniaguev, Idan Segev, Michael London

Article

# Single cortical neurons as deep artificial neural networks

## Graphical abstract

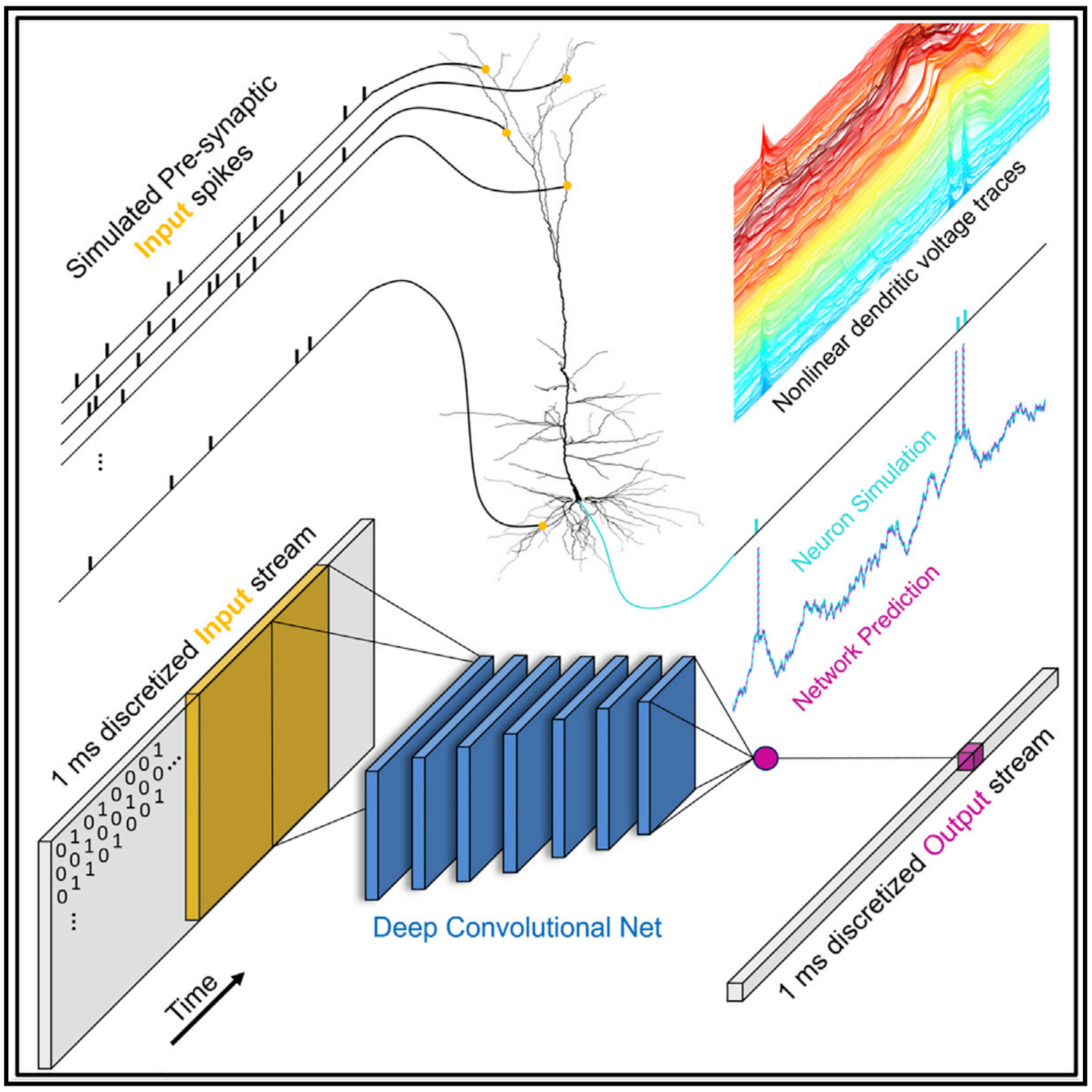

## Authors

David Beniaguev, Idan Segev, Michael London

## Correspondence

david.beniaguev@gmail.com

## In brief

Using a modern machine learning approach, we show that the I/O characteristics of cortical pyramidal neurons can be approximated, at the millisecond resolution (single spike precision), by a temporally convolutional neural network with five to eight layers. This computational complexity stems mainly from the interplay between NMDA receptors and dendritic morphology.

## Highlights

- Cortical neurons are well approximated by a deep neural network (DNN) with 5–8 layers
- DNN's depth arises from the interaction between NMDA receptors and dendritic morphology
- Dendritic branches can be conceptualized as a set of spatiotemporal pattern detectors
- We provide a unified method to assess the computational complexity of any neuron type

Article

# Single cortical neurons as deep artificial neural networks

David Beniaguev,[1,3,*] Idan Segev,[1,2] and Michael London[1,2]
[1]Edmond and Lily Safra Center for Brain Sciences (ELSC), The Hebrew University of Jerusalem, Jerusalem 91904, Israel
[2]Department of Neurobiology, The Hebrew University of Jerusalem, Jerusalem 91904, Israel
[3]Lead contact
*Correspondence: david.beniaguev@gmail.com

## SUMMARY

**Utilizing recent advances in machine learning, we introduce a systematic approach to characterize neurons' input/output (I/O) mapping complexity. Deep neural networks (DNNs) were trained to faithfully replicate the I/O function of various biophysical models of cortical neurons at millisecond (spiking) resolution. A temporally convolutional DNN with five to eight layers was required to capture the I/O mapping of a realistic model of a layer 5 cortical pyramidal cell (L5PC). This DNN generalized well when presented with inputs widely outside the training distribution. When NMDA receptors were removed, a much simpler network (fully connected neural network with one hidden layer) was sufficient to fit the model. Analysis of the DNNs' weight matrices revealed that synaptic integration in dendritic branches could be conceptualized as pattern matching from a set of spatiotemporal templates. This study provides a unified characterization of the computational complexity of single neurons and suggests that cortical networks therefore have a unique architecture, potentially supporting their computational power.**

## INTRODUCTION

Neurons are the computational building blocks of the brain. Understanding their input/output (I/O) transformation has therefore been a major quest in neuroscience since Ramon y Cajal's "neuron doctrine." With the recent development of sophisticated genetic, optical, and electrical techniques, it has become clear that many key neuronal types (e.g., cortical and hippocampal pyramidal neurons, cerebellar Purkinje cells) are highly complicated I/O information processing devices. They receive a barrage of thousands of synaptic inputs via their elaborated dendritic branches; these inputs interact with a plethora of local nonlinear regenerative processes, including the back-propagating ($Na^+$-dependent) action potential (Stuart and Sakmann, 1994), the multiple local dendritic NMDA-dependent spikes (Schiller et al., 2000; Polsky et al., 2004; Branco et al., 2010; Kastellakis et al., 2015), and the large and prolonged $Ca^{2+}$ spike at the apical dendrite of layer 5 (L5) cortical pyramidal neurons (Schiller et al., 1997; Larkum et al., 1999). The input synapses interact with these local nonlinear dendritic properties to eventually generate a train of output spikes in the neuron's axon, carrying information that is communicated, via synapses, to thousands of other (postsynaptic) neurons. Indeed, as a consequence of their inherent nonlinear mechanisms, neurons can implement highly complicated I/O functions (Bar-Ilan et al., 2013; Behabadi and Mel, 2014; Cazé et al., 2013; Doron et al., 2017; Häusser and Mel, 2003; Hawkins and Ahmad, 2016; Katz et al., 2009; Koch and Segev, 2014; Koch et al., 1982; London and Häusser, 2005; Mel, 1992; Moldwin and Segev, 2018; Poirazi et al., 2003b, 2003a; Shepherd et al., 1985; Tzilivaki et al., 2019; Zador et al., 1991; see recent work on nonlinear dendritic computations in human cortical neurons in Gidon et al., 2020).

A classical approach to study the I/O relationship of neurons is to construct a simplified model that omits many of their detailed biological mechanisms. These models present a highly reduced phenomenological abstraction of the neuron's I/O characteristics (Lapicque, 1907; McCulloch and Pitts, 1943). One such abstraction is the "perceptron" (Rosenblatt, 1958), which lies at the heart of some of the most advanced pattern recognition techniques to date (LeCun et al., 2015). However, the perceptron's basic function, a linear summation of its inputs and thresholding for output generation, ignores the nonlinear synaptic integration processes and the temporal characteristics of the output, which take place in real neurons. Some more recent modeling studies have addressed this gap (Gütig and Sompolinsky, 2006; Poirazi et al., 2003a; Polsky et al., 2004; Ujfalussy et al., 2018) but have either not considered fully diverse synaptic inputs distributed over the full nonlinear dendritic tree or did not aim to capture the I/O transformation of neurons at a millisecond temporal precision of output spikes. Attempts to predict the spiking activity of neurons in response to somatic input current/conductance, rather than dendritic synaptic input, can be found in Jolivet et al. (2008) and Naud et al. (2014), and attempts to predict the spiking activity of neurons in response to natural images can be found in Cadena et al. (2019) and Keat et al. (2001).

Another common approach to study the I/O characteristics of neurons is to simulate, via a set of partial differential equations,

the fine electrical and anatomical details of the neurons using the cable and compartmental modeling methods introduced by Rall (Rall, 1959, 1964; Segev and Rall, 1988). Using these models, it is possible to account for nearly all of the above experimental phenomena and explore conditions that are not accessible with current experimental techniques. While this is the only method to date to account for the full I/O transformation in a neuron, this success comes at a price. Compartmental and cable models are composed of a high-dimensional system of coupled nonlinear differential equations, which is notoriously challenging to understand (Strogatz, 2001). Specifically, it is a daunting task to extract general principles that govern the transformation of thousands of synaptic inputs to a train of spike output at millisecond precision from such detailed simulations (but see Amsalem et al., 2020; Larkum et al., 2009; Magee and Johnston, 1995; Rapp et al., 1992; Schiller et al., 2000; Spruston et al., 1995; Stuart and Sakmann, 1994; Stuart et al., 1997; Wybo et al., 2021).

Here, we propose a novel approach to study the neuron as a sophisticated I/O information processing unit by utilizing recent advances in the field of machine learning. Specifically, we exploited the capability of deep neural networks (DNNs) to learn very complex I/O mappings (Holden et al., 2019; Kasim et al., 2020; Senior et al., 2020), in this case, that of neurons. Toward this end, we trained DNNs with rich spatial and temporal synaptic input patterns to mimic the I/O behavior of a L5 cortical pyramidal neuron model with its full complexity, including its elaborated dendritic morphology, the highly nonlinear local dendritic membrane properties, and a large number of excitatory and inhibitory inputs that bombard the neuron. Consequently, we obtained a highly computationally efficient DNN model that faithfully predicted this neuron's output at millisecond (spiking) temporal resolution. We then analyzed the weight matrices of the DNN to gain insights into the mechanisms that shape the I/O function of cortical neurons. By systematically varying the DNN size, this approach allowed us to characterize the functional complexity of a single biological neuron, pin down the ion-channel-based and morphologically based origin of this complexity, and examine the generality of the resultant DNN to synaptic inputs that were outside of the training set distribution. We demonstrated that cortical pyramidal neurons, and the networks they form, are potentially computationally much more powerful and "deeper" than previously assumed.

## RESULTS

### Analogous DNN for integrate and fire (I&F) neuron model: Method overview and filters interpretation

Our goal is to fit the I/O relationship of a detailed biophysical neuron model by an analogous DNN. This DNN receives, as a training set, the identical synaptic input and the respective axonal output of the biophysical model. By changing the connection strengths of the DNN using a backpropagation learning algorithm, the DNN should replicate the I/O transformation of the detailed model. To accommodate the neuron's temporal aspect, we employed temporal-convolutional networks (TCNs) throughout the study.

Figure 1 illustrates this paradigm's feasibility and usefulness as a first demonstrative step, starting with the I/O transformation of a well-understood neuron model: the I&F neuron (Burkitt, 2006; Lapicque, 1907). This neuron receives a train of random synaptic inputs and produces a subthreshold voltage response as well as a spiking output (see STAR Methods). While this I/O transformation appears to be simple, it is unclear whether it can be learned from data by an artificial neural network using the backpropagation algorithm at millisecond temporal resolution with a compact architecture (this indeed has not been previously demonstrated). If successful in achieving a high degree of accuracy with a simple DNN for the I&F model, it will demonstrate that our approach, consisting of the specific way we represent the neuronal I/O data and subsequent fitting of an DNN on these data, is able to represent the functional relationship of the I&F neuron model compactly.

What is the simplest DNN that faithfully captures the I/O properties of this most basic single-neuron model? To answer this question, we constructed a minimal "DNN" consisting of one hidden layer with a single hidden unit (Figure 1A) and verified that it does indeed capture the complexity of this simple neuron model (Figure 1F). The time axis was divided into 1 ms bins in which only a single spike can occur in the I&F neuron model. The objective of the network is to predict the binary spike output of the I&F model at time $t_0$, based on the preceding input spike trains (the time-window history) up to $t_0$. This input is represented using a binary matrix of size $N_{syn} \times T$, where $N_{syn}$ is the number of input synapses and $T$ is the number of preceding time bins considered (Figure 1B). We used $N_{syn} = 100$, and trained a DNN with a single hidden unit on 7,200 s of simulated data. When using T = 80 ms, we achieved a very good fit, namely, a simple DNN with a single hidden unit that accurately predicted both the subthreshold voltage dynamics as well as the spike output of the respective I&F neuron model at millisecond precision (Figure 1C).

Figure 1D depicts the weights ("filters") of the single hidden unit of the respective DNN as a heatmap. It shows that the learning process automatically produced two classes of weights (filters), one positive and one negative, corresponding to the excitatory and inhibitory inputs impinging on the I&F model. In agreement with our understanding of the I&F model, the excitatory inputs contribute positively to output spike prediction (red), whereas the inhibitory inputs contribute negatively to it (blue). Earlier inputs, either inhibitory or excitatory, contribute less to this prediction (teal). Figure 1E depicts the temporal cross-section of those filters and reveals an exponential profile that reflects the temporal decay of postsynaptic potentials in the I&F model (in the reverse time direction). From these filters, one can recover the precise membrane time constant of the I&F model. These two temporal filters (excitatory and inhibitory) are easily interpretable, as they agree with our previous understanding of the temporal behavior of synaptic inputs that give rise to an output spike in the I&F model.

Figure 1F quantifies the model performance in terms of spike prediction using the receiver operating characteristic (ROC) curve (Figure 1F, left; see STAR Methods) and the area under it (area under the curve [AUC]). The AUC for the I&F case is 0.9973, indicating a very good fit. Figure 1F (right) shows an additional quantification of spike temporal precision using the DNN prediction by plotting the cross-correlation between the

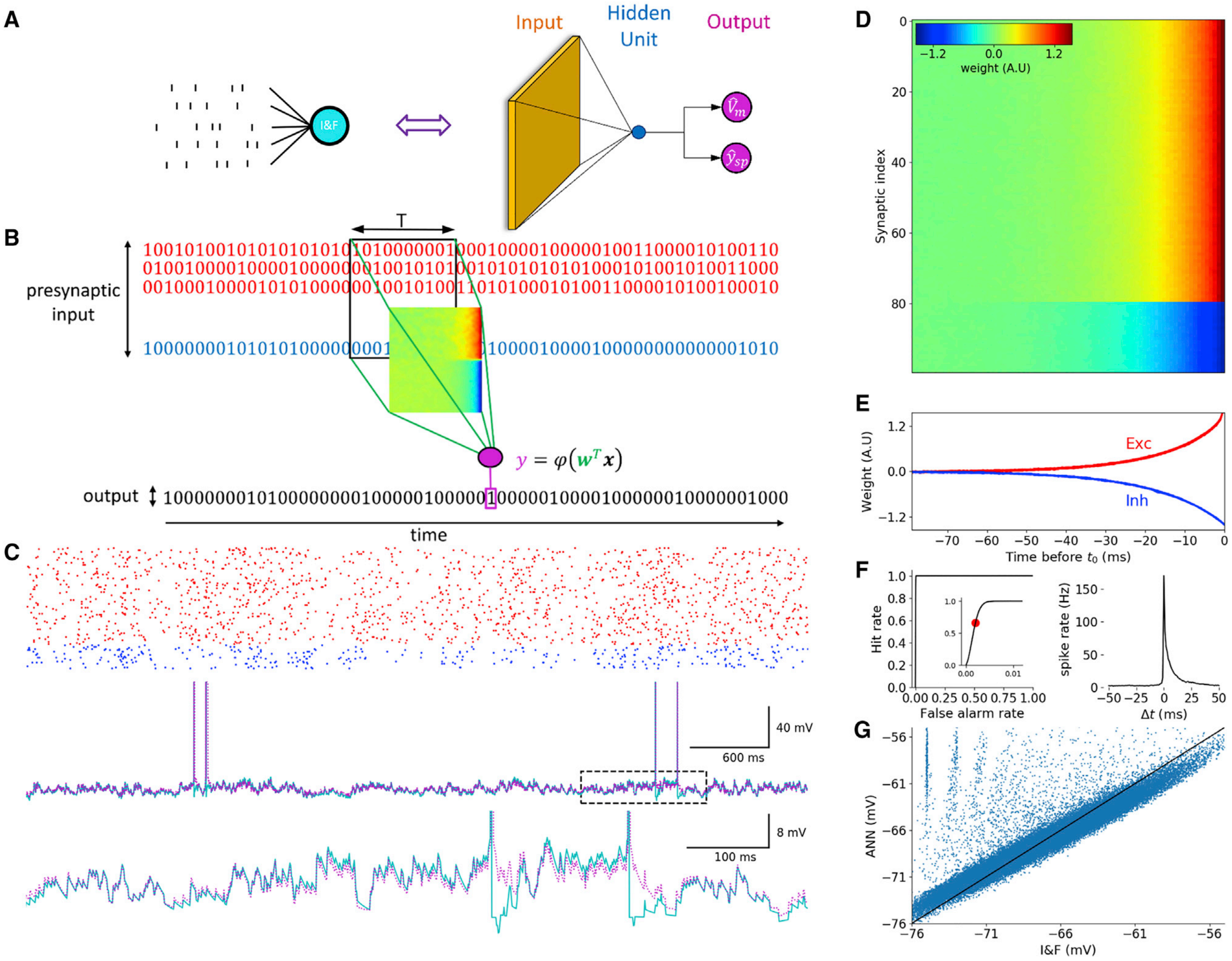

**Figure 1. The integrate and fire (I&F) neuron model is faithfully captured by a NN with one hidden layer consisting of a single hidden unit**

(A) Illustration of an I&F neuron model receiving a barrage of random synaptic inputs and generating voltage and spiking output (left) and its analogous DNN (right). Orange, blue, and magenta represent the input layer, the hidden layer, and the DNN output, respectively.

(B) Schematic overview of our prediction approach. The objective of the DNN is to predict the spike output of the respective I&F model based on its synaptic input. The binary matrix, denoted by x, represents the input spikes in a time window T (black rectangle) preceding $t_0$. x is multiplied by the synaptic weight matrix, w (represented by the heatmap image), and summed up to produce the activation value of the output unit, y. This value is used to predict the output (magenta rectangle) at $t = t_0$. Excitatory inputs are denoted in red and inhibitory in blue. Note that unlike the I&F, the DNN has no *a priori* information about the type of the synaptic inputs (excitation or inhibition).

(C) Top: example inputs (red, excitatory; blue, inhibitory) presented to the I&F neuron model. Middle: response of the I&F model (cyan) and the analogous DNN (magenta). Bottom: zoom in on the dashed-rectangle region in the top trace. Note the great similarity between the two traces.

(D) Learned weights of the DNN modeled synapses. The top 80 rows are excitatory synapses to the I&F model; the bottom 20 rows are its inhibitory synapses. Columns correspond to different time points relative to $t_0$ (rightmost time point). The prediction probability for having a spike at $t_0$ increases if the number of active excitatory synapses increases just before $t_0$ (red) and the number of active inhibitory synapses decreases (blue).

(E) Temporal cross-section of the learned weights in (D).

(F) Left: receiver operator characteristic (ROC) curve of spike prediction. The area under the curve (AUC) is 0.997, indicating high prediction accuracy at 1 ms precision. Inset: zoom in on up to 1% false alarm rate. The red circle denotes the threshold selected for the DNN model shown in (C). Right: cross-correlation between the I&F spike train (ground truth) and the predicted spike train of the respective DNN, when the prediction threshold was set to 0.2% false-positive (FP) rate (red circle in left plot).

(G) Scatterplot of the predicted DNN subthreshold voltage versus ground-truth voltage produced by the I&F model.

predicted spike train and the target I&F simulated spike train (the "ground truth"). The cross-correlation shows a sharp peak at 0 ms and has a short (~10 ms) half-width, suggesting high temporal accuracy of the DNN. We also quantified the DNN performance in predicting the subthreshold membrane potential by using standard regression metrics, and, in Figure 1G, depict the scatterplot of the predicted voltage versus the ground-truth simulated output voltage. The root mean square error (RMSE) is 1.73 mV (79.8% variance explained), indicating a good fit between the I&F and the respective DNN. Note that high accuracy on both binary spike prediction and continuous somatic voltage is a dual prediction attempt achieved with only a single hidden unit that is enforcing a strict bottleneck. This is possible only due to the strong relationship between outputs spikes and somatic voltages in the I&F model case.

In conclusion, as a proof of concept, we have demonstrated that a very simple DNN can learn the I/O transformation of an I&F model with a high degree of temporal accuracy. Importantly, the resulting weight matrix (the filter) obtained by the learning process is interpretable, as it represents known features of the I&F model, including the existence of two classes of inputs (excitatory and inhibitory), the convolution of the synaptic inputs with the exponential decay representing the passive membrane properties (resistance, capacitance), and the transformation from subthreshold membrane potential to spike output.

## Analogous DNN for the full complexity of the L5 cortical pyramidal neuron model

We next applied our paradigm to a morphologically and electrically complex detailed biophysical compartmental model of a 3D-reconstructed L5 cortical pyramidal cell (L5PC) from rat somatosensory cortex (Figure 2A). The model is equipped with complex nonlinear membrane properties, a somatic spike generation mechanism, and an excitable apical nexus capable of generating calcium spikes (Hay et al., 2011; Larkum et al., 1999; Schiller et al., 1997). The excitatory synaptic inputs are mediated through both voltage-independent AMPA-based conductance and voltage-dependent NMDA-type conductance (Jahr and Stevens, 1993); the inhibitory inputs are mediated through conductance-based $GABA_A$-type synapses. Both excitatory and inhibitory synapses are uniformly distributed across the dendritic tree of the model neuron (see STAR Methods and Figures S1 and S4 for more details). The training data consisted of a combined total of more than ~200 h of simulated time, with excitatory and inhibitory inputs randomly activated in time according to a Poisson distribution with a firing rate consisting of a piecewise constant temporal trajectory.

A thorough search of configurations of deep and wide fully connected neural network (FCN) architectures have failed to provide a good fit to the I/O characteristics of the L5PC model. These failures suggest a substantial increase in the complexity of I/O transformation compared to that of I&F model. Indeed, only a TCN architecture with seven layers (depth), 128 channels per layer (width), and T = 153 ms (history), provided a high precision fit (Figures 2B, 2C, and S2). The example in Figure 2C shows that this TCN can predict the somatic subthreshold voltage and spikes of a highly complex neuron with high precision when provided with a previously unseen input pattern from the test set. Although this was the first configuration of a network that met our criterion for a fit, we consequently managed to find other DNNs that provided comparable results. An extended analysis of the DNN depth, width, and time-window history required to replicate the I/O of this L5PC model faithfully is presented below.

It is important to note that the model's accuracy was relatively insensitive to the temporal kernel sizes of the different DNN layers when keeping the total temporal extent of the entire network fixed. Therefore, the first layer's temporal extent was selected to be larger than the subsequent layers, mainly for visualization purposes (see Figure 2G–2I). Figure 2H shows a filter from a unit in the first layer of the DNN. This filter is somewhat similar to the filter in Figure 1D but integrates only basal and oblique subtrees and ignores the apical tree's inputs. Moreover, the filters have different shapes, representing the differential contribution of inputs arriving at different distances from the soma as predicted by cable theory for dendrites (Rall 1967). Figure 2I, however, shows a filter of another unit that, in contrast to the filter in Figure 2H, has negligible weights assigned for basal and oblique dendrites but a very strong apical tuft dependency. By examining additional first layer filters (not shown), we found a wide variety of different activation patterns that the TCN utilized as an intermediate representation, including many temporally directionally selective filters (similar to those shown in Figure 5D below). Figures 2D–2F show the quantitative performance evaluation of this DNN model. For binary spike prediction (Figure 2D), the AUC is 0.9911. For somatic voltage prediction (Figure 2E), the RMSE is 0.71 mV, and 94.6% of the variance is explained by this model. Note that despite its seemingly large size, the resulting TCN represents a substantial decrease in computational resources relative to a full simulation of a detailed biophysical model (involving numerical integration of thousands of nonlinear differential equations), as indicated by a speedup of simulation time by several orders of magnitude.

## NMDA synapses are major contributors to the I/O complexity ("depth") of L5PCs

Now that we have obtained a DNN model that can replicate the I/O relationship of a detailed biophysical/compartmental model of a real neuron very accurately, can we learn from it what the essential features that contribute to neuron complexity are? Detailed studies of synaptic integration in dendrites of cortical pyramidal neurons suggested the primary role of the voltage-dependent current through synaptic NMDA receptors, including at the subthreshold and suprathreshold (NMDA spike) regimes (Polsky, Mel, and Schiller 2004; Branco, Clark, and Häusser 2010). As NMDA receptors (NMDARs) depend nonlinearly on the voltage, they are highly sensitive not only to the activity of the synapse in which they are located but also to the activity of (and the voltage generated by) neighboring synapses and their dendritic location. Moreover, the NMDA current has slow dynamics, promoting integration over a time window of tens of milliseconds (Doron et al., 2017; Jahr and Stevens, 1993; Major et al., 2013). Consequently, we hypothesized that removing NMDA-dependent synaptic currents from our L5PC model will significantly decrease the size of the respective DNN to achieve similar levels of accuracy, implying a reduction in the complexity of the I/O transformation.

In Figure 3, we present the results of a new set of simulations where the NMDA voltage-dependent conductances were removed, such that the excitatory input relies only on AMPA-mediated conductances. We compensated for the significant reduction of excitatory current resulting from the NMDA removal by adjusting the input firing rate of the AMPA-type synapses to maintain the same average output firing rate of the L5PC neuron model as in Figure 2 (see precise details in Figure S4). The figure shows that for this model, we have managed to achieve a similar quality fit as in Figure 2 with a much smaller (and shallower)

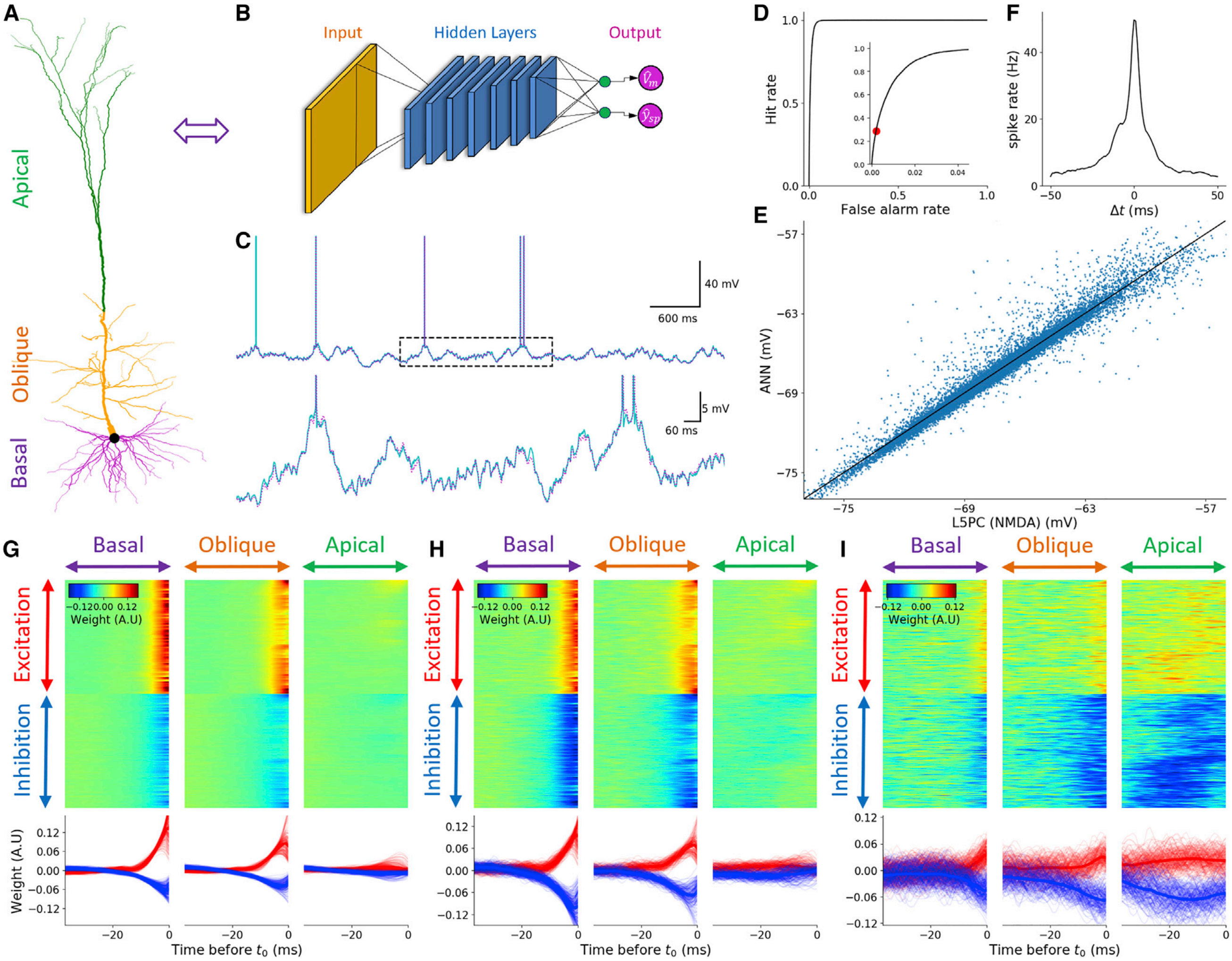

**Figure 2. A detailed model of an L5 cortical pyramidal neuron with AMPA and NMDA synapses is faithfully captured by a TCN with seven hidden layers consisting of 128 feature maps per layer and a history of 153 ms**

Figure360▷ For a Figure360 author presentation of this figure, see https://doi.org/10.1016/j.neuron.2021.07.002.

(A) The modeled L5PC. Basal, oblique, and apical dendrites are marked in purple, orange, and green, respectively.

(B) Analogous DNN with seven hidden layers. Orange, blue, and magenta circles represent the input layer, the hidden layer, and the DNN output, respectively. Green units represent linear activation units (see STAR Methods).

(C) Top: example voltage response of the L5PC model with AMPA and NMDA synapses (cyan) and the analogous DNN (magenta) to random synaptic input. Bottom: zoom in on the dashed-rectangle region in the top trace.

(D) ROC curve of spike prediction; the AUC is 0.9911, indicating high prediction accuracy at 1 ms precision. A zoom in on up to 4% false alarm rates is shown in the inset. The red circle denotes the threshold selected for the DNN model shown in (B).

(E) Scatterplot of the predicted DNN subthreshold voltage versus ground-truth voltage.

(F) Cross-correlation plot between the ground truth (L5PC model with AMPA and NMDA synapses) spike train, and the predicted spike train of the respective DNN, when the prediction threshold was set according to the red circle in (D).

(G) Learned weights of a selected unit in the first layer of the DNN. The top left, center, and right panels show inputs located on the basal dendrites, oblique dendrites, and the apical tuft, respectively. For each case, excitatory synapses are shown in the top half of the rows, whereas inhibitory synapses are shown in the bottom half. Different columns correspond to different time points relative to $t_0$ (rightmost time point). Bottom: temporal cross section of the learned weights above.

(H) Similar to (G), first layer weights for a different unit in the first layer but with a different spatiotemporal pattern.

(I) An additional unit that is weakly selective to whatever happens in the basal dendrites, weakly sensitive to oblique dendrites, but very sensitive to apical tuft dendrites. The output of this hidden unit is increased when there are apical excitation and a lack of apical inhibition in a time window of 40 ms before $t_0$. Note the asymmetry between the amplitudes of the temporal profiles of excitatory and inhibitory synapses, indicating that inhibition decreases the activity of this unit more than excitation increases it.

See also Figure S1 for more details about simulation.

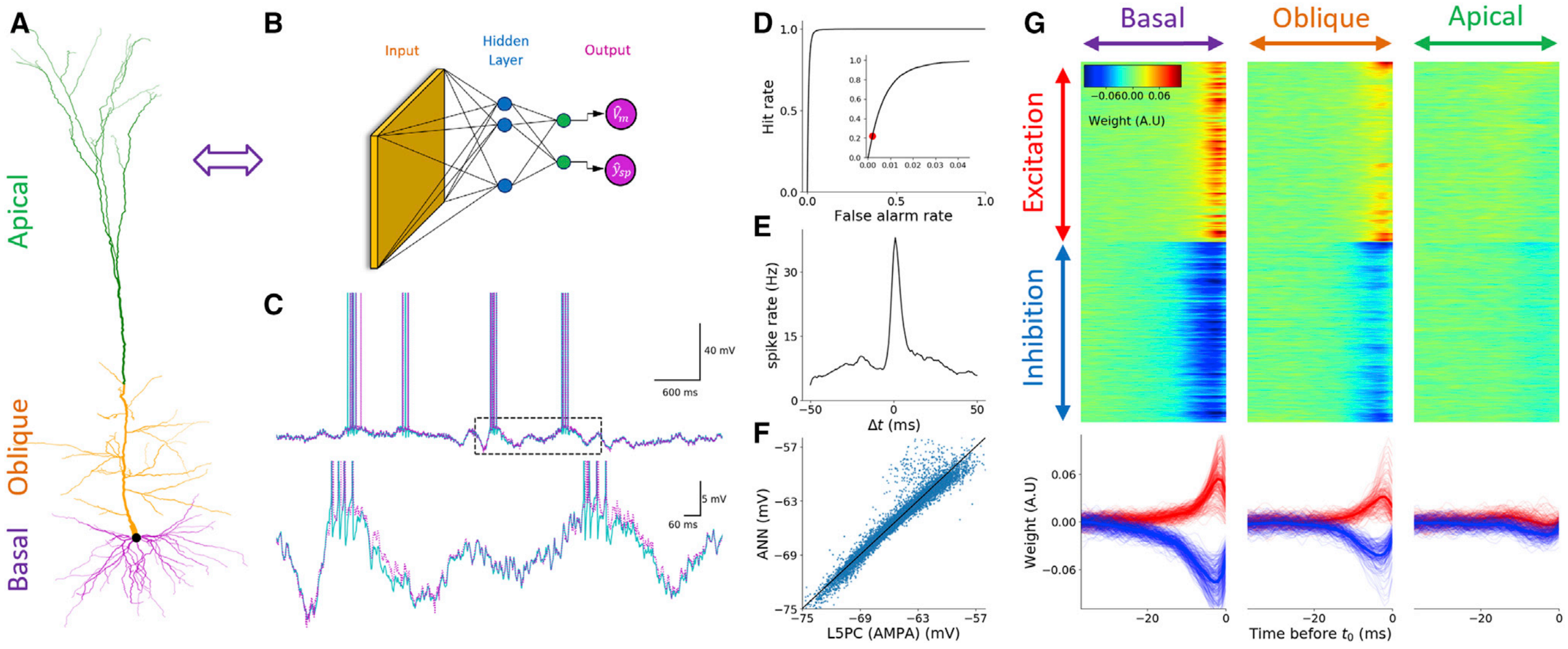

**Figure 3. A detailed model of an L5PC neuron with AMPA synapses is faithfully captured by a DNN with one hidden layer consisting of 128 hidden units**

(A) Illustration of the L5PC model. Excitation in this model is mediated only by AMPA conductances.

(B) Analogous DNN. As in Figure 2, orange, blue, and magenta circles represent the input layer, the hidden layer, and the DNN output, respectively. Green units represent linear activation units.

(C) Top: response of the L5PC model (cyan) and the analogous DNN (magenta) to random AMPA-based excitatory and $GABA_A$-based inhibitory synaptic input (see STAR Methods). Bottom: zoom in on the dashed-rectangle region in the top trace. Note the great similarity between the two traces.

(D) ROC curve of spike prediction; the AUC is 0.9913, indicating high prediction accuracy at 1 ms precision. Inset: zoom in on up to 4% false alarm rates. The red circle denotes the threshold selected for the DNN model shown in (B).

(E) Cross-correlation plot between the ground truth (L5PC model response) and the predicted spike train of the respective DNN for prediction threshold indicated by the red circle in the left plot.

(F) Scatterplot of the predicted DNN subthreshold voltage versus ground-truth voltage.

(G) Learned weights of selected units in the DNN, separated by their morphological (basal, oblique, and apical) location. Like in Figure 2, in each case, excitatory synapses are shown in the top half of the rows, and the bottom half shows inhibitory synapses. As in Figure 1D, different columns correspond to different time points relative to $t_0$ (rightmost time point). Note that just before $t_0$, the output of this hidden unit increases if the number of active excitatory synapses increases at the basal and oblique dendrites (red), whereas the number of active inhibitory synapses decreases (blue) at these locations. However, this unit is nonselective to activity at the apical tuft, indicating the lack of influence of the tuft synapses on the neuron's output by this unit.

See also Figures S3 and S4 for more details about simulation and fitting results with a larger TCN.

network. The network consists of a fully connected DNN (FCN) with 128 hidden units and only a single hidden layer (Figure 3B) and T = 43 ms (history). This significant reduction in complexity is due to the ablation of NMDA channels. Also, in our DNN training attempts, we have failed to achieve a good fit when using the smaller architecture that was successful for the I&F model neuron shown in Figure 1. This indicates that whereas the DNN model for L5PC is greatly simplified in the absence of NMDA conductance, additional neuronal mechanisms still contribute to the richness of its I/O transformation of the L5PC as compared to that of the I&F neuron model.

Figure 3C shows an exemplar test trace for the DNN illustrated in Figure 3B, whereas Figure 3H depicts a representative exemplar of the weight matrix for one of the hidden units of the DNN. By examining the filters of the hidden layer of the DNN, we observed that the weights representing inputs to the oblique and basal dendrites had profiles that resemble postsynaptic potentials (PSPs; mirrored in time). Interestingly, the weights associated with synapses of the apical tuft are essentially zero. This pattern remains consistent for all first layer filters of the network, implying that, for this model, the apical dendritic synapses had negligible information regarding predictions of the output spikes of the neuron, even in the presence of calcium spikes occasionally occurring in the nexus. Contrasting this filter with the one presented in Figure 2I suggests that the NMDA nonlinearity greatly assists in the activation of apical tuft dendrites. Figures 3D–3F show the quantitative performance evaluation. For binary spike prediction (Figure 3E), the AUC is 0.9913. For somatic voltage prediction (Figure 3F), the RMSE is 0.58 mV, and 95.0% of the variance is explained by this model.

In order to systematically compare the complexity of DNNs for the AMPA-only case (Figure 3) versus the AMPA and NMDA case (Figure 2), we selected a minimal approximation threshold (AUC = 0.9910) as a "good enough" performance threshold for spiking accuracy of the DNN as compared to the spiking of the respective biophysical model. DNNs that performed better than this threshold were considered to be "a good approximation." We then asked what is the minimal sized network that satisfies this threshold for AMPA-only case versus the AMPA and NMDA case. Toward this end, we trained a total of 137 DNNs (62 for AMPA only and 75 for AMPA and NMDA) while varying three main hyperparameters: depth, width [number of channels per

A

| ANNs that met minimal test set performance threshold (Spiking AUC > 0.9910) | | | | | |
|---|---|---|---|---|---|
| Criterion | Biophysical Model Description | Neural Network hyperparameters | | | Unseen test performance |
| | | Depth | Width | Time window history (ms) | Spiking AUC |
| Minimal Depth | L5PC w/ AMPA synapses | **1** | 128 | 43 | 0.9913 |
| | L5PC w/ AMPA & NMDA synapses | **5** | 128 | 205 | 0.9913 |
| Minimal Width | L5PC w/ AMPA synapses | 2 | **4** | 125 | 0.9919 |
| | L5PC w/ AMPA & NMDA synapses | 7 | **128** | 153 | 0.9911 |
| Minimal History | L5PC w/ AMPA synapses | 4 | 32 | **17** | 0.9930 |
| | L5PC w/ AMPA & NMDA synapses | 8 | 256 | **97** | 0.9914 |
| Best Performance | L5PC w/ AMPA synapses | 4 | 64 | 120 | **0.9959** |
| | L5PC w/ AMPA & NMDA synapses | 8 | 256 | 161 | **0.9924** |

B

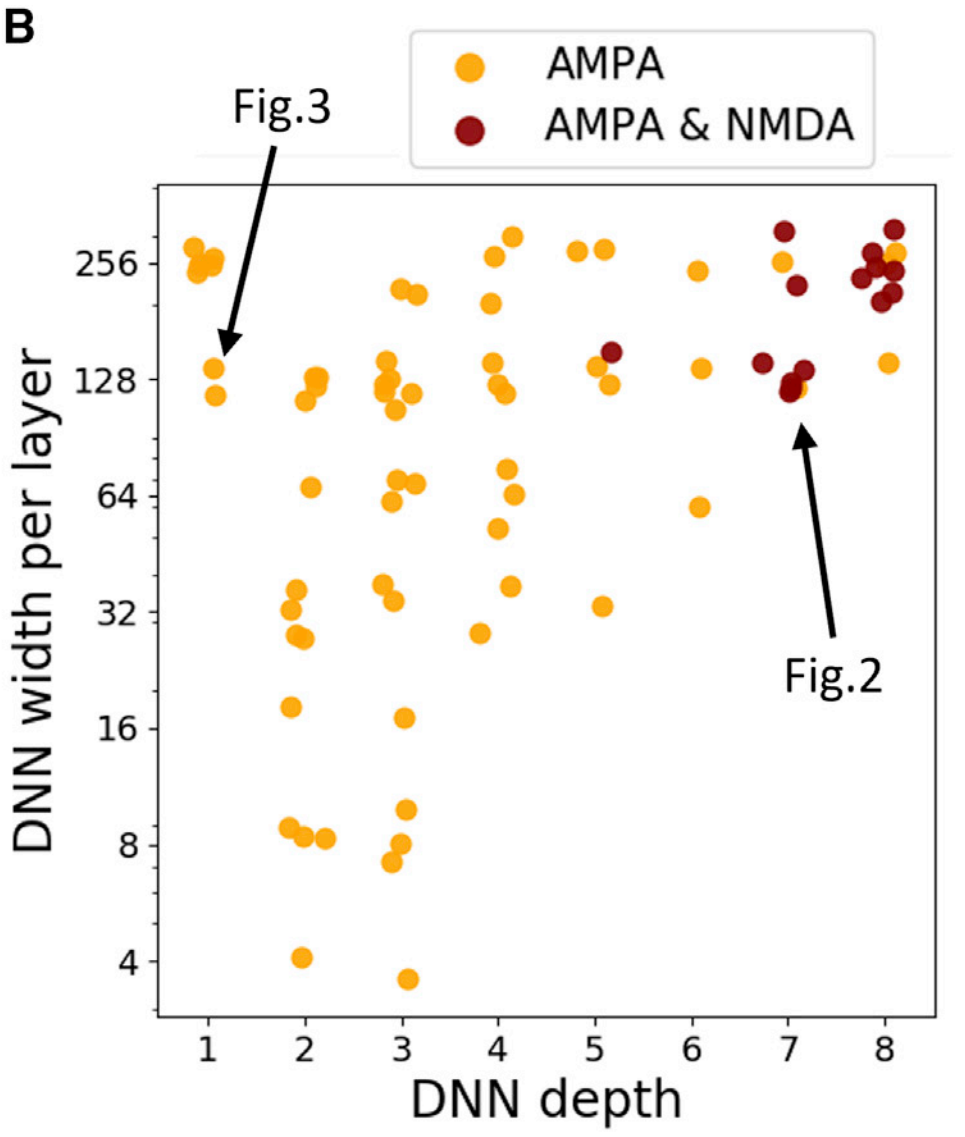

**Figure 4. The minimal DNN size required to achieve a good fit is larger for AMPA and NMDA synapses compared to AMPA-only synapses across all tested hyperparameters**

(A) Deep networks of various depths, widths, and maximal temporal history dependencies were fitted on two L5PC models, one with AMPA and NMDA synapses and the other with only AMPA synapses. The table summarizes the results by showing the minimal depth, width, and temporal dependence extent DNNs for each of the two biophysical model variations that successfully passed a minimal "good approximation" threshold of AUC = 0.9910. A noticeably clear trend arises that, with NMDA-based synapses, DNNs of depth 5–8 are required to achieve this threshold criterion, whereas shallower DNNs are required for the case of AMPA-only synapses.

(B) 2D scatterplot jointly depicting depth and width of DNNs that meet minimal performance criteria as in (A) for the two cases among all attempted networks for AMPA-only synapses (orange) and AMPA and NMDA synapses (dark red). A small Gaussian noise was added to depth and width axes for visualization purposes to avoid overlap of the respective points. Arrows indicate the networks utilized in Figures 2 and 3.

See also Figure S2 for more details.

layer], and temporal extent of the input history used to make the prediction. Figure 4A depicts a table that summarizes the results. It is clear that in order to reach similar levels of accuracy, much smaller DNNs are required for the AMPA case for all three hyperparameters considered. Figure 4B depicts a 2D scatterplot of both depth and width of DNNs that meet the minimum performance threshold among all attempted hyperparameter configurations. It is evident that only networks with large depths and widths can achieve this minimal level of accuracy for the AMPA and NMDA case (dark red dots), whereas much smaller (shallower) networks are sufficient for the AMPA-only case (orange dots). Note that although the precise details of the results will vary when selecting a different AUC threshold value, the overall trend stays the same. For an extended comparison, see Figure S2.

To investigate further the relationship between NMDARs and their interaction with dendritic morphology, as well as the influence of NMDAR density on I/O transformation complexity, we conducted a set of additional analyses. As shown in Figures S5A–S5C, we examined the I/O complexity when the AMPA and NMDA synapses were placed only on part of the L5PC morphology. We examined four such cases whereby the synapses impinge on different regions of the dendritic tree; these cases are strictly encompassed within each other so that we have a clear axis of complexity (full morphology, excluding tuft synapses, synapses on the basal tree only, and synapses placed only proximal compartments of the basal tree). Figure S5G shows a summary of these results, clearly indicating a reduction in I/O complexity as morphology containing synapses is progressively more restricted. In Figures S5D–S5F, we examine the interaction between the case where NMDA synapses are placed on proximal and distal parts of the basal tree as compared to the respective AMPA-only case. Figure S5H shows a summary of these results, which indicate that there exists an additive interaction between synapse type and morphology, specifically that segregated dendritic compartments with synapses are harder to model also in the AMPA-only case; the nonlinearities of NMDAR synapses in this case add an additional complexity to the modeling effort. In Figure S6, we examine the dependence of the I/O relationship on the maximal NMDA conductance. It is evident (Figure S6D) that even a small increase in NMDA conductance has a significant impact on I/O complexity and that the I/O complexity increases as the NMDAR density increases further, albeit with "diminishing returns."

## DNN analysis of a single dendritic branch provides new insights for the contribution of NMDA conductance to the computational complexity of neurons

To deepen the understanding of the contribution of NMDA synapses to the computational complexity of neurons, we next studied a simplified case. Here, the NMDA synapses were activated

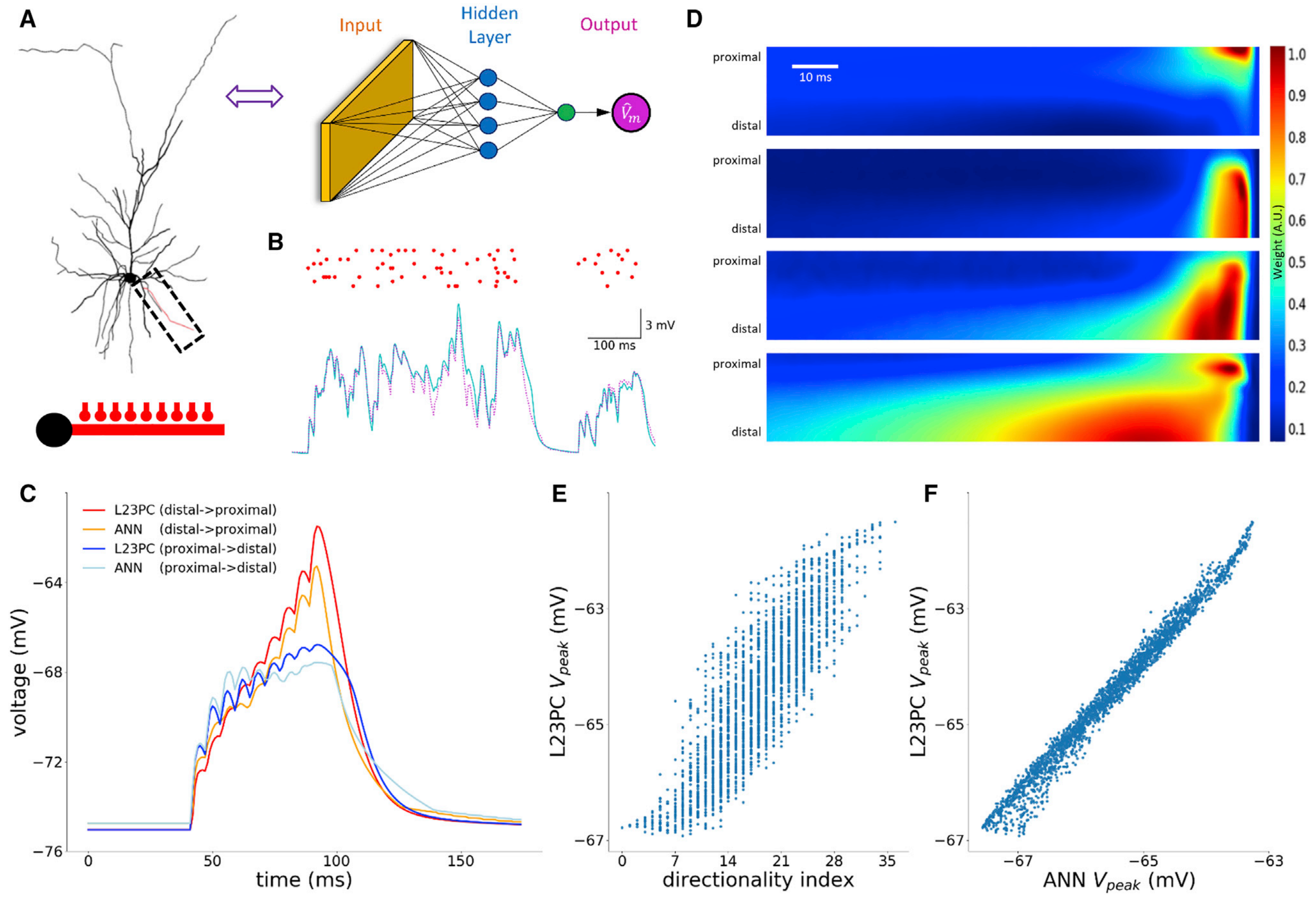

**Figure 5. Analysis of a DNN that successfully replicates the I/O of a single L23PC dendritic branch receiving NMDA synapses reveals spatio-temporal pattern matching with four distinct convolutional filters**

(A) Left: a layer 2/3 pyramidal neuron was used in the simulations with a zoom in on one selected basal branch (dashed rectangle). This modeled dendritic branch receiving nine excitatory synapses, depicted schematically by the "ball and stick" model at the bottom, was also used in a previous study (Branco et al., 2010). Right: illustration of the analogous DNN that was trained on random synaptic inputs impinging on this basal dendrite. Colors as in Figure 2A.

(B) Example of the somatic voltage response (cyan) and DNN predicted output (magenta) to a random input spike pattern impinging on the basal dendritic branch (red dots above).

(C) Example of somatic response to two spatiotemporal sequences of synaptic activation patterns (red, distal-to-proximal direction; blue, proximal-to-distal direction) and the DNN predicted output for these same sequences (orange and light blue traces, respectively).

(D) Learned weights of the four hidden units consisting the respective DNN model. Heatmaps are spatiotemporal filters, as shown in Figures 1D and 2C. Note the direction-selective shapes and long temporal extent of influence by distal synaptic activations.

(E) Scatterplot that shows the discrimination ability between different temporal orders of synaptic activations on the modeled basal branch. The vertical axis is the ground-truth maximum voltage at the soma during a specific synaptic order of activation. The horizontal axis is the directionality index proposed in Branco et al. (2010). The correlation coefficient is 0.86.

(F) Same as (E), but showing the DNN estimation of the maximum voltage of the respective order of activation. The correlation coefficient is 0.99.

See also Figure S7.

along a single basal branch of a layer 2/3 pyramidal cell (L23PC) from the mouse visual cortex, as in the experimental and modeling study of Branco et al. (2010). This modeled dendrite received a random activation of nine excitatory synapses uniformly distributed across the dendritic length. The temporal activation of each synapse followed a Poisson distribution, and the instantaneous firing rate was identical for all synapses (Figure 5A). We found that the output of this single dendritic branch is faithfully captured by a single layer of a fully connected DNN with four hidden units (Figures 5A and 5B). Examining the four filters of the first layer reveals interesting shapes that make intuitive sense, as first explored by the pioneering theoretical studies of Rall (1964). The topmost filter in Figure 5D appears to be summing only very recent and proximal dendritic activation. The second-from-top hidden unit sums up recent distal dendritic synaptic inputs. The third filter clearly shows a direction-selective hidden unit, preferring patterns in which synaptic activation is temporally activated sequentially from distal to proximal, whereas the last hidden unit responds to a prolonged distal dendrite summation of activity combined with precisely timed proximal input activation. Thus, the rich and complex integration of inputs on this dendritic branch, which involved the contribution of the nonlinear NMDA current, can be conceptualized as pattern matching of a set of four specific spatiotemporal templates.

To further examine the generalization capability of the DNNs, Figure 5C examines the special cases studied by Branco et al. (2010) of a sequential temporal activation of the nine synapses once in the distal-to-proximal direction and, conversely, in the proximal-to-distal direction. Importantly, our DNN network was trained on random synaptic activation patterns and was not exposed to these highly organized spatiotemporal input patterns during training. However, the DNN successfully replicated the response of the L23PC modeled branch to these spatiotemporal sequences. Figure 5E shows our reconstruction of the results of Branco et al. (2010), whereby a directionality index was suggested as a predictor for the peak somatic voltage for a random activation sequence of the nine input synapses. Figure 5F shows the much-improved prediction of the respective DNN for the same activation sequences as in Figure 5E. It is important to note that the special case of nine synaptic activations equally spaced in time is highly unlikely to occur during the random input stimulation regime that was used to train the DNN. Nevertheless, as shown in Figure 5, the network can generalize even to this new input regime with high precision.

These results illuminate the interpretability power of our approach. By examining the four kernels (filters), we provide an intuitive (Figure 5D), yet powerful (Figure 5F), interpretation for the complex process of nonlinear spatiotemporal synaptic integration in a single dendrite with NMDA synapses. It also further demonstrates the ability of our DNN models to generalize to previously unseen input patterns (out-of-distribution generalization), as will be discussed below.

## Generalization of single-neuron analogous DNNs to spatiotemporal structured inputs

The analogous DNN for the L5PC with NMDA synapses shown in Figure 2 was trained on a large set of synaptic inputs that were uniformly distributed across the dendritic trees and randomly activated in time (see STAR Methods). However, how well does this DNN capture the case of spatially clustered and temporally synchronized inputs that may give rise to highly nonlinear dendritic phenomena (e.g., NMDA spike)? Figure 6 shows that this DNN generalizes very well to a wide range of spatiotemporally structured stimulation protocols without re-training. Figure 6A depicts the case whereby excitatory and inhibitory synapses impinge on restricted subtrees of the modeled cell (purple dendritic regions); in this case, the temporal patterns of the instantaneous input rates are similar to those during the training of the DNN shown in Figure 2, and only the spatial pattern is altered. The voltage response traces of the L5PC model (in cyan) and the analogous DNN (magenta) are shown at right. Comparisons of the output of the L5PC model and that of the respective DNN for additional spatiotemporal structured synaptic input cases are depicted in Figures 6B–6D, including situations with synchronous temporal input patterns (Figures 6C and 6D) and without inhibition (Figures 6B and 6D). In all cases, a close similarity between the subthreshold voltage responses and spiking activity was found (see Figures S8B and S8C for aggregate details for all the above-mentioned conditions and combined simulation test time of $\sim$230 min). We conclude that, albeit trained on a set of input synapses that were activated randomly in time and uniformly distributed over the L5PC model dendrites, this training input set was sufficiently rich so that the respective DNN for this modeled cell successfully captures (generalizes well) the I/O of the L5PC model for a wide range of spatially and temporally structured inputs. We note that the ability to generalize to different input statistics, as depicted in Figure 6, is greatly dependent on the DNN size. Indeed, we have found that the deeper the analogous DNN is, the better it generalizes (see Figures S8E and S8F and Table S1 for a full comparative analysis).

# DISCUSSION

Recent advances in the field of DNNs provide, for the first time, a powerful general-purpose tool that can learn complex mappings from examples. In this study, we used these tools to study the I/O mappings of single complex nonlinear neurons at millisecond temporal resolution. We constructed a large dataset of pairs of (synaptic) input and (axonal) output examples by simulating a neuron model of L5PC receiving a rich repertoire of synaptic inputs over its dendritic surface and recorded its spike output at millisecond temporal resolution, as well as its somatic subthreshold membrane potential. We then trained networks of various configurations on these I/O pairs until we obtained an analogous ‘‘deep’’ network with close performance to that of the neuron’s detailed simulation. We applied this framework to a series of neuron models with various levels of morpho-electrical complexity and obtained new insights regarding the computational complexity of cortical neurons.

For simple I&F neuron models, our framework provides simple analogous DNNs with one hidden layer consisting of a single hidden unit that captures the full I/O relationship of the model. The respective DNN filters provided key biophysical insights that are consistent with our understanding of the parameters shaping the I/O relationship of I&F models (Figure 1). In the case where only a single basal dendritic branch of a cortical neuron, consisting of NMDA synapses, was modeled, a shallow (one hidden layer) DNN with only a few units was required to capture a different aspect of the spatiotemporal integration of synaptic inputs (Figure 5). Surprisingly, even a model of L5 cortical pyramidal neuron with the full complexity of its dendritic trees and a host of dendritic voltage-dependent currents and AMPA-based synapses is well captured by a relatively simple network with a single hidden layer (Figure 3). However, in a full model of an L5 pyramidal neuron consisting of NMDA-based synapses, the complexity of the analogous DNN is significantly increased; we found a good fit to the I/O of this modeled cell when using a TCN that has five to eight hidden layers (Figures 2 and 4; see also Figures S2 and S3). Furthermore, a seven-layer analogous DNN for a L5PC that was trained on random inputs successfully generalizes to new out-of-distribution set of clustered and synchronous inputs (Figure 6).

These results suggest that the single cortical neuron with its nonlinear synaptic inputs is already, on its own, a sophisticated computational unit. Consequently, cortical networks built from such units are deeper and computationally more powerful than they seem to be just based on their anatomical (pre- to post-synaptic) connections. Importantly, the implementation of the I/O

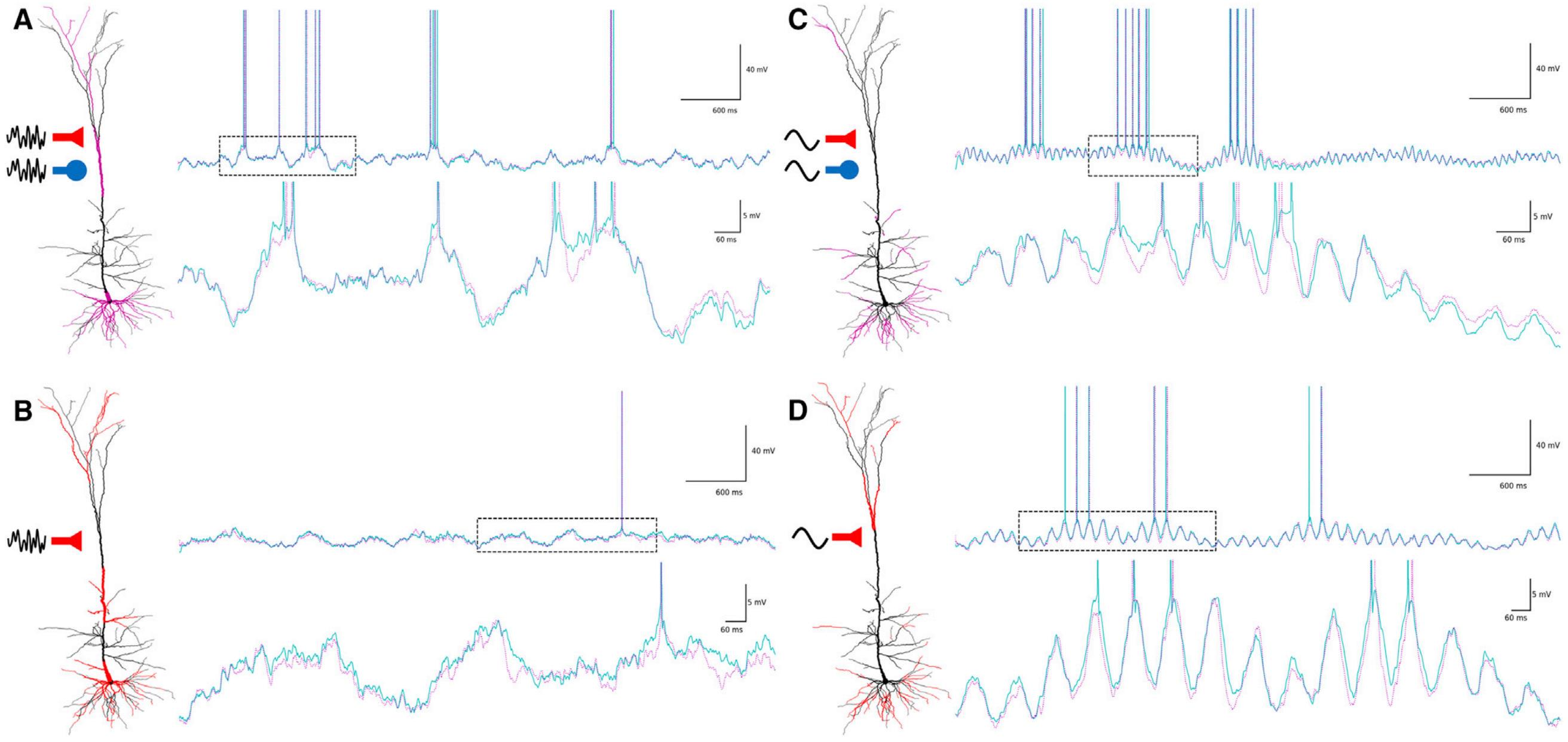

**Figure 6. The seven-layer analogous DNN for L5PC that was trained on random inputs successfully generalizes to new out-of-distribution clustered and synchronous inputs**

(A) Left: modeled L5PC with AMPA and NMDA synapses as in Figure 2 with excitatory (red) and inhibitory (blue) synapses impinging only on a specific subtree (purple dendritic regions). Instantaneous input rates for the excitatory (E) & inhibitory (I) synapses are similar to that during the training of the analogous DNN. Top right: voltage response traces of the modeled cell (cyan) and the analogous DNN (magenta). Bottom right: zoom in on the dashed-rectangle region in the top trace.

(B) As in (A), but this time with excitatory-only synapses impinging on subparts of the dendritic tree (red regions). The temporal profile of instantaneous input rates is like that given during training.

(C) As in (A), but with E&I synapses impinging on different subtrees (purple regions) of the modeled L5PC. Instantaneous rates of excitatory and inhibitory synapses are driven by a synchronous sinusoidal pattern that was not seen during training.

(D) As in (A), but here, only excitatory synapses are impinging on the dendritic tree (red regions) and are activated with synchronous sinusoidal instantaneous input rates.

See also Figure S8.

function of single neurons using a DNN also provides practical advantages. It is computationally much more efficient than the traditional compartmental model, which required the solution of many thousands of partial differential equations (PDEs) per neuron. Indeed, for the full model of L5PCs, we obtained a speedup of ~2,000× when using the DNN instead of its compartmental-model counterpart. However, we did not perform a comprehensive empirical or theoretical comparison of the computational advantage of our DNN method for different neuron models, different input conditions, or different hardware or software versions, as this lies beyond the scope of the present study. In general, the computational speedup of our approach results from (1) compression/simplification (via the computational nodes and layers of the DNN) of all dendritic computations performed in the cell, keeping only the computations that are relevant for output spike generation; and (2) utilization of specialized graphics processing unit (GPU) hardware and a corresponding software ecosystem, which are continuously developed for the increasingly growing needs of the deep-learning community. Together, our tool can potentially be useful for simulating large-scale realistic neuronal networks (Egger et al., 2014; Markram et al., 2015). Furthermore, the size of the respective DNN for a given neuron could be used (under certain assumptions [see below]) as an index for its computational power; the larger it is, the more sophisticated computations this neuron could perform. Such an index will enable a systematic comparison between different neuron types (e.g., CA1 pyramidal cell, cortical pyramidal cell, and Purkinje cell) or the same type of cell in different species (e.g., mouse versus human cortical pyramidal cells).

It is important to emphasize that for optimization reasons, the complexity of the analogous DNN described above is an upper bound of the true computational complexity of the I/O of the respective single neuron, i.e., it is possible that there exists a smaller DNN that mimics the biophysical neuron with a similar degree of accuracy but the training process we used did not find it. Additionally, we note that we have limited our architecture search space only to FCN and TCN neural network architectures. It is likely that additional architectural search could yield simpler and more compact models for any desired degree of prediction accuracy. Nevertheless, we stress that this upper bound is in fact several orders of magnitude computationally less intensive when compared to the detailed biophysical simulations. In addition, the results presented in Figure 1 clearly indicate that our proposed method would have been able to uncover simple I/O relationships if the I/O transformation of cortical neurons was in fact rather simple. This suggests that the DNN size that we have found might be a relatively tight upper bound. In order to facilitate this search in the scientific community, we have released our large readymade dataset of simulated inputs and outputs of a fully complex single L5 cortical neuron in an *in-vivo*-like regime so that the community can focus on modeling various aspects of this endeavor and avoid re-running the simulations themselves.

The analysis of DNNs is a challenging task and a rapidly growing field (Mahendran and Vedaldi, 2015; Mordvintsev et al., 2015; Olah et al., 2017). Nevertheless, observing the weight matrix of units (filters) in the first layer of the respective DNN is straightforward and can provide ample insights regarding the I/O transformation of the neuron. The full network can be interpreted as consisting of a basic set of filters that span the space of possible spatiotemporal patterns of synaptic inputs that will drive the original neuron to spike. The first layer defines this space, and the rest of the network mixes and matches within that space. For example, as shown in Figure 3, in the case of a pyramidal neuron without NMDA synapses, most filters have significant weights only for basal and apical oblique inputs, and the weight given for apical tuft synapses is negligible (despite the existence of voltage-dependent $Ca^{2+}$ and other nonlinear currents in this model, which results in occasional $Ca^{2+}$ spikes in the distal dendritic tuft). The picture is fundamentally different when NMDA synapses were included in the model. In this case, the weights assigned to apical dendrite synapses are significant (Figure 2; see also Figure S6). Moreover, the filters devoted to these apical inputs tend to have a temporal structure that is significantly broader (in time) than for that of the proximal synapses, suggesting that the temporal precision of input to the apical synapses is less important for spike generation in the soma. These are basic insights that could be drawn by observing first layer filters of the resulting analogous DNN.

This work opens multiple additional avenues for future research. One important direction is isolating the contribution of specific mechanisms to the computational power of the neuron in a similar way to that performed here for the NMDA-based receptors (see also Figure S2). Fitting a DNN while manipulating specific voltage-dependent dendritic currents (e.g., voltage-gated calcium channels [VGCCs], voltage-gated potassium [Kv] channels or hyperpolarization-activated cyclic nucleotide–gated [HCN] channels), will provide a deeper understanding of their contribution to the overall synaptic integration process and to the complexity of the respective DNN. An additional interesting direction is to utilize this work to explore how real neurons can use their rich biophysical repertoire in order to perform specific computations from the class computed by the equivalent DNNs. By taking advantage of gradient-descent optimization and specialized GPU hardware acceleration, one can efficiently train the DNN representing the neuron to compute an interesting, meaningful function (e.g., training it to classify images of handwritten digits or to classify sequences of auditory sounds). Then it might be possible to map this DNN back to the original biophysical neuron model. One can then both directly validate the hypothesis that single neurons could perform complex and useful computational tasks and investigate how these neurons and specific spatiotemporal distribution of synapses can actually implement such tasks.

If indeed one cortical neuron is equivalent to a multilayered DNN, then what are the implications for the cortical microcircuit? Is that circuit merely a deeper classical DNN composed of simple "point neurons"? A key difference between the classical DNN and a cortical circuit composed of deep neurons is that, in the latter case, synaptic plasticity can take place mainly in the synaptic (input) layer of the analogous DNN for a single cortical neuron, whereas the weights of its hidden layers are fixed (and dedicated to represent the I/O function of that single cortical neuron). It is important to note, however, that there are other forms of (non-synaptic) plasticity in neurons, such as branch-specific plasticity or intrinsic plasticity (Losonczy et al., 2008) that can, perhaps, shape the weights of deep layers of the analogous DNN in addition to synaptic efficacies. Taken together with the myriad of recurrent connections and network motifs between cortical neurons of different types (Markram et al., 2015), we hereby propose a concrete, biologically inspired network architecture for cortical networks that seamlessly incorporates single-neuron complexity. Focusing on architecture as a key element of research was recently advocated by (Richards et al., 2019). Indeed, the search for the appropriate architecture of artificial neural networks is one of the most rewarding avenues of machine learning today (He et al., 2015; Lin et al., 2014; Vaswani et al., 2017), and studying the specific architecture suggested in the present study may unravel some of the inductive bias hidden within the cortical microcircuit and harness it for future artificial intelligence (AI) applications.

## STAR★METHODS

Detailed methods are provided in the online version of this paper and include the following:

- KEY RESOURCES TABLE
- RESOURCE AVAILABILITY
  - Lead contact
  - Materials availability
  - Data and code availability
- METHOD DETAILS
  - I&F simulation
  - L5PC simulations
  - L23PC simulations
  - DNN fitting
  - Out-of-Distribution (OOD) simulations
- QUANTIFICATION AND STATISTICAL ANALYSIS
  - Model evaluation
  - Comparison of AMPA versus NMDA model complexity
  - Assessing the complexity resulting from the interaction between dendritic morphology and NMDA receptor density

## SUPPLEMENTAL INFORMATION

Supplemental information can be found online at https://doi.org/10.1016/j.neuron.2021.07.002.

## ACKNOWLEDGMENTS

We thank Oren Amsalem, Guy Eyal, Michael Doron, Toviah Moldwin, Yair Deitcher, Eyal Gal, and all lab members of the Segev and London Labs for many fruitful discussions and valuable feedback regarding this work. This work was supported by ONR grant N00014-19-1-2036, Israeli Science Foundation grant 1024/17 (to M.L.), and a grant from the Gatsby Charitable Foundation.

## AUTHOR CONTRIBUTIONS

D.B., conceptualization, methodology, investigation, visualization, software, validation, data curation, writing – original draft; I.S. and M.L., conceptualization, methodology, writing – review & editing, supervision, resources, funding acquisition.

## DECLARATION OF INTERESTS

The authors declare no competing interests.

## STAR★METHODS

### KEY RESOURCES TABLE

| REAGENT or RESOURCE | SOURCE | IDENTIFIER |
|---|---|---|
| Deposited data | | |
| Simulation data, pretrained models, and key results | This paper | https://doi.org/10.34740/kaggle/ds/417817 |
| Software and algorithms | | |
| NEURON 7.6.2 | NEURON | https://github.com/neuronsimulator/nrn |
| Python version 3.6 | Python | https://www.python.org |
| Custom Code for simulation, fitting and analysis | This paper | https://github.com/SelfishGene/neuron_as_deep_net |

### RESOURCE AVAILABILITY

#### Lead contact

Further information and requests for resources should be directed to and will be fulfilled by the lead contact, David Beniaguev (david.beniaguev@gmail.com)

#### Materials availability

This study did not generate new unique reagents.

#### Data and code availability

All data and pre-trained networks that were used in this work are available on Kaggle datasets platform (https://doi.org/10.34740/kaggle/ds/417817) at the following link:

https://www.kaggle.com/selfishgene/single-neurons-as-deep-nets-nmda-test-data

Additionally, the dataset was deposited to Mendeley Data (https://doi.org/10.17632/xjvsp3dhzf.2) at the link:

https://data.mendeley.com/datasets/xjvsp3dhzf/2

A github repository of all simulation, fitting and evaluation code can be found in the following link:

https://github.com/SelfishGene/neuron_as_deep_net.

Additionally, we provide a python script that loads a pretrained artificial network and makes a prediction on the entire NMDA test set that replicates the main result of the paper (Figure 2):

https://www.kaggle.com/selfishgene/single-neuron-as-deep-net-replicating-key-result.

Also, a python script that loads the data and explores the dataset (Figure S1) can be found in the following link: https://www.kaggle.com/selfishgene/exploring-a-single-cortical-neuron.

### METHOD DETAILS

#### I&F simulation

For Figure 1 simulations, membrane voltage was modeled using a leaky I&F simulation $V(t) = \sum_{i=1}^{N_{syn}} w_i \sum_{t_i} K(t - t_i)$, where $w_i$ denotes synaptic efficacy for each synapse, $t_i$ denotes presynaptic spike times, and $K(t - t_i)$ denotes the temporal kernel of each postsynaptic potential (PSP). We used a temporal kernel with exponential decay $K(t - t_i) = e^{-\frac{t-t_i}{\tau}} \cdot u(t - t_i)$ where $u(t)$ is the Heaviside function $u(t) = \begin{cases} 0, & |t < 0 \\ 1, & |t \geq 0 \end{cases}$ and $\tau = 20$ms is the membrane time constant. When the threshold was reached, an output spike was recorded, and the voltage was reset to $V_{rest} = -77$mV. As input to the simulated I&F neuron, $N_{exc} = 80$ excitatory synapses and $N_{inh} = 20$ inhibitory synapses were used. Synaptic efficacies of $w_{exc} = 2$mV were used for excitatory synapses and $w_{inh} = -2$mV for inhibitory synapses. Each presynaptic spike train was taken from a Poisson process with a constant instantaneous firing rate. Values used $f_{exc} = 1.4$Hz for excitatory synapses and $f_{inh} = 1.3$Hz for inhibitory synapses. The resulting output average firing rate for these simulation values was 0.9 Hz.

### L5PC simulations

For Figure 2 and Figure 3 simulations, we used a detailed compartmental biophysical model of cortical L5PC **as is**, modeled by Hay et al., 2011. For a full description of the model please see STAR Methods in the original paper. Briefly, this model contains in total 12 ion channels for each dendritic compartment. Some of the channels are unevenly distributed over the dendritic arbor. In Figure 3 double exponential conductances based AMPA synapses were used in simulations with $\tau_{rise} = 0.3$ms, $\tau_{decay} = 3$ms and $g_{max} = 0.4$nS. For Figure 2 and Figure 4, in related simulations we used the standard NMDA model (Jahr and Stevens, 1993), with $\tau_{rise} = 2$ms, $\tau_{decay} = 70$ms, $\gamma = 0.08$ mV$^{-1}$ and $g_{max} = 0.4$nS. For both Figure 2 and Figure 3, we also used double exponential $GABA_A$ synapses with $\tau_{rise} = 2$ms, $\tau_{decay} = 8$ms and $g_{max} = 1$nS on each independent dendritic segment, we placed a single AMPA (for Figure 3) or AMPA + NMDA (for Figure 2) synapse as well as a single $GABA_A$ synapse. In order to mimic uniform coverage of excitatory and inhibitory synapses on the entire dendritic tree, we stimulated each compartment with a firing rate proportional to the segment's length. Each presynaptic spike train was taken from a Poisson process with a smoothed piecewise constant instantaneous firing rate. The Gaussian smoothing sigma, as well as the time window of constant rate before smoothing were independently resampled for each 6 s simulation from the range 10ms to 1000 ms (Figure S1D). This was chosen, as opposed to a constant firing rate, to create additional temporal variety in the data in order to increase the applicability of the results to all possible situations. For Figure 2 simulations with NMDA synapses, the total amount of excitatory and inhibitory presynaptic spikes per 100 ms range between 0 and 800 spikes (Figure S1). This is equivalent to 8000 excitatory synapses with an average rate of 1 Hz and 2000 inhibitory synapses with an average rate of 4 Hz. The average output firing rate of the simulated cell across all simulations was 1.0 Hz. For Figure 3 simulations, with AMPA only synapses, the total amount of excitatory and inhibitory presynaptic spikes per 100 ms range were increased in order to account for the smaller amount of total current injected due to lack of NMDA current, with the purpose to achieve similar output firing rates of 1.0 Hz. See Figure S4 for detailed comparison.

### L23PC simulations

For Figure 4 simulations, we used a detailed compartmental biophysical model of cortical L23PC **as is**, modeled by Branco et al., 2010. In these experiments we stimulated a single branch with 9 dendritic segments with an NMDA synapse on each compartment, with parameters as in the simulation for Figure 4. The branch was selected as in Branco et al., 2010 in order to perform a comparison with the original paper. Similarly, to Figure 2 and Figure 3 simulations, each presynaptic spike train was taken from a Poisson process with a smoothed piecewise constant instantaneous firing rate. The number of presynaptic input spikes to the branch per 100 ms ranged between 0 and 15 in simulations used for training. In Figures 5C, 5E, and 5F, we repeated input stimulation protocol suggested by Branco et. al, 2010, consisting of single presynaptic spike per synapse with constant time intervals of 5 ms between subsequent synaptic activations, only randomly permuting the order of activation between trials.

### DNN fitting

In order to represent the input in a suitable manner for fitting with a DNN, we discretize time using 1 ms time bin $\Delta t$. Using this discretization, we can represent a spike train as a sequence of binary values $S[t]$, such that $S[t] \in \{0,1\}$, since the length of a spike is approximately 1 ms there cannot be more than a single spike in such a time interval. We denote the spike trains the neuron receives as input as $X[s,t], s \in \{1,2,\ldots,N_{syn}\}, t \in \{1,2,\ldots,T\}$, where $s$ denotes the synapse index, and $t$ denotes time. The spike trains a neuron emits as output we denote as $y_{spike}[t]$, The somatic voltage trance we denote as $y_{voltage}[t]$. For every point in time, we attempt to predict both somatic spiking $y_{spike}[t]$ and somatic voltage $y_{voltage}[t]$ based only a $T_{input}$ sized window of presynaptic input spikes. i.e., define the vector $\overrightarrow{x_{t_i}} = [X[s,t]], s \in \{1,2,\ldots,N_{syn}\}, t \in \{t_i, t_i-1, t_i-2, \ldots, t_i - T_{input}\}$ and a neural network that maps $\overrightarrow{x_t}$ to $\widehat{y}_{spike}[t]$ and $\widehat{y}_{voltage}[t]$. i.e., $\widehat{y}_{voltage}[t], \widehat{y}_{spike}[t] = DNN(\overrightarrow{x_t}, \theta)$. We treat spike prediction as a binary classification task and use standard log loss and treat voltage prediction as regression task and use standard MSE loss. We wish to find a model's parameters $\theta$ such that we minimize a combined loss $L(\theta) = L_{LogLoss}(y_{spike}, \widehat{y}_{spike}) + w_{voltage} \cdot L_{MSE}(y_{voltage}, \widehat{y}_{voltage})$, where $w_{voltage}$ is the relative importance of the spike prediction loss with respect to the somatic voltage prediction loss. For most of the experiments we set $w_{voltage}$ to be about half the size of the spike loss. The DNN architecture we used was a temporally convolutional network (TCN) (Bai et al., 2018) and we applied it in a fully convolutional manner on all possible time points. Note that when the temporal filter size after the first layer is 1 in a TCN applied as described, this is effectively a fully connected neural network. In most of our experiments we used fully connected neural networks, except for Figure 2 in which we used a proper TCN with a hierarchical convolutional structure. After every convolutional layer, a batch normalization layer immediately follows. We employed a learning schedule regime in which we lowered the learning rate and increased batch size as we progressed through training. Full details of the learning schedule in each case are in the attached code repository. For the generation of Figure S2 we trained many networks with different hyperparameters and trained each network for 2-14 days on a GPU cluster consisting of several V100, K80 and 2080Ti Nvidia GPUs. All results of the different hyperparameters and results can be found in the data link on the Kaggle platform. The total amount of single GPU years needed to fit all DNNs throughout the entire study was ~3.4 years.

### Out-of-Distribution (OOD) simulations

To find spatial clusters in a data-dependent manner, we applied a K-means clustering algorithm on the cross-correlation matrix of all dendritic voltages. We use k to be 64 different spatial clusters. During the simulations used to create the data for Figures 4 and S8, we randomly selected up to 30% of all spatial clusters for each simulation and stimulated only these clusters during that simulation. In 50% of the simulations, the temporal profile of instantaneous rates for all synapses was modulated by a sinusoidal function. The period of this temporal modulation was randomly selected to be between 15 ms and 300 ms for the duration of the simulation. Additionally, in 50% of the simulations we canceled all inhibition and kept only the excitation. The full OOD test set consisted of 2688 simulations, 6 s of simulation time each. The full details of this process can be found in the code repository on github and results of our spatial clustering can be found on the dataset on Kaggle. Together, these 3 manipulations (spatial clustering, temporal synchronization, and excitation only input) consist of large deviations from the statistics of the input during training and thus provide a robust OOD test.

## QUANTIFICATION AND STATISTICAL ANALYSIS

### Model evaluation

We divided our simulations to train, validation, and test datasets. We fitted all DNN models on the training dataset, and all reported results are on an unseen test dataset. A validation dataset was used for modeling decisions, hyperparameter tuning and snapshot selection during the training process (early stopping). We evaluated binary spike prediction results using the receiver operator characteristic (ROC) curve and calculated the area under the curve (AUC). We note that due to the relatively low firing rate of the neuron, the binary classification problem of the instantaneous spike prediction problem is highly unbalanced. For every second of simulation there was on average 1 positive sample (spike) for every 999 negative samples (non-spikes). Therefore, we used a very conservative threshold over the binary spike probability prediction output of the DNN in order to create the final spike train prediction and examine the cross-correlation plot. Note also that a prediction without a single True Positive on the 1 ms time binning binary spike prediction problem can still be in fact, a very good solution, e.g., if our model outputs, as its prediction, the exact same spike train as the original but offset by 1 ms in time. In this case, there will be no True positives and many False positives, but the predicted spike train is quite good nonetheless. To summarize: the temporal cross-correlation between the original and predicted spike trains is not directly related to binary prediction metrics used and therefore we display it as it's not redundant. For creating a binary prediction, we chose a threshold over the continuous model prediction that corresponds to 0.2% false-positive rate on the validation set after training (a different threshold was used for each model trained). In order to evaluate the temporal precision of the binary spike prediction we plotted the normalized cross-correlation between the predicted output spike train and the ground truth simulated spike train for a 50 ms temporal offset in either direction. To quantify the width of cross correlation plot, we fit a Gaussian to it and use the sigma parameter (see Table S1). In order to evaluate the voltage prediction, we calculated the RMSE and plot the scatterplot between predicted voltage and the ground truth simulated voltage.

### Comparison of AMPA versus NMDA model complexity

In order to systematically compare the complexity of DNNs for AMPA case (Figure 3) versus NMDA case (Figure 2), we select a minimal approximation threshold (AUC = 0.9910) for spiking accuracy of the DNN as compared to the spiking of the respective biophysical model. We consider DNNs that perform better than this threshold to be a good approximation. We then ask what is the minimal sized network that satisfies this threshold for AMPA versus NMDA cases. For this, we trained a total of 137 DNNs (62 for AMPA, 75 for NMDA) while varying 3 main hyperparameters - depth, width (number of channels per layer), and temporal extent of the input history required to make a prediction. For an extended comparison of these data please see Figures 4 and S2. For full fitting results with many additional various metrics for all 137 DNN fits, see released accompanying data. Although the quantitative results are dependent on the precise threshold used (AUC = 0.9910), they are mostly insensitive to small changes in this threshold.

### Assessing the complexity resulting from the interaction between dendritic morphology and NMDA receptor density

To investigate further the relationship between NMDAR and its interaction with morphology as well as the influence of NMDA receptor density on I/O transformation complexity, we conducted a set of additional analyses (Figures S5 and S6). When assessing the I/O complexity of NMDA synapses when placed only on part of the morphology, we perform simulations similar to those described previously, with the only difference being the part of the dendritic tree receiving synaptic input. In order to conduct a proper comparison, we tune the input firing rates such that the average output firing rate will be identical for all simulations (both test and train datasets). In this case we train DNNs of {1,2,3,7} layers for a fixed width (128) and fixed history dependence (100 ms). We limit training time to ~24 hours since our goal is to perform a quantitative comparison and not achieve maximally performance models like for the case shown in Figures 2 and 3. To establish a "morphological complexity" axis we select different regions of the dendritic tree of the modeled L5PC that are strictly contained within each other. These regions are: full morphology, full morphology excluding tuft synapses, the basal tree only, and only proximal compartments of the basal tree. Results depicted in Figures S5A–S5C, and summary results in Figure S5G. summary results depict the spiking AUC accuracy measure for the case of DNNs with 3 layers for all 4 morphological complexities. Higher I/O complexity is interpreted as corresponding to lower fit accuracy. Similarly, In Figures S5D–S5F and S5H we examine the interaction between the existence of NMDARs on proximal and distal parts of the basal tree as compared to AMPA only case. The only difference from what was described above for the morphological case is that now we fitted DNNs with a smaller fixed width (32 filters/layer). In a similar way, In Figure S6 we examine the dependence of the I/O relationship on varying the NMDAR max conductance level (which models receptor density). Here again we use 32 wide DNNs for this analysis.

## Supplemental information

# Single cortical neurons as deep artificial neural networks

**David Beniaguev, Idan Segev, and Michael London**

## SUPPLEMENTAL INFORMATION

## INVENTORY OF SUPPLEMENTAL INFORMATION

## SUPPLEMENTAL FIGURES

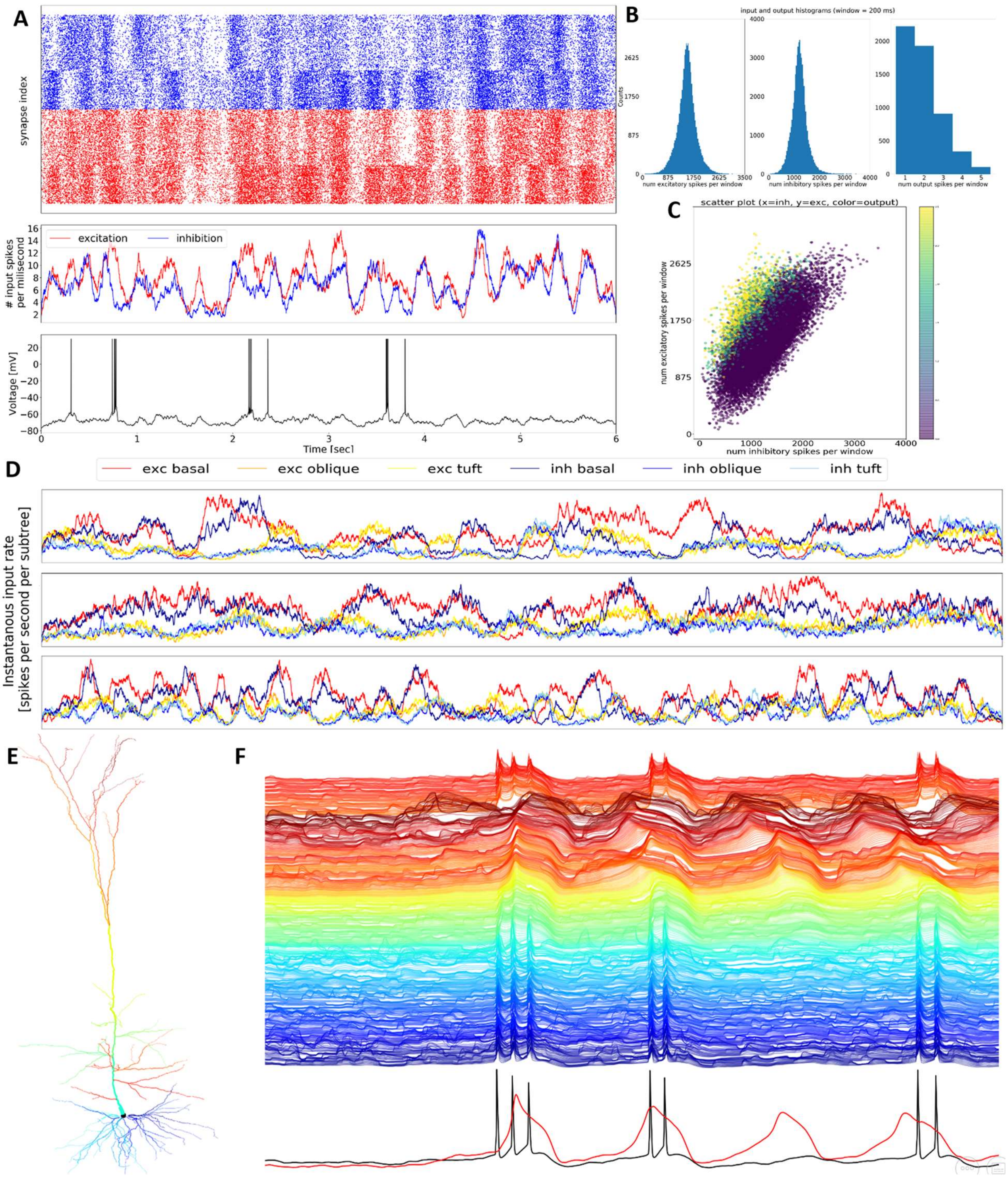

**Figure S1. Simulations Input-Output of an in vivo like regime of a detailed L5PC neuron model with NMDA synapses, Related to Figure 2**

(A) Inputs and outputs of a sample simulation. Top, raster plot of presynaptic input excitatory spikes (red) and inhibitory spikes (blue). Middle, total number of excitatory spikes per 1 millisecond. Smoothed with a Gaussian kernel with tau=20ms. Bottom, somatic voltage trace and spiking output of the same simulation (B) Histograms of number of excitatory input spikes (left), inhibitory input spikes (middle) received as input on the entire dendritic tree in a 200ms time window for the L5PC model with NMDA synapses. Right. Histogram of number of output spikes in the same time period.
(C) Scatter plot expanding the information presented in (B): horizontal axis is the number of inhibitory spikes, the vertical axis is the number of excitatory spikes and the color of each dot represents the number of output spikes for the same time window of 200ms.
(D) Example illustration of inputs, in units of spikes per millisecond per subtree and synapse type, for 5 different simulations. We can see a wide amount of variability in the input regimes of our simulations.
(E) Illustration of the simulated cell morphology.
(F) Top. Local dendritic voltage traces of all 639 simulated compartments. The colors of each trace are color-coded as the colors of the morphology illustration in (A). Bottom. Somatic (black) and nexus (red) voltage traces.

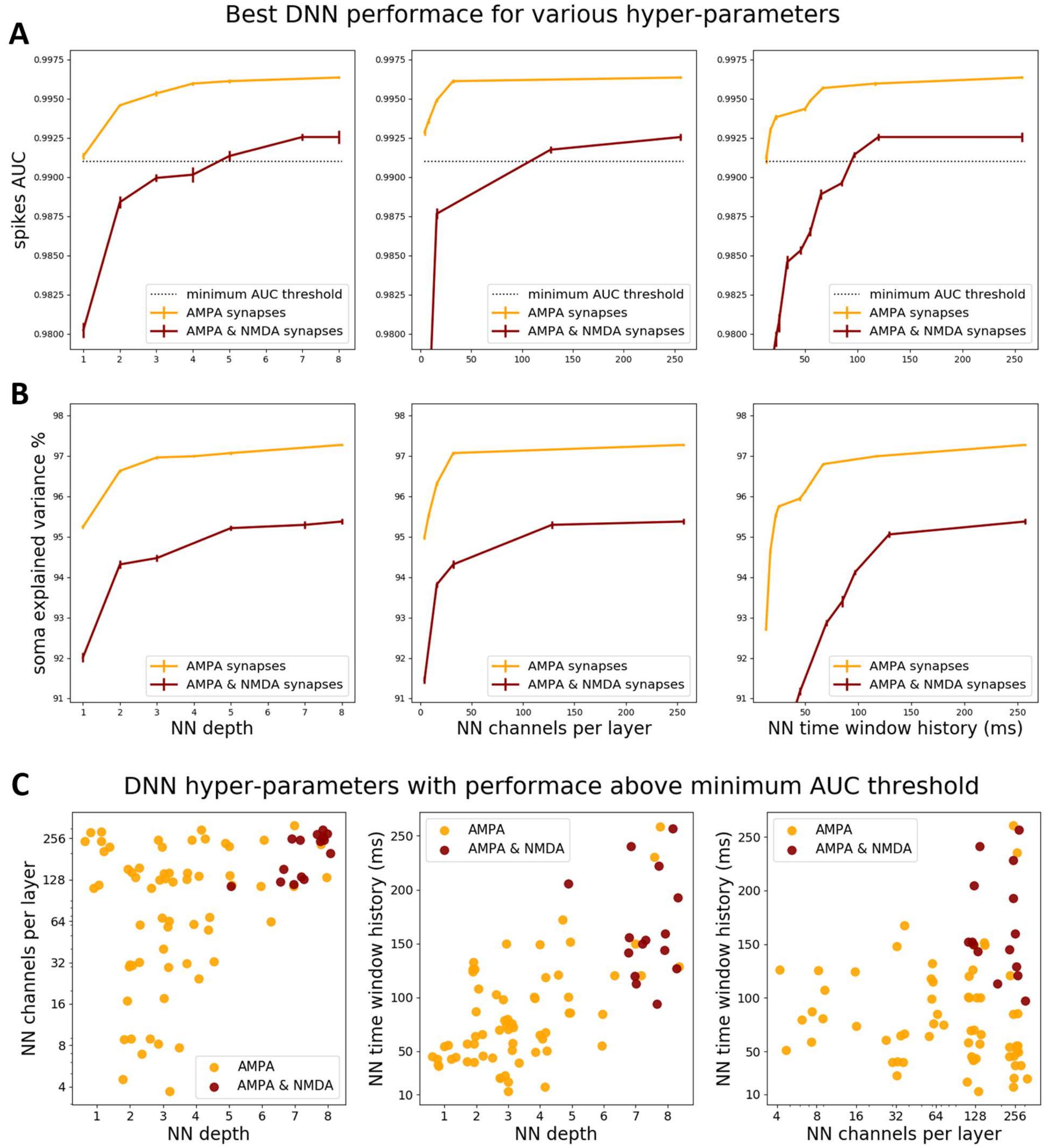

**Figure S2. The complexity of DNNs required to fit L5PC with NMDA synapses is substantially greater than L5PC with AMPA only synapses, Related to Figures 2, 3 & 4**

A hyper-parameter sweep of 137 network configurations was performed over 3 main hyperparameters that related to the functional complexity of the I/O relationship of a single neuron – depth, width, and the total history extent (the maximal receptive field size in the temporal domain of the DNN)

(A) Left, spiking prediction accuracy measure (AUC) as a function of the depth of the fitted DNN for the case of AMPA only (orange) synapses and AMPA & NMDA synapses (Dark red). The dashed line indicates the minimum AUC performance threshold. Error bars indicate +/- 2 standard deviations of fitting performance across different test data subsets. Middle, similar plot to the one on the left but for the width of the DNN (i.e., number of channels/feature maps). Right, like the previous two plots, but for the temporal extent of the input time window (i.e., effective "memory" of the neuron)
(B) Similar plots in (A), but now the vertical axis represents the somatic subthreshold prediction accuracy as depicted by the percent of variance explained.
(C) 2D scatter plots, this time only of models that were above the minimum AUC threshold. Left. Depth X Width scatter. Middle. Depth x History scatter. Right. Width x History scatter. It is evident that much smaller DNNs were able to pass the minimum performance threshold for the AMPA-only case (orange dots) compared to the NMDA case (dark red dots) for all 3 2D cross-sections of the data.
Note: each point of each of the individual curves in (A) and (B) represents the **best** network across all other architecture hyperparameters.
Also Note: Error bars in (A) and (B) represent variance associated with different subsets of test data (2 standard deviations) for the best network, and **not** random network initialization (which is awfully expensive to compute). The full data regarding all networks (137 networks total) trained and their performance details can be found in the accompanying released dataset on Kaggle (see **Code and data availability**).

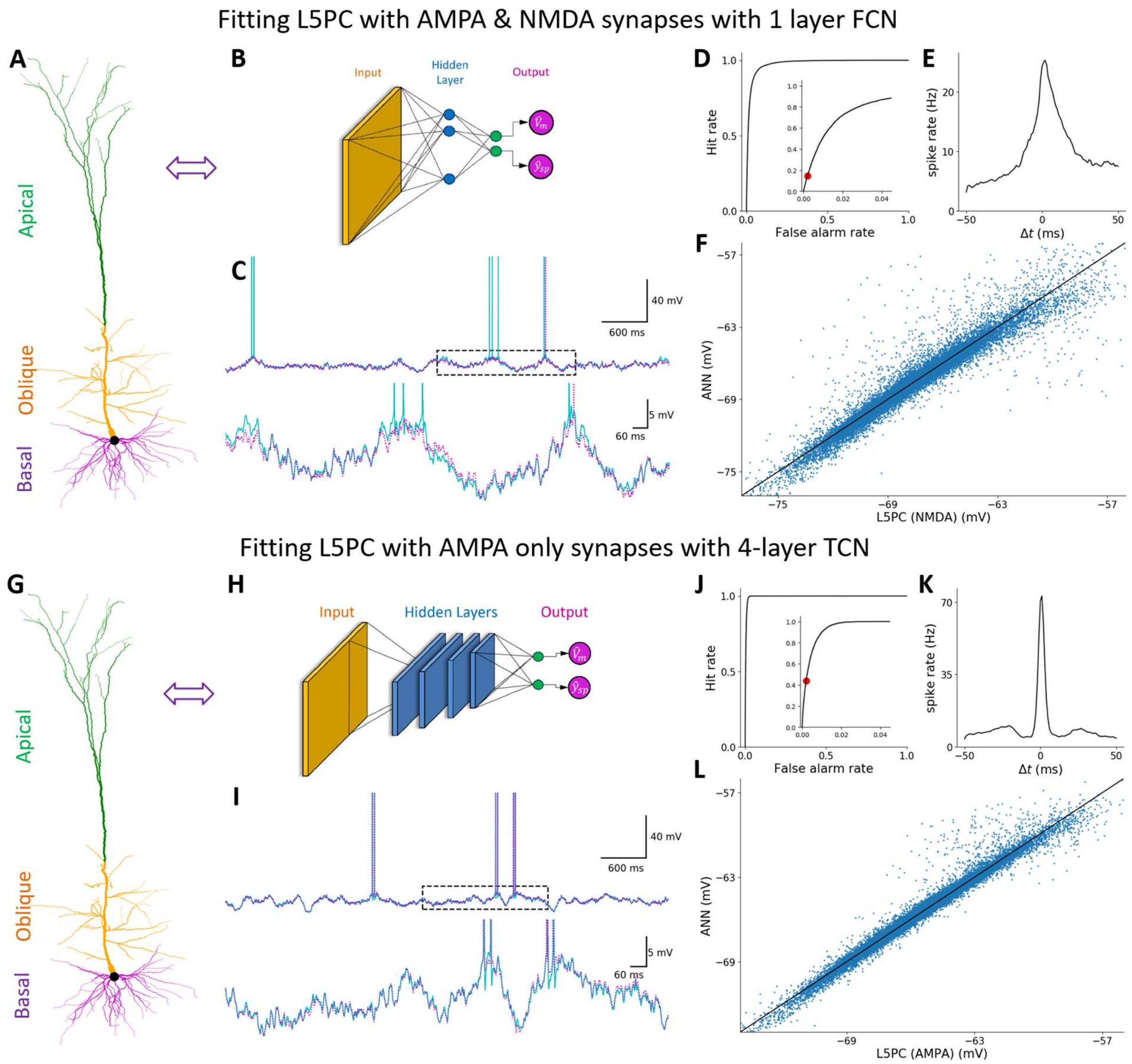

**Figure S3. A 1-layer FCN consisting of 128 hidden units fails to capture the spiking output of a detailed L5PC neuron model with NMDA synapses, while moderately large TCN fits a detailed L5PC neuron model with AMPA synapses with an extremely high degree of prediction accuracy, Related to Figures 2,4 & S2**

(A) Illustration of the L5PC model. Basal oblique and apical dendrites are marked by respective purple, orange and green colors.

(B) analogous DNN. Orange, blue and magenta circles represent the input layer, the hidden layer and the DNN output, respectively. Green units represent linear activation units. This is the best model among all 1-layer FCNs we attempted. (Detailed architecture: 1 hidden layer, 128 units per layer, time window extent of 43ms)

(C) Top. An exemplar voltage response of the L5PC model with NMDA synapses (cyan) and of the analogous DNN (magenta) to random synaptic input stimulation. Bottom. Zoom in on the dashed rectangle region in the top trace.

(D) ROC curve of spike prediction; the area under the curve (AUC) is 0.9769. Zoom in on up to 4% false alarm rates is shown in the inset. The red circle denotes the threshold selected for the model shown in (B).

(E) Scatter plot of the predicted DNN subthreshold voltage versus ground truth voltage.

(F). Cross-Correlation plot between the ground truth (L5PC with NMDA synapses) spike train and the predicted spike train of the respective DNN, when prediction threshold was set according to the red dot in (D).
(G-L). similar to (A-F) only for L5PC model with AMPA only synapses fitted by a moderately large neural network (detailed architecture: 4 temporally convolutional layers, 64 features maps per layer, time window extent of 120ms). The area under the ROC curve (AUC) is 0.9959.

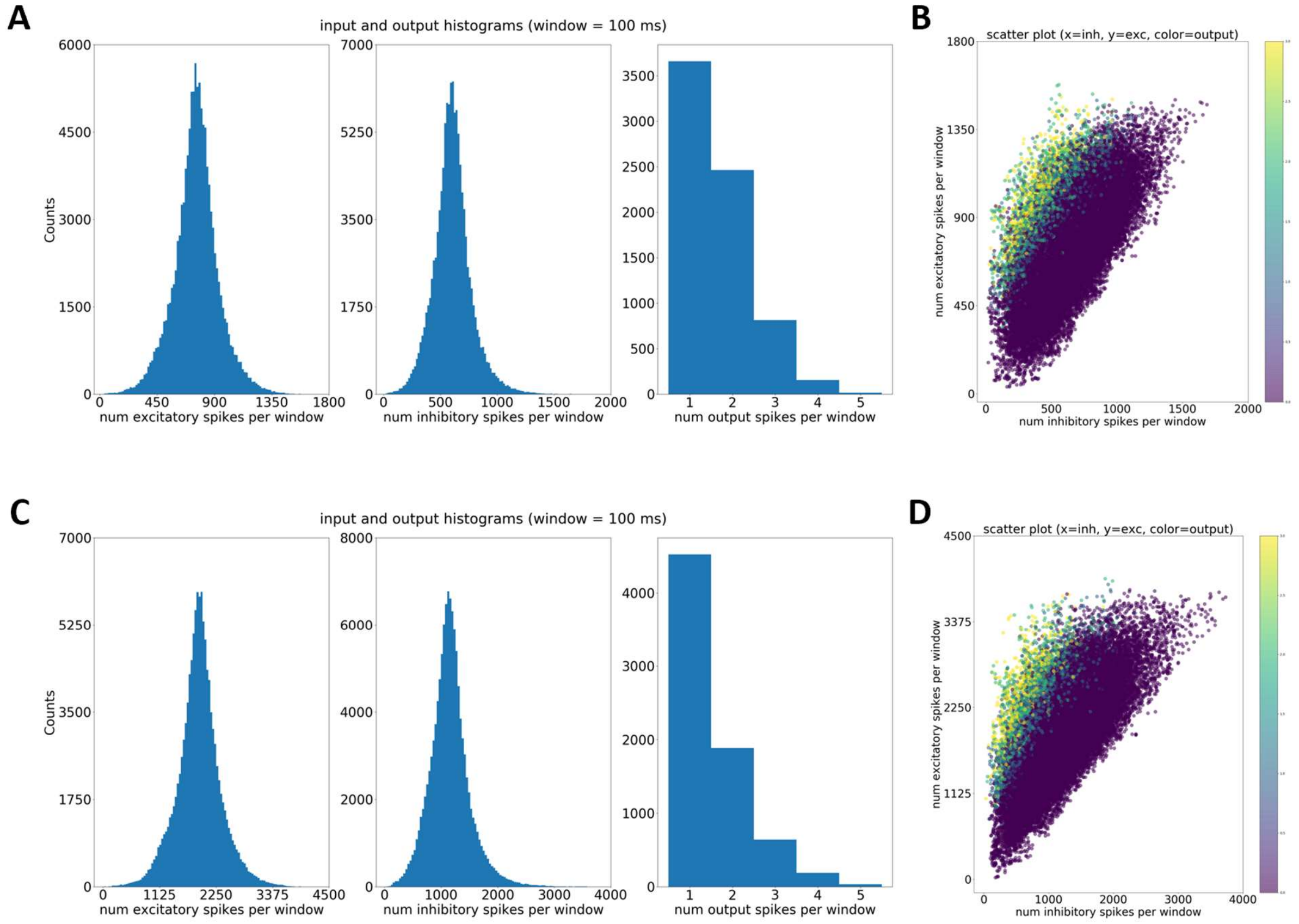

**Figure S4. Input-Output regimes of detailed L5PC model in the two different simulation cases: NMDA synapses and AMPA synapses, Related to Figures 2,3,4 and S2**

(A) Histograms of the number of excitatory input spikes (left), inhibitory input spikes (middle) received as input on the entire dendritic tree in a 100ms time window for the L5PC model with NMDA synapses. Right. Histogram of number of output spikes in the same time period.

(B) Scatter plot expanding the information presented in (A): horizontal axis is the number of inhibitory spikes, the vertical axis is the number of excitatory spikes and the color of each dot represents the number of output spikes for the same time window of 200ms.

(C-D) Similar to (A-B), but for the case of the L5PC model with AMPA only synapses.

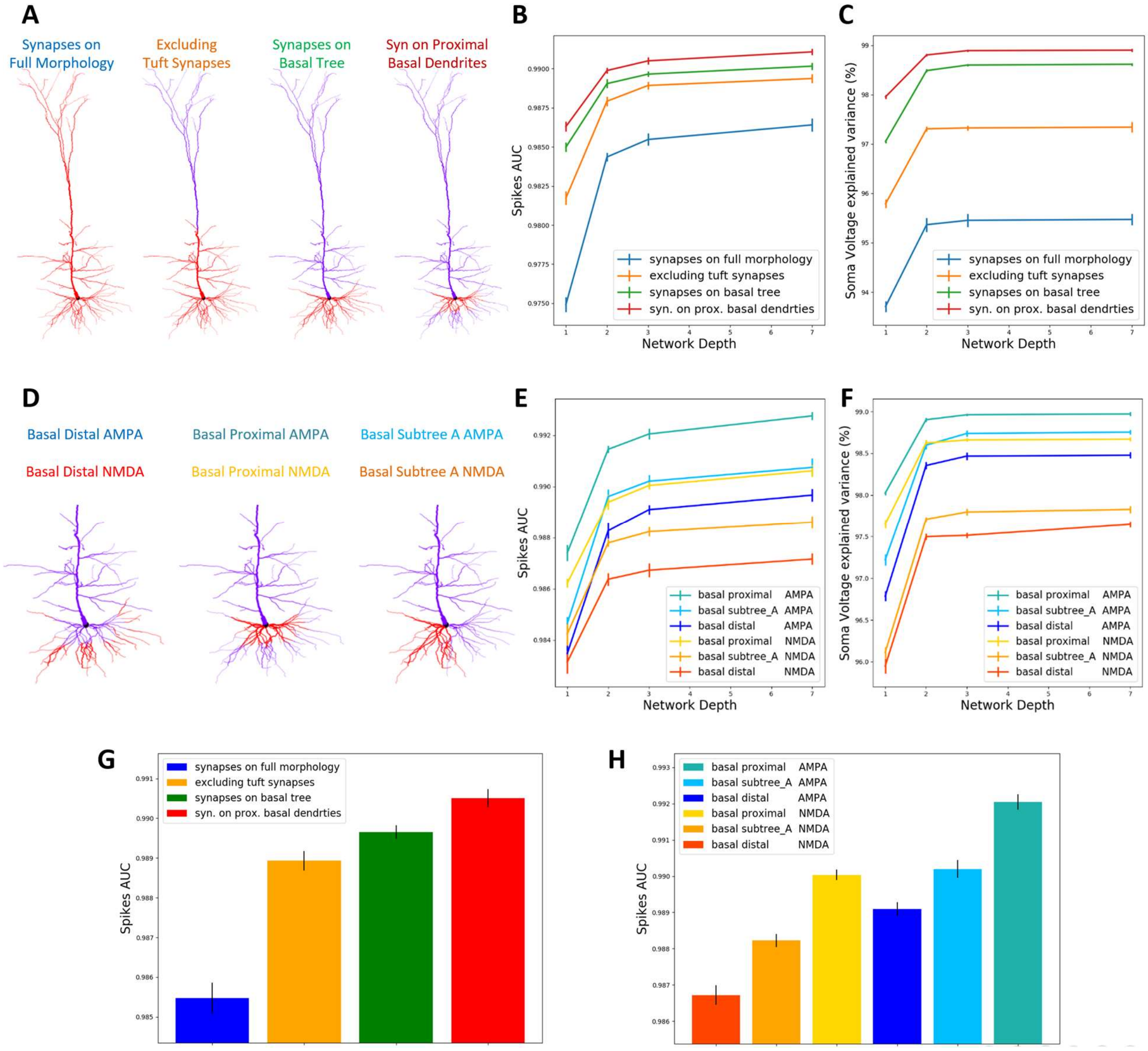

**Figure S5. Varying Morphological Complexity and Interaction Between morphology and synapse Type, Related to Figures 2,3,4 & S2**

(A) Illustration of 4 morphological regions (highlighted in red). For each region, we repeated the procedure discussed throughout the paper, this time stimulating only the highlighted regions. From left to right, the regions are basal and proximal sections only, the entire basal tree, the basal tree and the oblique dendrites but excluding the tuft, and lastly the full morphology of the neuron.

(B) For each morphological section depicted in (A) we fitted DNNs of depths {1,2,3,7} (and fixed width of 128 hidden feature maps per layer) to the data to access the complexity of the I/O mapping of that particular region. Results are shown for spike prediction. There is a clear trend of an increase in functional complexity that goes along with the increase in morphological complexity. Removal of the apical tuft appears to reduce complexity by a significant amount.

(C) Same as (B) only for somatic voltage prediction. Note that the full tree is present in the simulations, the figure zooms in on the relevant sections for illustration purposes.

(D) Illustration of 3 additional morphological regions (highlighted in red). Only zoomed in and relevant part of morphology is depicted. Left, the distal parts of the basal tree. Middle, proximal sections only of the basal tree. Right, basal subtree A. These three morphological sections have almost identical total dendritic lengths. Like in (A), for each illustration, we stimulate only the highlighted regions. In this experiment we additionally stimulated each region once with AMPA only synapses, and once with AMPA & NMDA synapses to assess the interaction between morphology and synapse type.

(E-F) Like (B-C), for each morphological section depicted in (D) we fitted DNNs of depths {1,2,3,7} (and fixed width of 32 hidden feature maps per layer) to the data to assess the complexity of the I/O mapping of that region. Cool colors depict cases with AMPA-only synapses. Warm colors depict cases with both AMPA and NMDA synapses. We see an effect of both synapse type and morphology, and an interesting interaction case for proximal sections and NMDA synapses for which although NMDA synapses exist, the complexity appears to be like other cases with AMPA only synapses. This is evident particularly in (F). It appears that NMDA synapses require segregated morphological sections for them to increase the I/O complexity.
(G) Summary bar plot of results depicted in (A-C). The vertical axis is the AUC performance of a 3-layer DNN taken from (B). The different bars correspond to different morphological sections. Color coding is identical to that in (A-C). There is a clear increase in accuracy as the morphology is reduced.
(H) Similar to (G), but for results depicted in (D-F). It is evident that distal dendrites as compared to proximal dendrites are more difficult to predict for both AMPA and NMDA cases accurately. Interestingly, NMDA synapses when present only on proximal synapses (yellow bar) are predicted to similar levels of accuracy as the AMPA cases (cool shaded bars). This indicates an interaction between synapse type and morphology. i.e., that segregated dendritic compartments are necessary to utilize the nonlinearities introduced by NMDARs.

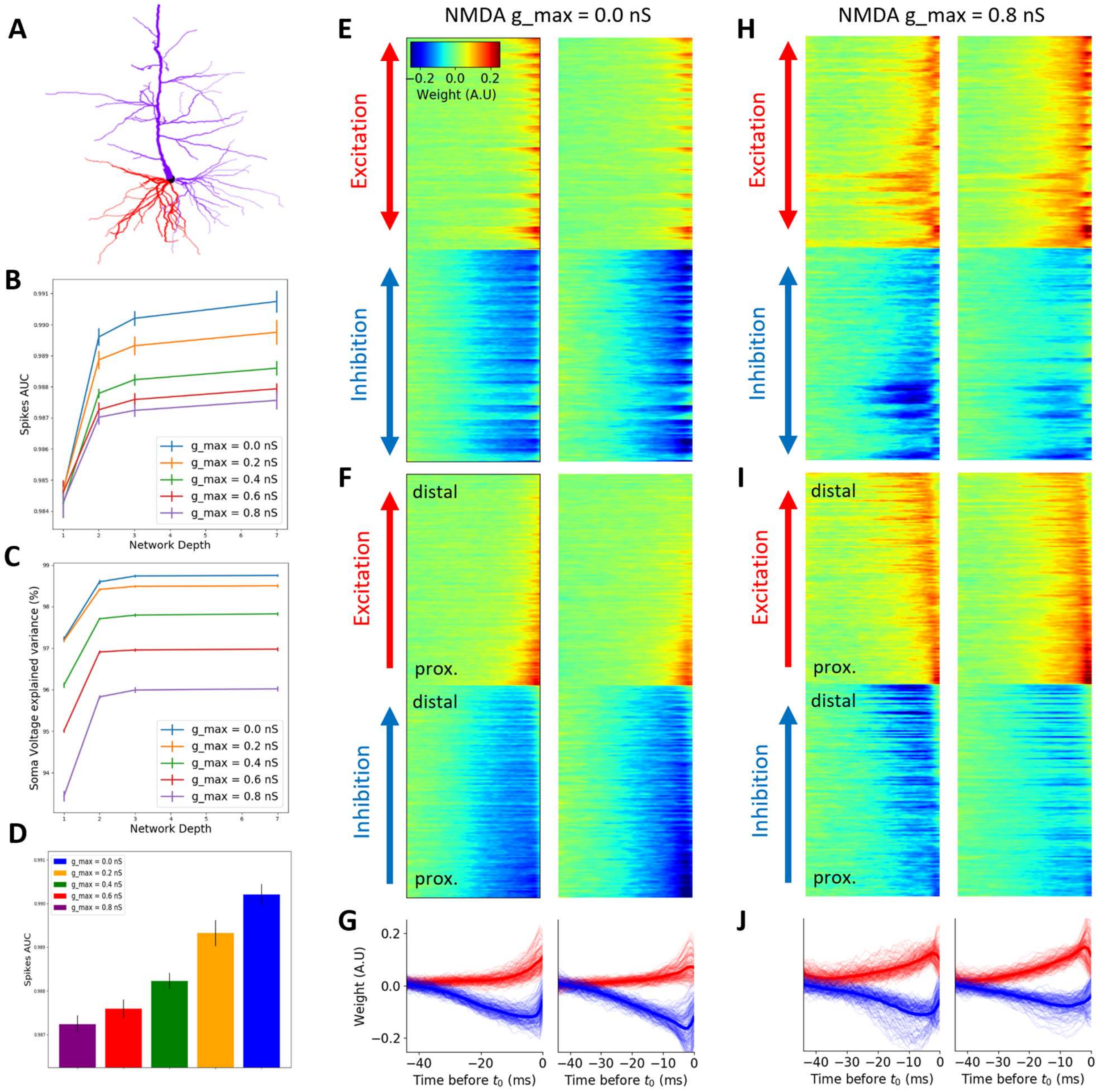

**Figure S6. Varying NMDA conductance over a basal subtree, Related to Figures 2,3,4 & S2**

(A) Illustration of the basal segments we stimulate in this experiment. Only zoomed in and relevant part of morphology is depicted.

(B) In this experiment we vary the NMDA g_max conductance while keeping the AMPA conductance fixed (g_max = 0.4nS). For NMDA g_max value in the set {0.0nS, 0.2nS, 0.4nS, 0.6nS, 0.8nS}, we fitted DNNs of depths {1,2,3,7} (and fixed width of 32 hidden feature maps per layer) to the data to access the complexity of the I/O mapping of that conductance level. Results are shown for spike prediction. A clear trend of increase in functional complexity appears as we increase the NMDA g_max conductance level.

(C) Like (B) only for somatic voltage prediction. A similar trend is shown.

(D) Summary bar plot of results depicted in (A-C). The vertical axis is the AUC performance of a 3-layer DNN taken from (B). The different bars correspond to different conductance. Color coding is identical to that in (A-C). There is a clear decrease in accuracy as the conductance of NMDA is increased indicating that even small NMDAR density already has a meaningful impact on I/O.

(E) two learned weights of the model with two layers for the case when g_max = 0.0nS (i.e., AMPA only synapses). The weights are organized according to the in-order tree traversal of the morphology depicted in (A). we can clearly discern horizontal lines that are related to different branches of the subtree.
(F) Same weights shown in (E), only this time organized according to distance from the soma. It is now clear that these weight matrices predominantly summate proximal inputs and less distal inputs for the excitatory synapses and are rather uniform for the distal inputs.
(G) Temporal cross-sections of the weights depicted in (E) and (F).
(H-J) Similar to (D-F), depicting two weights matrices for the case when g_max = 0.8nS (high NMDA conductance). We see a different pattern compared to (D-F), now both the proximal and the distal synapses are summated. Similar to Figure 6 in the main text, we can see the distal branches playing an important part in the I/O transformation.

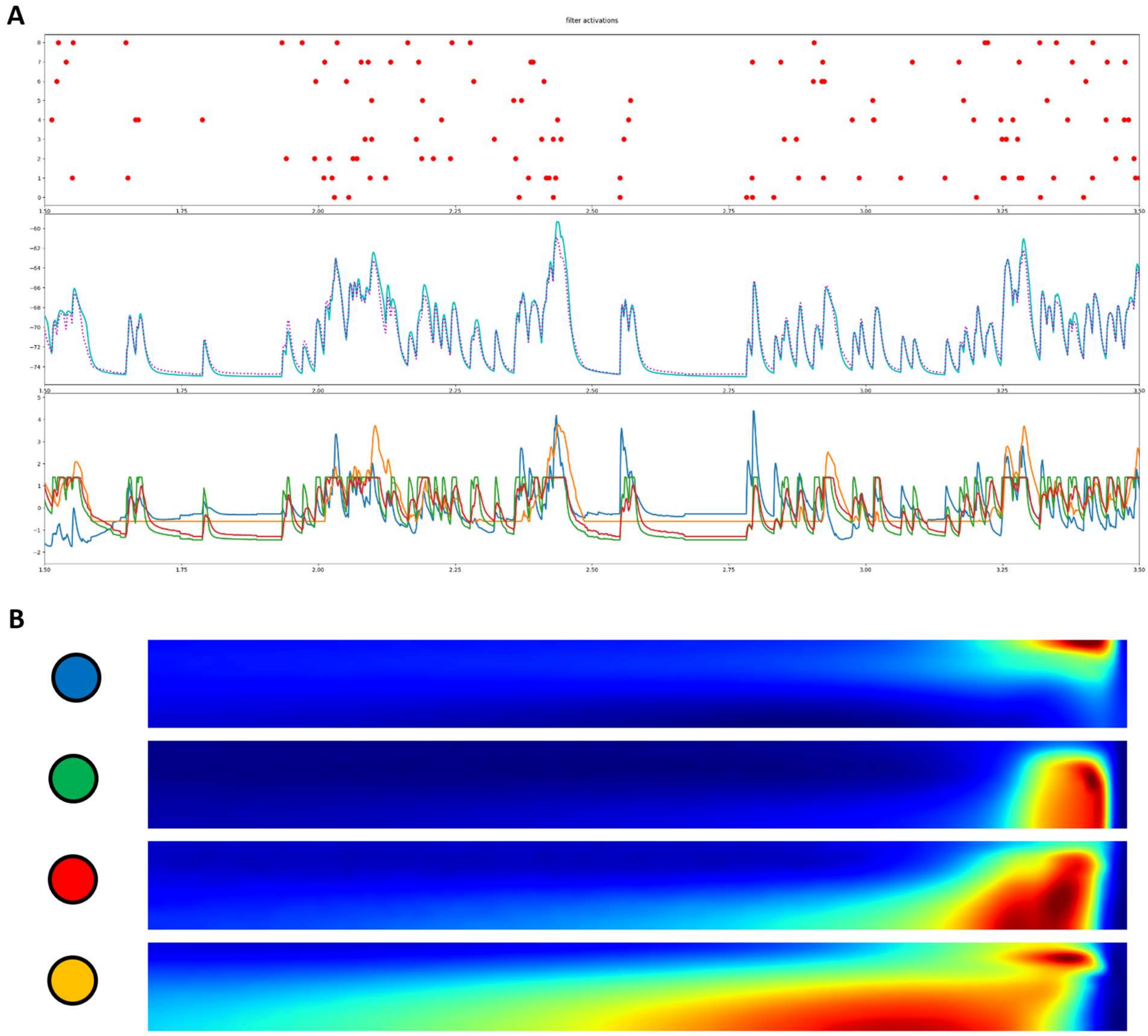

**Figure S7. Interpretation of first layer filter activations of DNN representing a single dendritic branch, Related to Figure 5**

(A) Top. Raster plot of presynaptic input spikes to the branch from Figure 5. Middle. Somatic voltage trace (cyan) and predicted voltage by the DNN from Figure 5. (magenta). Bottom. Activations of the DNNs 4 units

(B) Illustration of the DNN first layer weights and the colors they are represented by in the bottom plot of (A) (circles on the left of each filter).

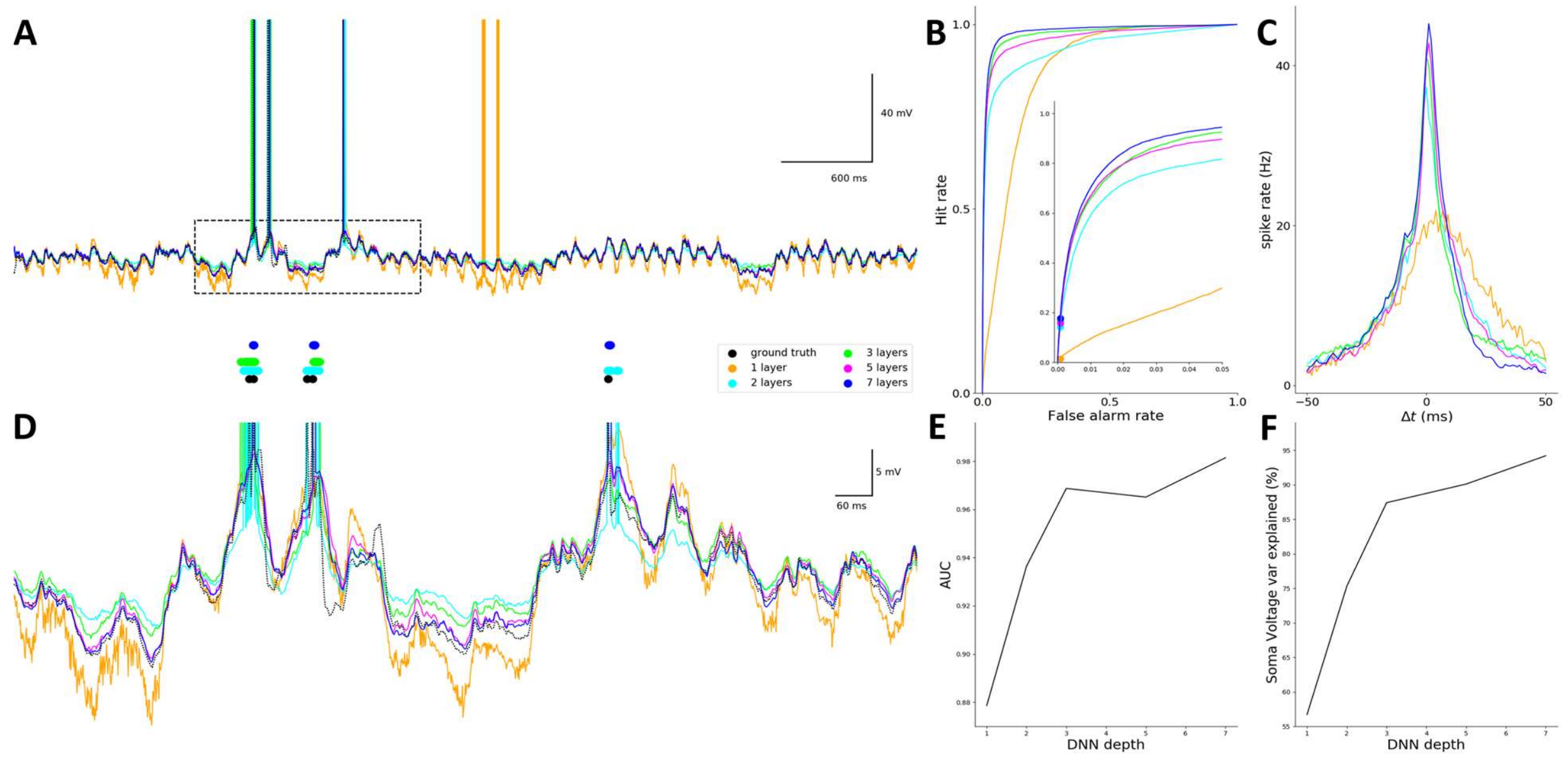

**Figure S8. Comparison of different DNN models performance on out-of-distribution generalization, Related to Figure 6**

(A) Exemplar output DNN prediction trace for five different models with various depths when the input statistics did not match the statistics during training.

(B) Overlaid ROC plot for the five models. Inset depicts a zoom in on the low False Alarm region of the ROC plot

(C) Overlaid cross-correlation plot of the DNN.

(D) Bottom. Zoom in on the dashed rectangle area in (A). Top. Dots represent the times where an output spike was predicted by each of the models. Color code is shown on the right. (Black: ground truth simulation data, Orange: best 1-layer model, cyan: best 2-layer model, Green: best 3-layer model, Magenta: best 5-layer model, Blue: best 7-layer model)

(E) Area under ROC curve (AUC) as a function of model depth for the models shown in (A). We see a large decline in model performance similarly to when model size is reduced.

(F) Somatic voltage explained variance (%) as function of model depth for the models shown in (A). Similarly to (E), we see a large decline in model performance as to when model size is reduced. See Table S1 for more details.

## SUPPLEMENTAL TABLES

| Model | Unseen test | | | Out-of-Distribution test | | |
|---|---|---|---|---|---|---|
| | Spiking AUC | Somatic voltage variance explained (%) | Cross-Correlation std. (ms) | Spiking AUC | Somatic voltage variance explained (%) | Cross-Correlation std. (ms) |
| Best 1 Layer Model (FCN) | 0.9791 | 91.9 | 7.7 | 0.8786 | 56.7 | 7.3 |
| Best 2 Layer Model (TCN) | 0.9877 | 94.3 | 5.2 | 0.9364 | 75.3 | 5.7 |
| Best 3 Layer Model (TCN) | 0.9893 | 94.4 | 3.7 | 0.9687 | 87.4 | 5.2 |
| Best 5 Layer Model (TCN) | 0.9913 | 95.2 | 3.5 | 0.9651 | 90.1 | 4.8 |
| Best 7 Layer Model (TCN) | **0.9917** | **95.3** | **3.1** | **0.9814** | **94.2** | **4.3** |

**Table S1. Comparison of different DNN models performance on the unseen test set and out-of-distribution test set, Related to Figure 6**
Out of distribution performance summary table for the selected set of models of various depths. Out of distribution test set was constructed by averaging performance on simulations coming from 2304 simulations of 6 seconds each (see **Methods**) with 8 district scenarios that account for all combinations of the cases illustrated in Figure 6: {clustered input +/-, synchronous input +/-, no inhibition +/-}. We see a clear drop in performance for all models, but for the larger models the drop is much less substantial compared to the shallower models.

# Chapter 2:

## Multiple Synaptic Contacts Combined with Dendritic Filtering Enhance Spatio-Temporal Pattern Recognition of Single Neurons

David Beniaguev, Sapir Shapira, Idan Segev, Michael London

# Multiple Synaptic Contacts combined with Dendritic Filtering enhance Spatio-Temporal Pattern Recognition of Single Neurons

David Beniaguev[1], Sapir Shapira[1,] Idan Segev[1,2] and Michael London[1,2]
[1]Edmond and Lily Safra Center for Brain Sciences (ELSC), The Hebrew University of Jerusalem, Jerusalem 91904, Israel
[2]Department of Neurobiology, The Hebrew University of Jerusalem, Jerusalem 91904, Israel.
Correspondence: david.beniaguev@gmail.com

## Abstract

A cortical neuron typically makes multiple synaptic contacts on the dendrites of its postsynaptic target neuron. The functional implications of this apparent redundancy are unclear. Due to dendritic cable filtering, proximal dendritic synapses generate brief somatic postsynaptic potentials (PSPs) whereas distal synapses give rise to broader PSPs. Consequently, with multiple synaptic contacts, a single presynaptic spike results in a somatic PSP composed of multiple temporal profiles. We developed a "Filter-and-Fire" (F&F) neuron model that incorporates multiple contacts and cable filtering; it demonstrates threefold increase in memory capacity as compared to a leaky Integrate-and-Fire (I&F) neuron, when trained to emit precisely timed spikes for specific input patterns. Furthermore, the F&F neuron can learn to recognize spatio-temporal input patterns, e.g., MNIST digits, where the I&F model completely fails. We conclude that “dendro-plexing” single input spikes by multiple synaptic contacts enriches the computational capabilities of cortical neurons and can dramatically reduce axonal wiring.

## Introduction

In recent decades it has been shown experimentally that when cortical neurons are synaptically connected they typically connect to each other via multiple synaptic contacts rather than a single contact (Holler et al. 2021; Silver et al. 2003; Feldmeyer, Lübke, and Sakmann 2006; Shepherd et al. 2005; Markram et al. 1997). Multiple synaptic contacts originating from a single presynaptic axon often impinge on different parts of the dendritic tree of the post-synaptic neuron (Feldmeyer et al., 2006; Holler et al., 2021; Silver et al., 2003). Interestingly, if the formation of synaptic contacts were based purely on the proximity of the axon-to-the-dendrite and independent of each other ("Peters' rule" (Peters and Feldman, 1976)) then one would expect that the distribution of the number of such multiple contacts would be geometric, with one contact per axon being the most frequent case (Fares and Stepanyants 2009; Markram et al. 2015; Rees, Moradi, and Ascoli 2017). This is far from what was empirically observed where, e.g., around 4-8 synaptic contacts are formed between pre- and postsynaptic layer 5 cortical pyramidal neurons (Markram et al. 1997). This deviation from the distribution predicted by Peters' rule suggests that the number of synaptic contacts between two connected neurons is tightly controlled by some developmental process and is thus likely to serve a functional purpose.

Several phenomenological models have attempted to explain how multiple synaptic contacts between the presynaptic and postsynaptic neurons are formed (Fares and Stepanyants, 2009), but only a few studies have attempted to tackle the question of how they might be beneficial from a computational perspective. It is typically thought that this redundancy overcomes the problem of probabilistic synaptic vesicle release, which results in unreliable signal transmission between the connected neurons (Rudolph et al., 2015). However, the same effect using a simpler mechanism could be achieved by multiple vesicles release (MVR) per synaptic activation (Holler et al., 2021; Rudolph et al., 2015) and does not require multiple synaptic contacts (which is more “expensive” to establish). Hiratani and Fukai (Hiratani and Fukai, 2018) demonstrated that multiple synaptic contacts might allow synapses to learn quicker. However, faster learning, although beneficial, fundamentally does not endow the neuron with the ability to perform new kinds of tasks. Recently, Zhang et al. (Zhang et al., 2020) incorporated multiple contacts in the context of deep artificial neural networks but did not demonstrate tangible computational benefits. Jones et al. (Jones and Kording, 2021) also model dendrites in the context of artificial neural networks demonstrating that, when using a threshold linear dendrite model, the classification performance of a single neuron with multiple contacts is improved compared to the single contact case. Several other studies use multiple synaptic contacts in the context of artificial neural networks, demonstrating some computational benefits (Acharya et al., 2021; Camp et al., 2020; Sezener et al., 2021).

In the present study we propose two key functional consequence of multiple contacts. Towards this end, we developed a simplified spiking neuron model that we termed the Filter & Fire (F&F) neuron model. This model is based on the Integrate and Fire (I&F) neuron model (Burkitt, 2006; Lapicque, 1907) but incorporates two key additional features: (i) It takes into account the effect of the dendritic cable filtering on the time course of the somatic PSP, whereby proximal synapses generate brief somatic PSP while identical synapses that are located at distal locations on the dendritic tree give rise to broad somatic PSPs (Rall 1964; Rall 1967); (ii) Each presynaptic axon makes multiple synaptic contacts on the F&F neuron model. Consequently, a single presynaptic spike results with PSP that is composed of multiple temporal profiles (we term this phenomenon “dendro-plexing” of the presynaptic spike).

To analyze the memory capacity of this model, we used the formulation of Memmesheimer et al. (Memmesheimer et al., 2014) developed for the I&F model. In this approach, the model is trained (via changes in synaptic strengths) to emit precisely-timed output spikes for a specific random input pattern; the capacity is defined as the maximal number of precisely timed output spikes during some time period divided by the number of incoming input axons. We show that the memory capacity of the F&F model is three-fold larger than that of the I&F model. We next show the F&F neuron can solve real-world classification tasks where the I&F model completely fails. We further explored the effect of unreliable synapses on the memory capacity of the F&F model and, finally, we demonstrate that multiple synaptic contacts dramatically reduce axonal wiring requirement in cortical circuits.

## Results

### Mathematical description of the filter and fire (F&F) neuron model with multiple synaptic contacts

We propose hereby a Filter and Fire (F&F) neuron model, which is similar to the standard current-based Leaky Integrate and Fire (I&F) spiking neuron model, but with two added features. The first feature incorporates the temporal characteristics of a dendritic cable as initially demonstrated by Rall (Rall 1964; Rall 1967) in which, due to cable filtering, synaptic inputs that connect at distal dendritic locations exhibit prolonged PSPs at the soma (Fig. 1**B** top traces), whereas proximal synaptic inputs generate brief PSP profiles (Fig. 1**B** bottom traces). The second feature is that each input axon connects to multiple locations on the dendritic tree, sometimes proximal and sometimes distal (Fig. 1**A**,**B)**.

Formally, consider $N_{axons}$ the number of input axons (Fig. 1**A**), denoted by index i, and their spike trains denoted by $X_i(t)$. Each axon connects to the dendrite via $M$ synaptic contacts ($M = 3$ is illustrated in Fig. 1). Each contact connects to the dendrite at a location denoted by index j and filters the incoming axon spike train with a specific synaptic kernel $K_j(t)$. The voltage contribution trace (PSP) of this contact is: $V_{c,j}^{i}(t) = X_i(t) * K_j(t) = \sum_{t_i} K_j(t - t_i)$. There is a total of $M \cdot N_{axons}$ such contacts and therefore the same number of overall synaptic contact voltage contribution traces (Fig. 1**C**). In vector notation we denote $\boldsymbol{V_c}(t) = [V_{c,1}^{1}(t), V_{c,2}^{2}(t), \cdots, V_{c,M \cdot N_{axons}}^{N_{axons}}(t)]$. Each synaptic contact has a weight, $w_j$. In vector notation, $\boldsymbol{w} = [w_1, w_2, \cdots, w_{M \cdot N_{axons}}]$. The contribution of each contact is multiplied by its corresponding weight to form the somatic voltage trace: $V_s(t) = \boldsymbol{w}^T \cdot \boldsymbol{V_c}(t) = \sum_j w_j \cdot V_{c,j}(t)$ (Fig. 1**D**). When the spike threshold is reached, a standard reset mechanism is applied. Please note that the "dendrites" in this model are linear and therefore retain the analytic tractability of the I&F neuron model.

Here we model the temporal ramifications of the effect of adding a passive dendritic cable; we did not consider the effect of nonlinear dendrites (see Discussion). The kernels we used are typical double exponential PSP shapes of the form: $K_j(t) = A \cdot \left(e^{-t/\tau_{decay,j}} - e^{-t/\tau_{rise,j}}\right)$, where A is a normalization constant such that each filter has a maximum value of 1, and $\tau_{decay,j}$, $\tau_{rise,j}$ are randomly sampled for each synaptic contact, representing randomly connected axon-dendrite locations. Note that for mathematical simplicity we do not impose any restrictions on synaptic contact weights, each weight can be both positive or negative regardless of which axon it originates from. Indeed, the goal of the study is not to replicate all particular biological details, but rather to specifically explore the computational benefit that arises due to two key general dendritic/connectivity features: temporal filtering of synaptic potentials due to dendritic cable properties and multiple synaptic connections between pairs of cortical neurons.

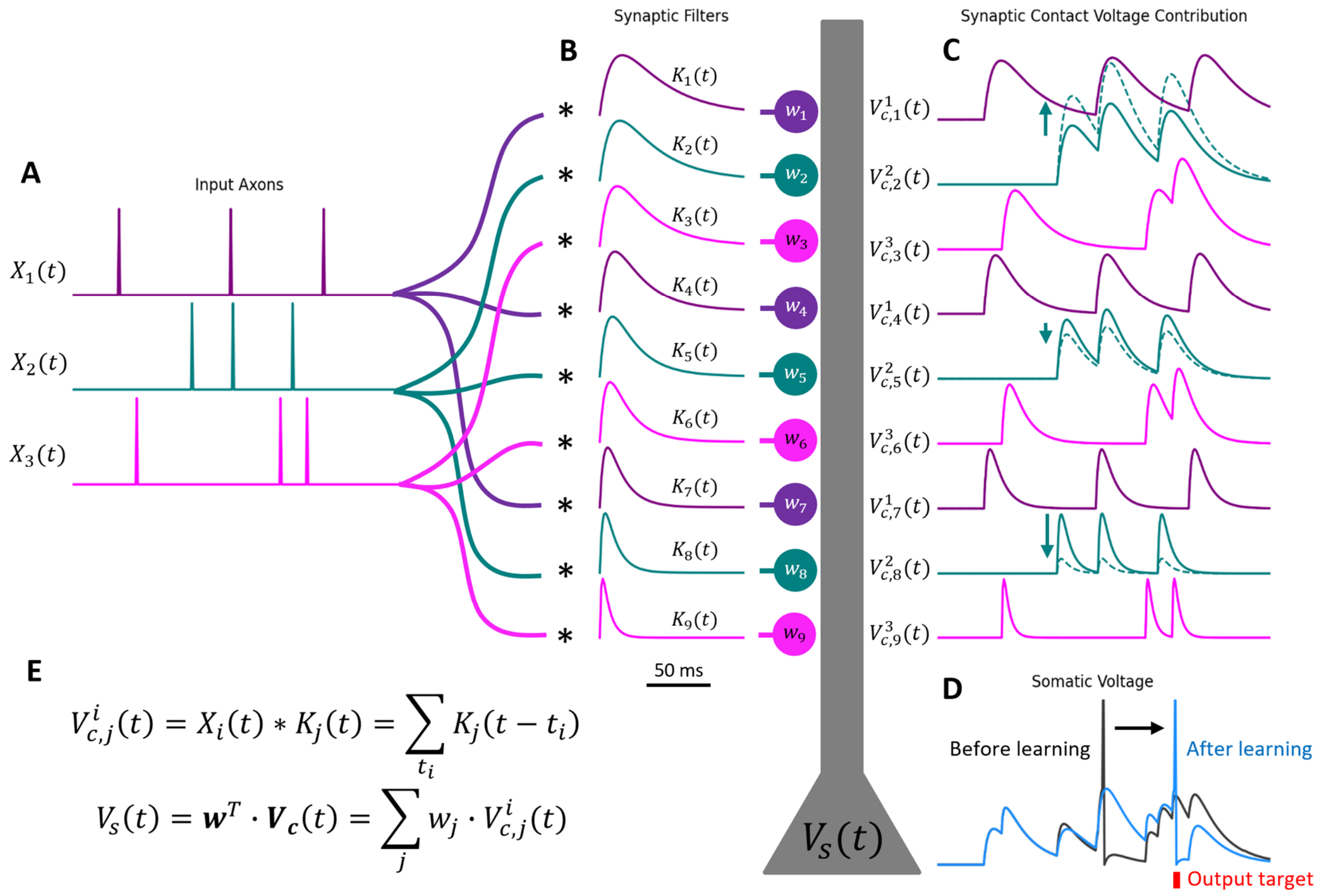

**Fig. 1. The Filter and Fire (F&F) neuron receives input through multiple synaptic contacts per axon and filters each contact with a different synaptic kernel ("dendritic multiplexing")**
(**A**) Example for three incoming input axons, each making three synaptic contacts onto the postsynaptic neuron. (**B**) Various synaptic filters (kernels) representing their respective locations on the dendritic tree (shown schematically in grey). Proximal synaptic filters are brief, whereas more distal synaptic filters have broader temporal profiles. Kernel colors are according to the source axon. (**C**) Individual contact voltage responses at the synaptic loci that result from the convolution (* symbol) of an axonal spike train with the respective synaptic filter. Continuous line is the voltage contribution of the contact before learning and dashed lines after learning. In this example, the weight of the second-from-top green synaptic contact, $w_2$, was increased, whereas $w_5$ and $w_8$ decreased following learning. The weight of each contact can change independently during learning. (**D**) Somatic voltage is a weighted sum of the contributions of each synaptic contact. In black is the somatic trace before learning and in blue after learning. Standard I&F reset mechanism was applied for the spike generation mechanism at the soma. (**E**) Equations formally describing the synaptic contact voltage contributions and the somatic voltage.

## Increased memory capacity of the F&F neuron with multiple synaptic contacts

We first tested the memorization capacity of the F&F neuron model as a function of the number of multiple connections per axon. We utilized the framework proposed by Memmesheimer et al. (Memmesheimer et al., 2014) to measure memory capacity, and used their proposed local perceptron learning rule for that task. In short, this capacity measure indicates the maximal number of precisely timed output spikes in response to random input stimulation during some time period divided by the number of input axons. Fig. 2**A** shows random spiking activity of 100 input axons for a period of 60 seconds (top). Below the output of the postsynaptic neuron is shown before learning (black), after learning (blue) and the desired target output spikes (red). In this example, we used $M = 5$ multiple contacts per axon. For the given set of random input spike trains, it is possible to find a synaptic weight vector to perfectly place all output spikes at their precisely desired timing. In Fig. 2**B** we repeat the simulation shown in Fig. 2**A** for various values of multiple contacts (M) while re-randomizing all input spike trains, desired output spike trains, and the synaptic filter parameters of each contact. We repeat this both for the I&F neuron model (i.e., a single synaptic PSP kernel for all synapses/axons) and the F&F neuron model with a randomly selected synaptic kernel for each synapse (see **Methods** for full details of the kernel shapes used). The y axis represents our success in placing all of the output spikes accurately, as measured by the area under the receiver operating characteristic (ROC) curve (AUC) for the binary classification task of placing each spike in 1ms time bins. Error bars represent the standard deviation of the AUC over multiple repeats (18), while re-randomizing the input, re-randomizing the synaptic kernels and re-randomizing the desired output spike trains.

Fig. 2**B** shows that all the curves obtained from the I&F neuron models cluster together, independent of M (the number of multiple contacts). This is to be expected as this is the classic case of synaptic redundancy when using a single temporal kernel for all presynaptic axons. For example, for a single axon and two contacts one can see that, for the case of a single synaptic kernel (as is the case in I&F model), the somatic voltage can be written as $V_s(t) = w_1 \cdot \sum K(t - t_i) + w_2 \cdot \sum K(t - t_i) = (w_1 + w_2) \cdot \sum K(t - t_i) = w_{eff} \cdot \sum K(t - t_i)$. Meaning, that the different weights $w_1, w_2$ associated with the same input axon are redundant and equivalent to a single effective weight $w_{eff}$, and therefore this difference is not utilized in the case of an I&F neuron model where all synaptic kernels are identical.

The case of the F&F neuron with $M = 1$, is identical to the case of an I&F model only that it has different kernels for different synapses. This change on its own does not affect the capacity of the model, as the number of learnable and utilizable parameters is identical to the I&F case with $M = 1$, and thus this curve lies with the other curves of the I&F models. However, for the F&F models with multiple contacts ($M = 2{,}3{,}5{,}10{,}15$) there is an increase in accuracy, demonstrating that some of the additional synaptic weights are utilized (Fig. 2B). Fig. 2**C** displays the maximal number of output spikes that can be precisely timed as a function of M, for both I&F and F&F models. We measure the number of precisely timed spikes as the maximal number of spikes that is above a high accuracy threshold (AUC > 0.99), normalized by the number of axons. This provides the number of precisely timed output spikes per input axon on the y-axis. The figure shows that, for the I&F model, the capacity is around 0.15, whereas for the F&F neuron with a large number of synapses per axon, it saturates at about 0.45 (~3-fold increase in capacity). Already for 3-5 contacts/axon (as is the case found for connected pairs of cortical pyramidal neurons) the capacity approaches the saturation level (see **Discussion**). Note that the number of degrees of freedom (tunable parameters) scales linearly with the number of multiple contacts, so there is no obvious explanation for the observed saturation in the F&F model. We will explain the origin of this 3-fold increase compared to the I&F model in Figure 4.

In Fig. 2D we vary the number of input axons and observe linear scaling of the number of precise spikes achieved with different slopes for different number of multiple connections. This indicates that increasing the number of multiple connections increases the effective number of parameters utilized per axon. To better illustrate the results of Fig. 2 more intuitively, we show in Fig. S1**A** a simple case whereby the I&F neuron can emit temporally precise output spikes by employing a spatial strategy, effectively selecting the input axon traces that happen to correlate with the desired output trace. In Fig. S1**B** we show how a F&F neuron can employ a temporal strategy by weighting differently the synaptic kernels with different time scales, allowing it to select individual input spikes that correlated with the desired output spikes instead of the entire axonal spiking activity.

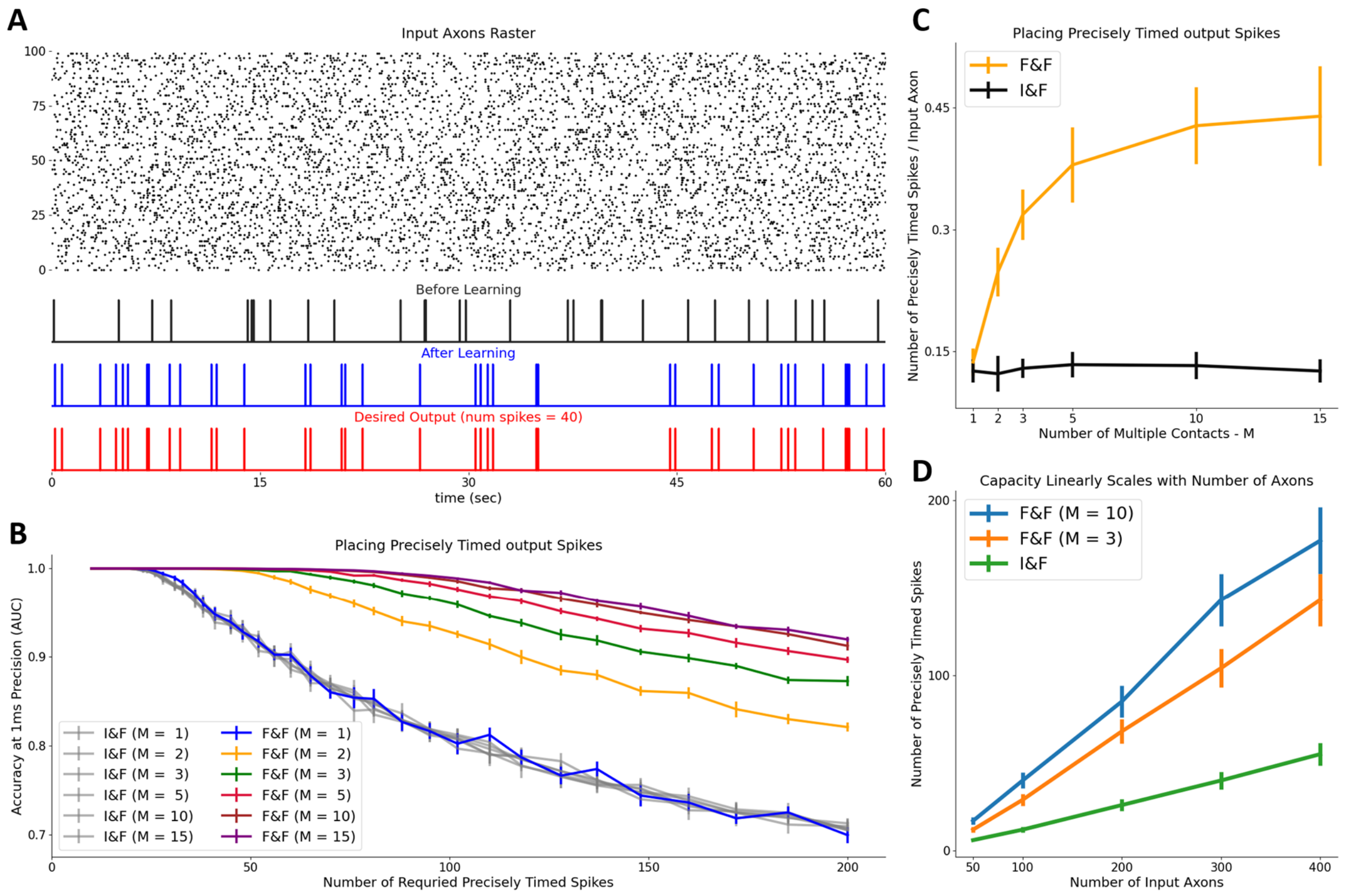

**Fig. 2. Increased memory capacity of precisely timed output spikes of the F&F neuron versus the I&F neuron.** (**A**) Learning to place precisely timed output spikes for randomly generated input with F&F neuron model with M=5 multiple contacts. **Top**. Random axonal input raster plot. **Bottom**. Output spikes of the F&F model before learning (top, black), after learning (middle, blue) and the desired output spikes (bottom, red). (**B**) Binary classification accuracy at 1ms temporal resolution as measured by area under ROC curve (AUC) as a function of the number of required output spikes for input with 200 input axons. The capacity of the F&F models increases as the number of multiple contacts increases, whereas no such increase is observed for the I&F case. (**C**) Capacity as a function of the number of multiple connections per axon. For this plot we use the maximal number of spikes that achieves accuracy above AUC threshold of 0.99. The vertical axis depicts the fraction of successfully timed spikes for each input axon. Note that the saturation in the capacity for many multiple contacts is ~3x compared to the respective I&F capacity. (**D**) The capacity scales linearly as a function of the number of input axons and exhibits no saturation.

## The F&F neuron can learn real-life spatio-temporal tasks that the I&F neuron cannot

Next, we demonstrate new capabilities of the F&F neuron model with multiple synaptic connections that are beyond that of the I&F neuron model. For this purpose, we developed a new spatio-temporal task derived from MNIST task, which is a large database of handwritten digits (Deng, 2012). Towards this end, we converted the horizontal spatial dimension of the image (image width) into a temporal dimension (Fig. 3**A** top) with a uniform time warping, such that 20 horizontal pixels will be mapped onto T milliseconds. T is the duration of presentation of the respective pattern (say digit “3”). The vertical spatial dimension of the image (image height) is simply replicated five times so that 20 vertical pixels will be mapped onto 100 axons. We then sample spikes for each axon according to the time varying Poisson instantaneous firing rate with additional background noise. For example, in this way, an axon corresponds to a row in a handwritten digit image, and this axon fires spikes with increased probability at times when the pixels in that row are active. An example of the resulting input spike trains representing different digits is shown by the raster plot in Fig. 3**A** middle frame.

Next, we trained the F&F neuron to produce a spike at the end of a specific digit. We trained the model on the full MNIST train subset of digits, and present results on an unseen test set. Before learning (black), after learning (blue), and the desired output (red) are presented at the bottom of Fig. 3**A** for the case where the selected digit to be recognized was "3". We then repeat this process for all digits for three models - I&F neuron, F&F neuron, and a spatio-temporal, temporally sliding, logistic regression (LR) model (Cox, 1958). We note that the LR model is not biologically plausible and cannot be considered as a model of a neuron. Importantly, the LR model has $N_{axons} \cdot T$ learnable and fully utilizable parameters ($N_{axons}$ parameters for each 1ms time bin) which is much greater than $N_{axons} \cdot M$ parameters that are used in the F&F and I&F models (which are also not fully utilizable as we have seen in Fig. 2).

The MNIST training set consists of 60,000 images, approximately 6,000 for each digit. The positive class for each digit classification in the experiments in Fig 3 consisted of 6.000 images, and the negative class around 54,000 images. When attempting to detect a single digit the baseline accuracy of a neuron that never spikes is around 90% accuracy, we therefore measure our accuracies compared to that baseline. The test accuracies following training for all models are depicted in Fig. 3**B**. For this plot a successful true positive (hit) is achieved if at least 1 spike has occurred in the time window of 10 ms around the ground truth desired spike. The temporal duration of each pattern T was 40 ms and the number of multiple contacts M was 5, for the plot in Fig. 3**B**. This figure clearly shows that the I&F neuron model is at chance level for almost all digits and is incapable of learning the task. In contrast, the F&F model is consistently better than chance, and sometimes approaches the "aspirational" spatio-temporal logistic regression model. In Fig. 3**C** we visually display the learned weights of all models when attempting to learn the digit 3. The weight matrix of the logistic regression model clearly depicts what appears to be an average-looking digit 3. The F&F neuron model depicts a temporally smoothed version of the logistic regression model, and the I&F model clearly cannot learn temporal patterns and therefore cannot recognize this digit at above chance level. For precise details of how the weights for the F&F and I&F models were visualized, see **Methods**. For a more simplified pattern classification case, see Fig. S1**C**. Fig. S1**D** shows how a F&F neuron can solve the task shown in Fig. S1**C**. Fig. S1**E** explains why this task cannot be solved by an I&F neuron.

The effect of T, the presentation duration of each digit, can be seen in Fig. 3**D**, which displays summary statistics of test accuracy, averaged across all digits, for the three models as a function of T. The interval between presented patterns was 70 ms and the decay time constant for the I&F model was 30 ms, to match the maximal decay time constant for all synaptic kernels in the F&F model. Fig. 3**D** shows that there is an optimal pattern presentation duration for the F&F model in the range of 40-50 ms, which is ~1.5 times the maximal decay time constant in the model.

Next, we explored the effect of unreliable synaptic transmission and its interaction with multiple synaptic contacts. As explained in the **Introduction**, multiple synaptic contacts are often considered as a mechanism to overcome synaptic transmission unreliability. In Fig. 3**E** we display the test accuracy as a function of the number of multiple contacts for all 3 models for fully reliable synapses that we have displayed so far; the dashed line shows the accuracy under the unreliable synapse regime with release probability of p = 0.5 per contact. Note the consistent drop in test accuracy that does not go away even with a large number of multiple contacts. In this graph, the positive digit used was "7" and T = 30 ms. The number of positive training samples used in this case was 2048 patterns.

In principle, unreliable synaptic transmission could be considered as a mechanism for implementing "drop-connect", a method used in training artificial neural networks that has a known regularization effect (Srivastava et al., 2014; Wan et al., 2013). In Fig. 3**F** we test if this is also applicable in our case. We show the test accuracy as a function of the number of training samples and demonstrate a regularization effect in the case of F&F that increases test accuracy for a low number of training input patterns. Note that each training sample was shown to the model multiple times (15 times in this case), and for each spike and each contact an independent probability of release was applied, effectively resulting in 15 noisy patterns that were presented to the neuron during training for each original training pattern. This suggests that unreliable synapses can also be viewed as a "feature" rather than a "bug" for the regime of a small number of training data points and can help avoid overfitting (see **Discussion**).

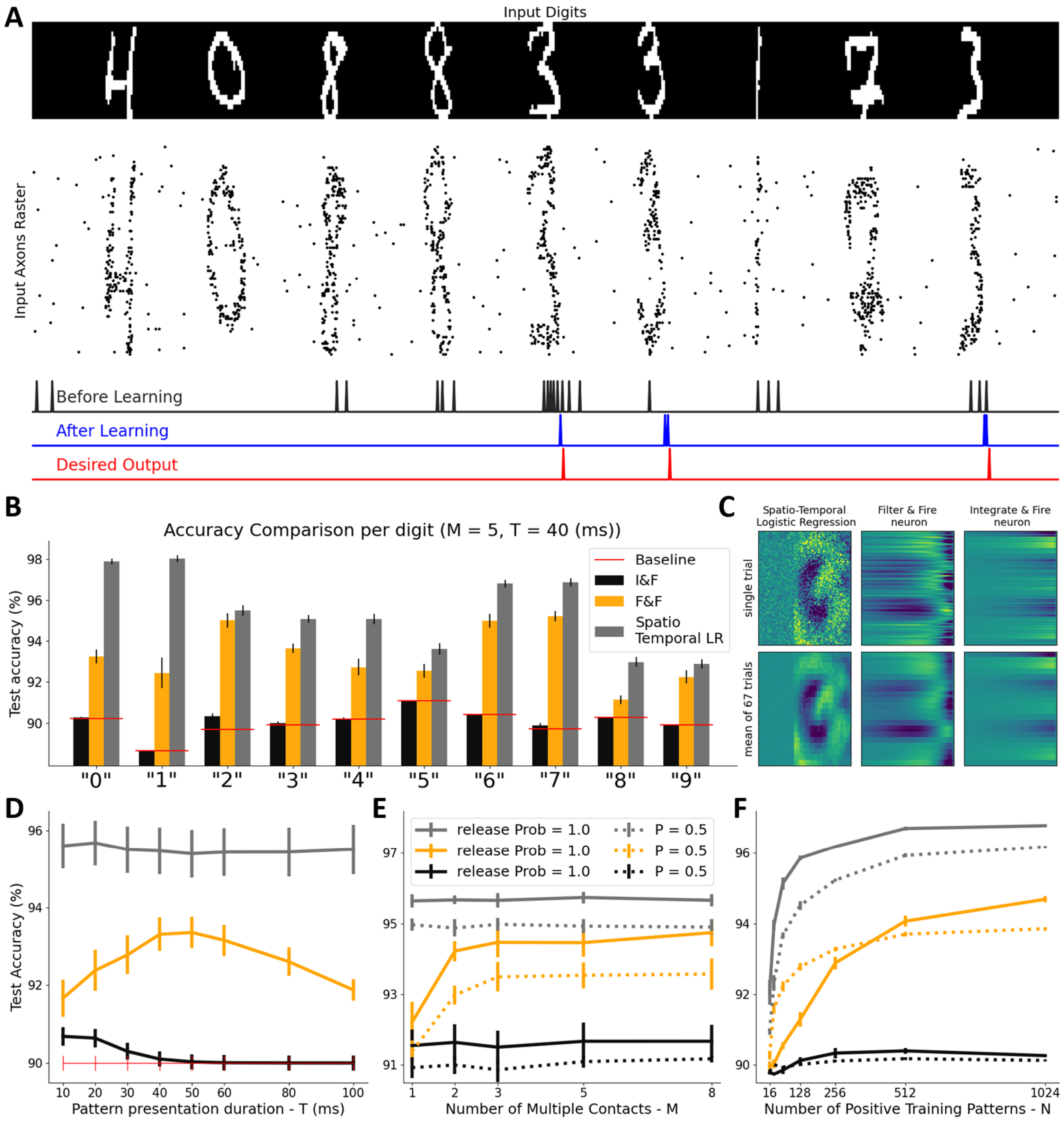

**Fig. 3. The F&F neuron can learn to recognize spatio-temporal patterns whereas the I&F neuron cannot.** (**A**) A F&F neuron was trained to recognize digits in a spatio-temporal version of the MNIST task. Top: Original input digits; Middle: Raster plot of the axonal spikes (100 axons, 5 contacts per axon with duration of digit representation, T = 40 msec) representing the respective digits (with some additional background noise, see **Methods**). Bottom: Output of the F&F neuron (with 5 multiple contacts) before learning (black), after learning, (blue) and the desired output spike train (red). In this case the neuron was trained to detect the digit "3". The depicted traces are from a new test set. (**B**) Full test set classification accuracy of the F&F neuron (orange), the I&F neuron (black) and a non-biologically plausible temporally sliding spatio-temporal Logistic Regression model (gray) for the spatio-temporal digit classification task, when each digit was used as the positive class and the rest of the digits were considered the negative class. The random chance baseline for each case is shown by the red horizontal lines (baseline fluctuates

near 90% due to slightly unequal number of images in the MNIST dataset for each digit). The I&F models cannot learn the task whereas the F&F neurons sometimes approach the (non-biological) spatio-temporal logistic regression model. (**C**) Spatio-temporal representation of the learned weights of the three models when attempting to visually detect the digit "3". The digit "3" can be clearly discerned in the logistic regression case, a faint "3" in the F&F case, whereas for the I&F model, an attempt to use spatial-only information to detect the digit is visible and the detection of the digit is poor, at chance level. The top three images are for a single trial with a single random axonal wiring, and Poisson sampling of the inputs. Bottom three images are an average of 67 trials. (**D**) Summary of test accuracy, averaged across all digits, for the three models as a function of the temporal duration (presentation time) of the digit. Time between digit presentations was 70 ms. The decay time constant of the synaptic kernel for the I&F model (black curve) was 30 ms, as is the maximal decay time constant for all synaptic kernels for the F&F model. The performance of the F&F neuron peaks at T = 40-50 ms (orange curve). In this case, the F&F model shown had M=5 multiple contacts. (**E**) Test accuracy as a function of the number of multiple contacts for all 3 models. The dashed line is the accuracy under the unreliable synapse's regime with a release probability of 50% per contact. Note the consistent drop in accuracy. In this graph the positive digit used was "7" and the pattern temporal duration was T=30 ms. The number of positive training samples used in this case was subsampled to 2048 patterns. (**F**) Test accuracy as a function of the number of training samples. Each training sample was shown to the model 15 times and for each spike and each contact an independent probability of release was applied. Note the regularization benefit of unreliable synaptic release for a small number of training samples (dashed orange line above solid orange line).

## Maximal capacity is explained by the effective 3-dimensional subspace spanning all synaptic kernels

Next, we wish to pinpoint the mathematical origin of the properties depicted in Figures 2 and 3 of the proposed F&F neuron model with multiple synaptic connections. We observed that all the PSP kernels we used have similar shapes and therefore the PSPs generated by the same axon will produce correlated inputs at the local synaptic responses vector $\boldsymbol{V_c}(t)$. Any input correlation will limit the number of degrees of freedom available for learning by modification of the synaptic weights $\boldsymbol{w}$. Fig. 4**A** shows all the PSPs used as heatmaps organized according to increasing values of $\tau_{rise}$ within each block and increasing $\tau_{decay}$ between the blocks. In Fig. 4**C** we show all the PSPs as temporal traces overlayed on each other. Both Fig. 4**A** and 4**C** clearly show that the shapes of the PSP kernels, although selected randomly and have some variance, are overall very similar to each other. We therefore apply singular matrix decomposition (SVD) on all PSP shapes (Fig. 4**B**) and found that 99.93% of the variance in all PSP shapes is explained by the first three singular vectors (Fig. 4**E**). This means that all synaptic kernels depicted in Fig. 4A and 4C are effectively spanned by a basis set of three orthogonal PSP-like shapes. For the sake of presentation and to avoid negative values in the trace shapes, we display the three independent kernels that are the result of non-negative matrix factorization (NMF) in Fig. 4**D**.

It is easy to see that these PSP shapes basically filter the input signal with various time constants and various delays. These are very intuitive shapes that we can easily interpret. This enables one to understand both the temporal smoothing aspect of the learned weights by the F&F neuron model (Fig. 3**C**) and also suggesting that the number of independent PSP shapes that span al the PSPs is a good candidate to explain the 3-fold increase in capacity shown in Fig. 2**C**. To verify this, we repeat the same experiment as in Fig. 2, but now each axon is connected to the post synaptic neuron via only three multiple connections, but this time we use the optimal PSPs shapes depicted in Fig. 4**D**. These results are depicted in Fig. 4**F**. The model with three orthogonal kernels has effectively identical results to those of the F&F neuron with randomly selected filters. First, to explain the resultant capacity, when we randomly sample more and more connections from the set of all possible kernels (e.g., increasing M), we slowly approach to span the entire three-dimensional space by each contact and therefore we have a saturation effect at three times the I&F baseline level. This means that we have reached the three independent kernels, each corresponding to an independent learnable parameter. Second, the specific shapes of these kernels explain the temporal smoothing effect we have observed in the learned weights of the F&F neuron model in Fig. 3**C**. Importantly, although we used somewhat artificial double exponential synaptic kernels, we verified that our results also hold for PSPs in a simulation of a highly realistic detailed L5 cortical neuron model. Fig. S2, shows that, despite some quantitative differences, qualitatively the results presented in our work using a reduced F&F neuron model are also valid for the case of a neuron with a complex dendritic tree receiving synaptic inputs over its entire dendritic tree. Namely, a basis set of three temporally distinct kernels can effectively span all synaptic kernels also in realistic models of cortical pyramidal neurons.

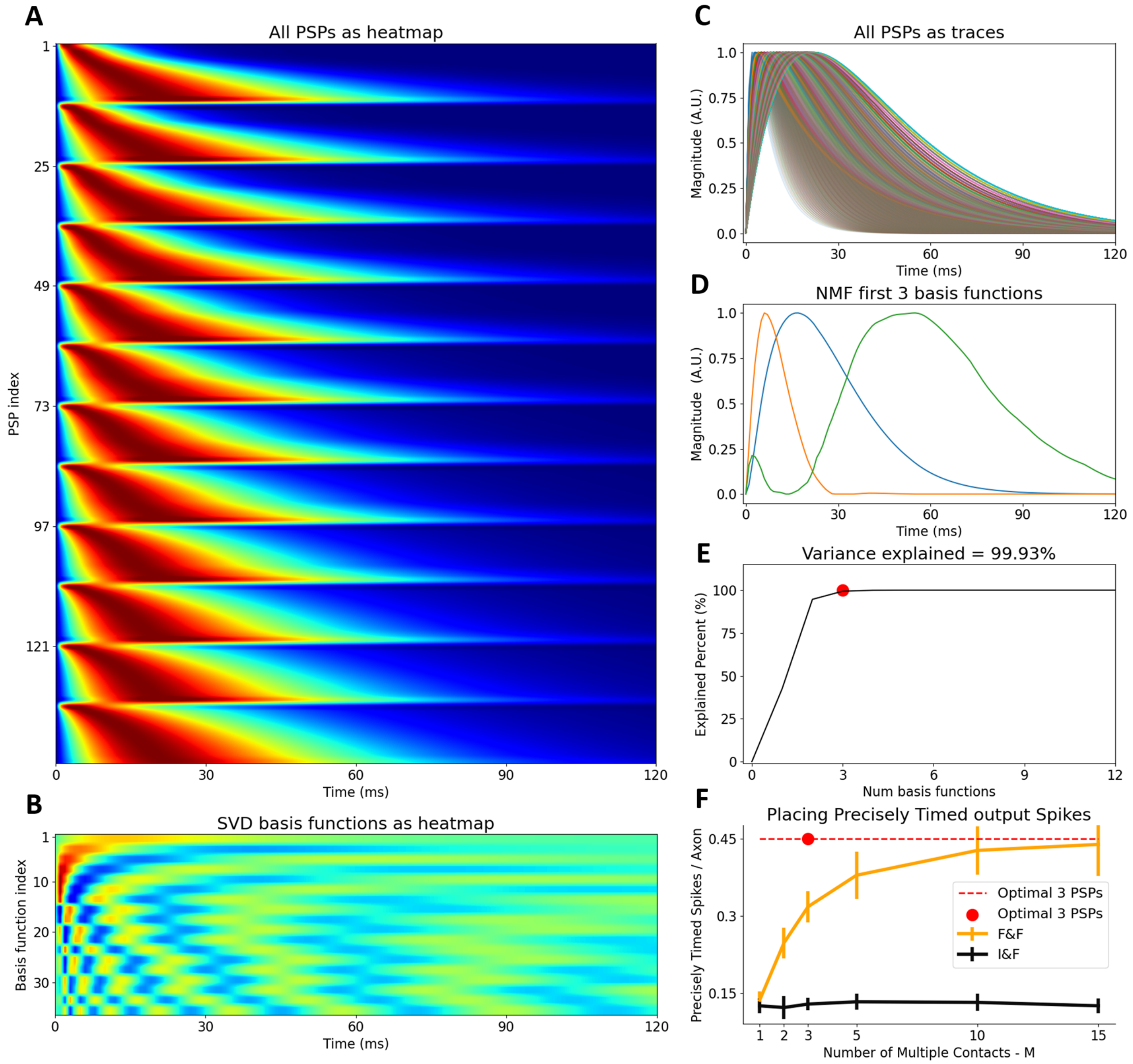

**Fig. 4. The dendritic filters are spanned by a three-dimensional basis set of PSPs accounting for the three-fold increase in the F&F capacity for a large number of multiple contacts as compared to the I&F model.** (**A**) All possible postsynaptic potentials (PSPs) that were used in this study shown as heatmaps. Every row has a different rise time and decay time**.** (**B**) The Singular Value Decomposition of all the PSPs shown in (A) as heatmaps. This provides a Fourier-like basis set. (**C**) Traces of all post synaptic potentials (PSPs) that were used in the study. (**D**) First three basis functions of the non-negative matrix factorization (NMF) of all PSPs shown in (A & C). NMF instead of SVD was used for ease of interpretation and visualization purposes (see **Methods**). (**E**) Cumulative variance explained by each basis component. The three basis functions shown in (**D**) can span all PSPs shown in (C), red circle. (**F**) Direct verification that the three orthogonal basis set in (D) can be used as three optimal multiple contact filters achieving the same capacity as do 15 randomly selected multiple contacts.

**Multiple dendritic synapses per axon significantly reduce axonal wiring or network size**

Fig. 5 shows that multiple dendritic synaptic contacts per axon allow cortical circuits to transmit axonal information in a more hardware-efficient way, by relying on downstream dendritic decoding. Three computationally equivalent alternatives are depicted in Fig 5**C**,**D**&**E**. Fig. 5**C** illustrates the case in which N axons transmit information that is readout by a downstream F&F neuron model with multiple synaptic contacts. In Fig, 5**D**, we illustrate a scenario of an equivalent I&F neuron. If we wish for a downstream I&F point neuron to produce the same output as that of the F&F in Fig. 5**C**, the spikes present in the N axons the F&F recives should spread out to 3N axons for the I&F neuron to receive as input (the "spreading" could even be random, as long as the information content is kept identical). This will allow the I&F in Fig.5**D** to precisely replicate the I/O transformation of the F&F in Fig.5**C**. Alternatively, each of the N axons could be replicated three times, each with some additional time delay, to account for the temporal dendritic integration properties implemented by the F&F neuron model and then fed to an I&F neuron. This last scenario is illustrated graphically in Fig. 5**E.** It is also verified in simulations in Fig. 5**A** for learning to produce precise spike time output and in Fig. 5**B** for solving the spatio-temporal MNIST digit recognition task. The three equivalent alternatives shown in Fig. 5**C, D&E**, highlight the general statement that a system can tradeoff between putting recourses into encoding or decoding. In this particular case the three alternatives are (a) an encoder with 3N transmitting axons + 1 simple I&F neuron decoder (b) simple encoder with N transmitting axons + 3N simple delay decoding neurons + 1 simple I&F final decoder neuron and (c) simple encoder with N transmitting axons + 1 complex F&F decoder neuron.

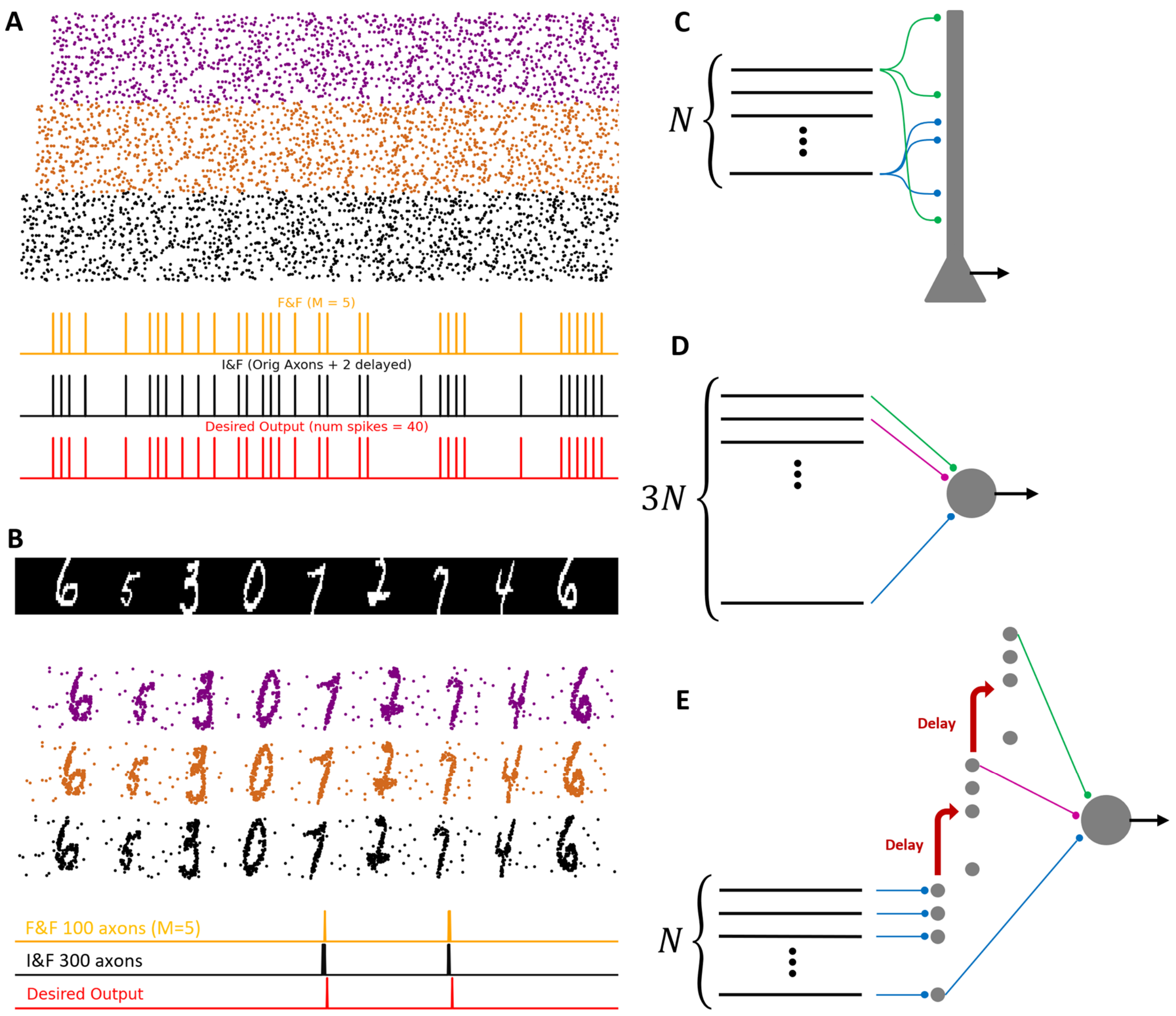

**Fig. 5. Multiple dendritic synaptic contacts per axon reduce the number of transmitting axons and/or decoding neurons. Illustration of three computationally equivalent implementations.** (**A**) Input axonal raster composed of three sets of axons (black, brown and purple axons). Raster of the brown axons is a delayed version of the black raster, and the purple raster is delayed once more. In the first scenario, only the black axons are fed to a downstream F&F neuron, whose output is shown in orange (with 5 contacts per axon). In the second scenario, the full set of axons is fed into an I&F neuron (1 contact per axon), whose output is shown in black. The desired output spike train is shown in red. This panel demonstrates that an I&F neuron with 3N axons (with delayed input) replicates the precise timing capacity of the F&F model with N axons. (**B**) As in (A) but for the spatio-temporal MNIST digit recognition task. Here again an I&F neuron requires additionally delayed input axons to replicate the performance of the F&F neuron with N axons on the spatio-temporal MNIST digit recognition task. (**C**) Illustration of N axons that feed into a single F&F neuron model receiving multiple synaptic contacts. (**D**) For an I&F point neuron model to have identical memorization capacity as the F&F neuron illustrated in (C), the same information needs to be separated into 3N axons. (**E**) In order for an I&F point neuron to have identical spatio-temporal patterns separation capabilities with N axons as in (C), it needs to have a feed-forward "delay line" decoding network prior to being fed into the I&F “readout” neuron.

## Discussion

Cortical neurons typically connect to each other via multiple synaptic contacts. The computational implications of this apparent redundancy are unclear. To explore this question, we presented in the present study the Filter and Fire (F&F) neuron model that augments the commonly used Leaky Integrate and Fire (I&F) neuron model by incorporating in it multiple synaptic contacts per axon as well as the effect of dendritic filtering that transforms the incoming spike train to a set of multiple post synaptic potential (PSP) filters. Each filter corresponds to a particular synaptic contact, with varied time constants. This directly relates to the location-dependent filtering of the dendritic synaptic inputs by the dendritic cable properties (Rall 1964; Rall 1967). In the F&F model, a single presynaptic spike impinging on several dendritic locations results in a somatic PSP composed of multiple temporal profiles. We termed this phenomenon "dendro-plexing" of the presynaptic spike (Figure 1).

We have demonstrated that the capacity of the F&F neuron to memorize precise input-output relationships is increased by a factor of ~3 compared to that of the regular I&F. The capacity is measured as the ratio between the number of precisely timed output spikes and the number of incoming input axons. This ratio is ~0.15 spikes per axon for the I&F case as was shown by Memmesheimer et al. (Memmesheimer et al., 2014), and ~0.45 spikes per axon for the F&F model (Fig. 2C). We showed in Figure 4, that the origin of this threefold increase in capacity is due to the fact that all possible PSP shapes are spanned by a subspace consistent of 3 basis shapes. This subspace sets the effective upper limit on the capacity when using any number of multiple synaptic contacts. i.e., even if using numerous synaptic contacts, they effectively serve as only 3 independent contacts.

Next (Fig. 3), we constructed a new spatio-temporal pattern discrimination task using the MNIST dataset and demonstrated that the F&F model can learn to detect single digits at well-above chance level performance on an unseen test set, whereas an I&F neuron model cannot learn the task at all, because in the specific way we chose to represent each digit – the task does not contain enough spatial-only information suitable for I&F neuron discrimination. Our specific task design was deliberately chosen to highlight this temporal aspect of pattern discrimination that is possible when taking into account the temporal filtering due to cable properties of dendrites. We show that multiple synaptic connections with different PSP profiles allow the neuron to effectively parametrize the temporal profile of the PSP influence of each pre-synaptic axon on the somatic membrane potential. This is enabled by modifying the weight of the various (multiple) contacts made between the axon and the post synaptic cell. We show that, for the range of PSP filters we have used, as well as for the range of somatic PSP in a detailed model of L5 cortical pyramidal cell (Fig. S2) all PSPs can be spanned by a 3 basis PSP filters, each with a different temporal profile. Taken together, this suggests that the F&F neuron model provides a low temporal frequency approximation to a spatio-temporal perceptron that assigns independent weights to each point in time. An alternative description that is mathematically equivalent is that the F&F model effectively bins the membrane integration time into 3 non-uniform time bins and can learn to assign independent weights for each temporal bin.

Our study demonstrates that even when considering highly simplified neuron model as used here, one that only implements the passive temporal filtering aspect of dendrites, a computational role of dendrites is unraveled for the seeming redundancy of multiple synaptic contacts between pairs of neurons in the brain. Dendrites therefore allow us to salvage some of the redundant connectivity and put those synaptic weight parameters to good use. This allows for increased memorization capacity, but perhaps more importantly to detect specific spatio-temporal patterns in the input axons. The spatio-temporal filtering properties of dendritic processing are prominently featured in our recent study describing how single neurons produce output spikes in response to highly complex spatio-temporal patterns (Beniaguev et al., 2021). It is important to note that there is an additional nonlinear amplification aspect in dendrites (Bicknell and Häusser, 2021; Poirazi and Mel, 2001; Poirazi et al., 2003; Polsky et al., 2004) that we did not consider in this study and that is likely to provide additional computational benefits also in the context of multiple synaptic contacts, as we previously suggested (Beniaguev et al., 2021). We have therefore added in the present study a new perspective on the growing literature regarding the computational function of individual neuron (Golkar et al., 2020; Gütig and Sompolinsky, 2006a; Häusser and Mel, 2003; Hyvarinen and Oja, 1998; McCulloch and Pitts, 1943; Mel, 1992; Moldwin and Segev, 2018; Moldwin et al., 2021; Oja, 1982; Pehlevan et al., 2020; Poirazi and Mel, 2001; Poirazi et al., 2003; Rosenblatt and F., 1958; Zador et al., 1991)

It is notable that it is typically believed that the phenomenon of multiple synaptic contacts is primarily attributed to noise reduction related to unreliable synaptic transmission, but our work strongly suggests that, from a computational standpoint, this is only a small part of the story. In fact, in principle each synapse can reduce its own unreliability by employing a multiple vesicle release (MVR) mechanism, as was also recently suggested by (Rudolph et al., 2015). Here we suggest that unreliable synaptic transmission might play a computational role, a “feature” rather than being a "bug". In Fig. 3**F** we show that synaptic “noise” might play a role that is similar to “dropout” (Srivastava et al., 2014) or more precisely drop-connect (Wan et al., 2013) mechanism which is commonly used in present day artificial neural networks paradigm. There, drop-connect is typically employed as a regularization technique that reduces overfitting and usually improves generalization.

The question of how trains of spikes represent information in the nervous system has been a long-standing question in neuroscience, since its inception. A major debate revolves around whether information is largely carried by firing rates averaged over relatively long time periods, or rather that precisely timed spikes carry crucial bits of information. Evidence for both alternatives has been found for both sensory systems and motor systems and, consequently, much theoretical work on this key topic has been conducted (Abeles, 1982; Abeles et al., 1993; Castelo-Branco et al., 2000; Christopher Decharms and Merzenich, 1996; DeWeese et al., 2003; Florian, 2012; Gütig and Sompolinsky, 2006b; Hopfield, 1995; Johansson and Birznieks, 2004; Kara et al., 2000; London et al., 2002, 2010; Maass and Schmitt, 1999; Meister et al., 1995; Memmesheimer et al., 2014; Neuenschwander and Singer, 1996; Schneidman et al., 1998; Thorpe et al., 2001; Wehr and Laurent, 1996). Since we have shown that dendrites and multiple synaptic connections per axon play a crucial role in decoding incoming spike trains and increase the neuronal repertoire in emitting precisely timed output spikes in response to spatio-temporal input patterns (and they might do this using a simple biologically plausible learning rule), we wish to suggest that dendritic hardware should be considered when discussing the question of the neural code and what information is transmitted via axons. Indeed, (Perez-Nieves et al., 2021) show that a diversity of time constants helps increasing the computational repertoire of spiking networks.

Here we suggest that diversity of time constants, that helps in increasing the computational repertoire, already happens at the neuronal level. The fact that a single neuron can decode complex spatio-temporal patterns on its own and does not require a highly coordinated decoding network of neurons to extract temporal information from incoming spike trains, not only allows for potential "hardware" (wiring) savings as we illustrated in Figure 5, but also suggest that information might be ubiquitously encoded by precise spike times throughout the central nervous system. A single neuron can emit precisely timed output spikes in response to spatio-temporal inputs, as we showed here and was previously shown in simpler I&F models by (Memmesheimer et al., 2014). It is therefore not required to have a large and highly coordinated network of neurons to encode temporally precise patterns transmitted via axons. We believe that if already a single neuron can learn to generate such precise spike timing without relying on network mechanisms, this might increase the likelihood that neuronal information is encoded by precise spike timing as opposed to average firing rates (over relatively long periods of time) throughout the CNS.

Additionally, the three equivalent alternatives shown in Fig. 5**C,D&E**, highlight the general scenario in which the evolutionary pressure to reduce axonal wiring in an ever-increasing brain volume (Chen et al., 2006; Chklovskii et al., 2002) could be solved by compressing information on a limited number of axons and by relying on sophisticated dendritic integration to decode these signals. Indeed, our study shows that employing multiple synaptic contacts between the axon and its postsynaptic neuron improves the decoding capability of spatio-temporal patterns by neurons and, at the same time, save "hardware" (reduced total axon length and/or number of neurons) at the cost of a small increase in single neuron complexity.

Lastly, in recent years, multiple groups around the world have started to generate dense reconstruction of rodent and human neuronal (cortical) circuits at the EM level, and report neuronal connectivity maps (Kasthuri et al., 2015; Motta et al., 2019; Shapson-Coe et al., 2021). We would like to suggest that analyzing these EM datasets, focusing on the number of multiple contacts and their locations on the dendritic tree, might shed additional light on the extent and role of multiple synaptic contacts between different cell types and brain regions, and hint to the possible “style” of information processing in these networks based solely on EM data. As illustrated in our work, if two neurons form a connection on distal dendrites, or if they form a connection on proximal dendrites, these will result in completely different time course of the somatic voltage and, therefore, on the temporal coding capabilities of the neuron. Indeed, simply reporting connectivity maps and even the size of post synaptic density (PSD) areas is not enough to determine the temporal influence of the connection on the post synaptic cell, as the dendritic location of the synapse is key, as was shown by Rall (Rall 1964; Rall 1967) and recently also suggested in (Liu et al., 2021). Furthermore, due to the nonlinear amplification of dendrites, such EM-based data will be crucial in assessing whether two presynaptic neurons connect to nearby dendritic locations as, in this case, these synapses are much more likely to undergo nonlinear amplification (and might produce additional broad/slow NMDA or calcium-dependent temporal filters) if activated at similar times. This will add additional broader/long-lasting filters to neurons and increase their capability to learn to recognize spatio-temporal input patterns beyond what we have shown in this study.

## Materials and Methods

### F&F neuron simulation

A F&F neuron receives as input $N_{axons}$ input axons, their spike trains will be represented by $X_i(t) = \sum_{t_i} \delta(t - t_i)$ and they will be denoted by index $i$. Each axon connects to the dendrite via $M$ contacts Each contact connects on the dendrite at a location that will be denoted by index $j$, and filters the incoming axon spike train with a specific synaptic kernel $K_j(t)$. The kernels are typical double exponential PSP shapes of the form $K_j(t) = A \cdot \left(e^{-t/\tau_{decay,j}} - e^{-t/\tau_{rise,j}}\right)$ where A is a normalization constant such that each filter has a maximum value of 1, and $\tau_{decay,j}$, $\tau_{rise,j}$ are different for each contact, sampled randomly and independently for each contact from the ranges $\tau_{decay,j} \in [12ms, 30ms]$, $\tau_{rise,j} \in [1ms, 12ms]$. Different kernel parameters represent a randomly connected axon-dendrite location.

The result of the kernel filtering of the corresponding input axons forms the contact voltage contribution trace $V_{c,j}^{i}(t) = X_i(t) * K_j(t) = \sum_{t_i} K(t - t_i)$. There are overall a total of $M \cdot N_{axons}$ such contact voltage contributions traces. In vector notation we denote $\boldsymbol{V_c}(t) = [V_{c,1}(t), V_{c,2}(t), \cdots, V_{c,M \cdot N_{axons}}(t)]$. Each synaptic contact has a weight, $w_j$. In vector notation we write $\boldsymbol{w} = [w_1, w_2, \cdots, w_{M \cdot N_{axons}}]$. Each local synaptic response is multiplied by its corresponding weight to form the somatic voltage $V_s(t) = \boldsymbol{w}^T \cdot \boldsymbol{V_c}(t) = \sum_j w_j \cdot V_{c,j}(t)$. When threshold is reached, the voltage is reset, and a negative rectifying current is injected that decays to zero with a time constant of 15ms. Note that due to mathematical simplicity we do not impose any restrictions on synaptic contact weights, each weight can be both positive or negative regardless of which axon it comes from.

### I&F neuron simulation

The I&F simulation details are identical in all ways to the F&F neuron, except that all of its contact kernels are identical $K_j(t) = K_{I\&F}(t) = A_{I\&F} \cdot \left(e^{-t/\tau_{decay,I\&F}} - e^{-t/\tau_{rise,I\&F}}\right)$, where $\tau_{rise,I\&F} = 1ms$ and $\tau_{decay,I\&F} = 30ms$.

### F&F learned weights visualization

In the F&F model a single input axon is filtered by multiple contact kernels $V_{c,j}^{i}(t) = X_i(t) * K_j(t)$. In order to display the effective linear model weights, we can group together all kernels that relate to the same axon $V^i(t) = \sum_{j_i} w_{j_i} \cdot X_i(t) * K_{j_i}(t) = X_i(t) * \sum_{j_i} w_{j_i} \cdot K_{j_i}(t)$. Therefore, the term $\sum_{j_i} w_{j_i} \cdot K_{j_i}(t)$ is a composite kernel that filters each axon. This is a function that can be visualized for each input axon, and therefore related directly to the input space.

**Capacity calculation details**

To measure capacity, we sample $N_{axons}$ random spike trains to serve as axons with a Poisson instantaneous firing rate of 4Hz for a period of 120 seconds. We randomly distribute $N_{spikes}$ output spikes throughout the 120-second time period to generate $y_{GT}(t)$. For the sake of mathematical simplicity and not dealing with reset issues, we make sure that the minimal distance between two consecutive spikes is at least 120ms $(= 4 \cdot \tau_{decay,I\&F})$. We bin time to 1ms time bins. We calculate $\boldsymbol{V_c}(t)$ for the entire trace. Our task is to find a set of weights such that $y_{GT}(t) = \varphi(\boldsymbol{w}^T \cdot \boldsymbol{V_c}(t))$, where $\varphi(\cdot)$ is a simple thresholding function. This is in essence a binary classification dataset with 120,000 timepoints (milliseconds), for each of those timepoints the required output of a binary classifier is either 1 (for time points that should emit an output spike), or 0 for all other timepoints. We have a total $M \cdot N_{axons}$ weights that need to fit the entire 120K sample dataset. We calculate the AUC on the entire 120K datapoint dataset. We declare that the fit was successful when AUC > 0.99. We repeat the procedure for various values of $N_{spikes}$. The maximal value of $N_{spikes}$ that we still manage to fit with AUC > 0.99 is termed $N_{spikes,max}$ and the capacity measure is this number normalized by the number of axons used $Capacity(M) = N_{spikes,max}/N_{axons}$. Note that due to $N_{spikes} \ll 120{,}000$, the capacity of the problem effectively doesn't depend on the length of the time period we use, but by rather the number of spikes we wish to precisely time.

**Spatio-temporal MNIST task and evaluation details**

The MNIST dataset contains 60,000 training images and 10,000 test images. The images are of size 28x28 pixels. We crop the images at the center to be 20x20 pixels, and binarize the values. We then convert the horizontal spatial image dimension (width) into a temporal dimension by uniformly warping the time such that 20 horizontal pixels will be mapped onto T milliseconds. T is the pattern presentation duration. The vertical spatial image dimension (height) is simply replicated 5 times so that 20 vertical pixels will be mapped onto 100 axons. The entire training set is the concatenated sequentially (in random ordering), with 70ms of zeros between every two patterns. We then sample spikes for every 1ms time bin according to each axon's instantaneous firing rate to generate the raster for the entire training set. On top of that we add an additional background noise rate. The output ground truth is a single output spike 1ms after the pattern is presented for the positive class digit, and no spikes for negative patterns. The fitting of the model is identical to that described in the capacity sections. i.e., we wish to find a set of weights such that $y_{GT}(t) = \varphi(\boldsymbol{w}^T \cdot \boldsymbol{V_c}(t))$, where $\varphi(\cdot)$ is a simple threshold function. In this case, we allow for wiggle room when evaluating performance on the test set. A successful true positive (hit) is achieved if at least 1 spike has occurred in the time window of 10ms around the ground-truth desired spike. A failed false positive (false alarm) is considered if a spike has occurred during the time window of 10ms around the end of the pattern presentation. We measure the classification accuracy under these criteria. We sometimes train only on a part of the training dataset and use only $N_{positive\ samples}$ as the positive class. In the regime of unreliable synaptic transmission with some synaptic release probability $p$, we sample the spikes of the input patterns once, and then present it $N_{epochs} = 15$ times, each time with random release probability samples for each presynaptic spike for each contact. In those cases, we perform the same procedure also for the test set (i.e., display the same pattern $N_{epochs} = 15$ times, each time with different synaptic release sampled for each presynaptic spike for each contact)

**Spatio-temporal Logistic Regression (LR) evaluation details**

A spatio-temporal temporally sliding logistic regression (LR) model is a non-biologically plausible that is specifically used to serve as an aspirational model used for comparison for our proposed F&F neuron. It has a two-dimensional weight matrix $W_{LR}(s,t)$, where $s$ spans the spatial dimention and $t$ spans the temporal dimension. Mathematically, the equation describing the relationship between the input and the output of the model is given by $y_{LR}(t) = \varphi(\sum_{s=0}^{N_{axons}} \sum_{\tau=0}^{T_{LR}} W_{LR}(s,\tau) \cdot X_s(s,t-\tau))$. Note that this model has in total $N_{axons} \cdot T_{LR}$ weights (time is discretized into 1ms time bins here as well, as throughout this study).

## Supplementary Materials

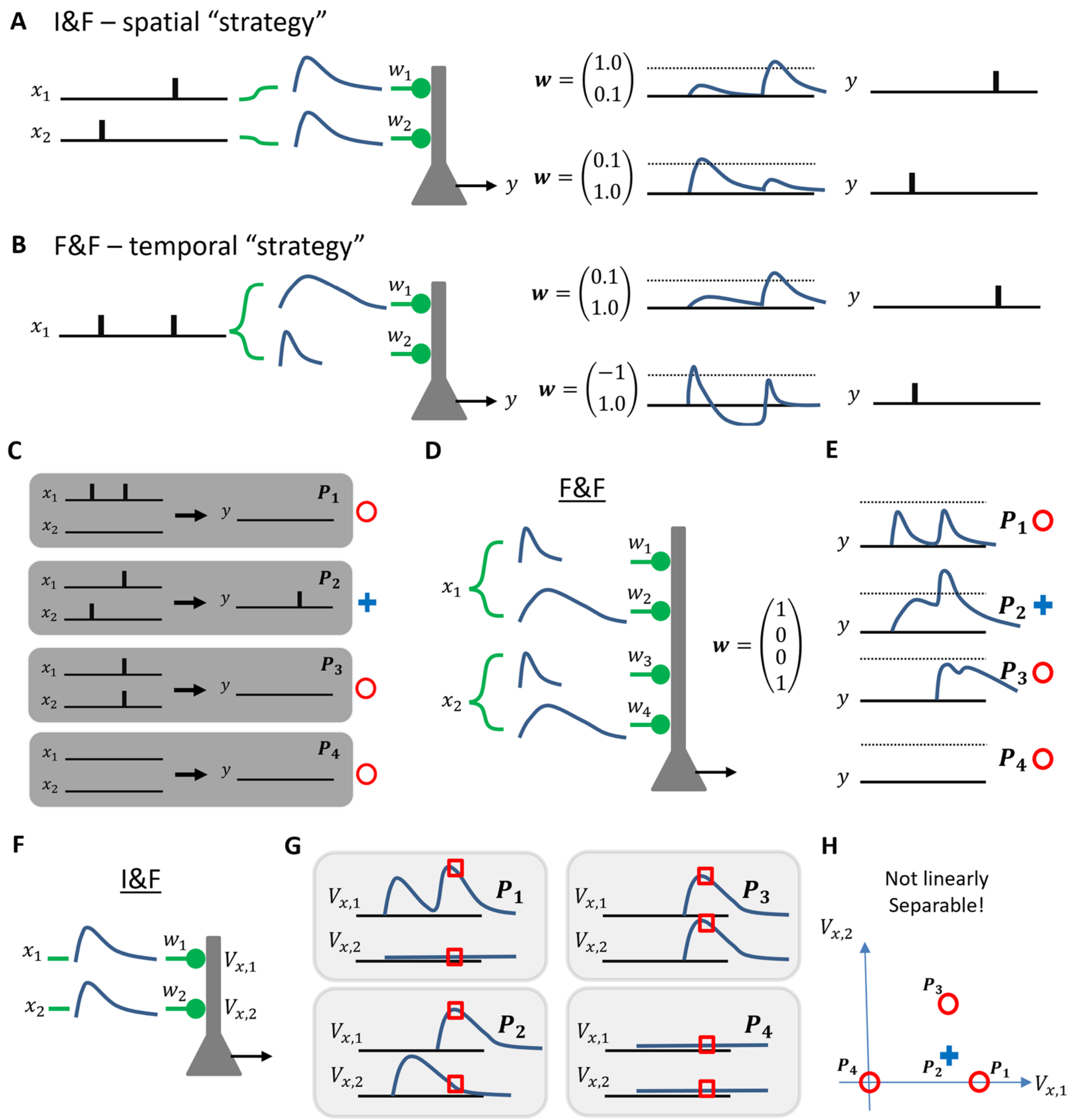

**Fig. S1. Toy illustration to highlight differences in computational capabilities between the F&F neuron and I&F neuron** (**A**) Illustration of the spatial strategy the I&F model employs. Left. Input axons x1, x2 are fed into an I&F model to emit an output y. Right, two output for different weight configurations that result in two different output patterns y. Dashed line represent spike threshold. (**B**) Illustration of the temporal strategy the F&F model employs. Left. A single input axon x1 is fed into an F&F neuron with two different PSP filters to produce output y. Right, two output for different weight configurations that result in two different output patterns y. note that although the output patterns are identical to those in (A), this was achieved by receiving a single input axon. (**C**) Example of a temporal task we wish to teach the two neuron models. For the four different input patterns (x1, x2) we require to produce the output y on the right.

(Patterns are denoted as P1, P2, P3, P4). Only P2 is required to emit an output spike (marked with blue plus sign), the other patterns are required to emit no spikes (marked with red circle) (**D**) Illustration of a F&F model with fast and slow PSP filters for each input axon that can solve the task. (**E**) The somatic responses for each pattern with the weights vector w that solves the problem. The dotted lines represent the spike threshold. (**F**) Illustration of the I&F neuron model, its inputs and synaptic contact voltage contribution illustrated. (**G**) The normalized unweighted synaptic contact contribution voltage traces for all 4 patterns. The red squares mark the point in time where an output spike is expected. (**H**) The I&F neuron model cannot classify the output patterns as the four patterns are not linearly separable in the dendritic representation space (Vx,1 Vx,2) as seen in the illustration, and therefore a satisfactory weight vector for the synapses does not exist.

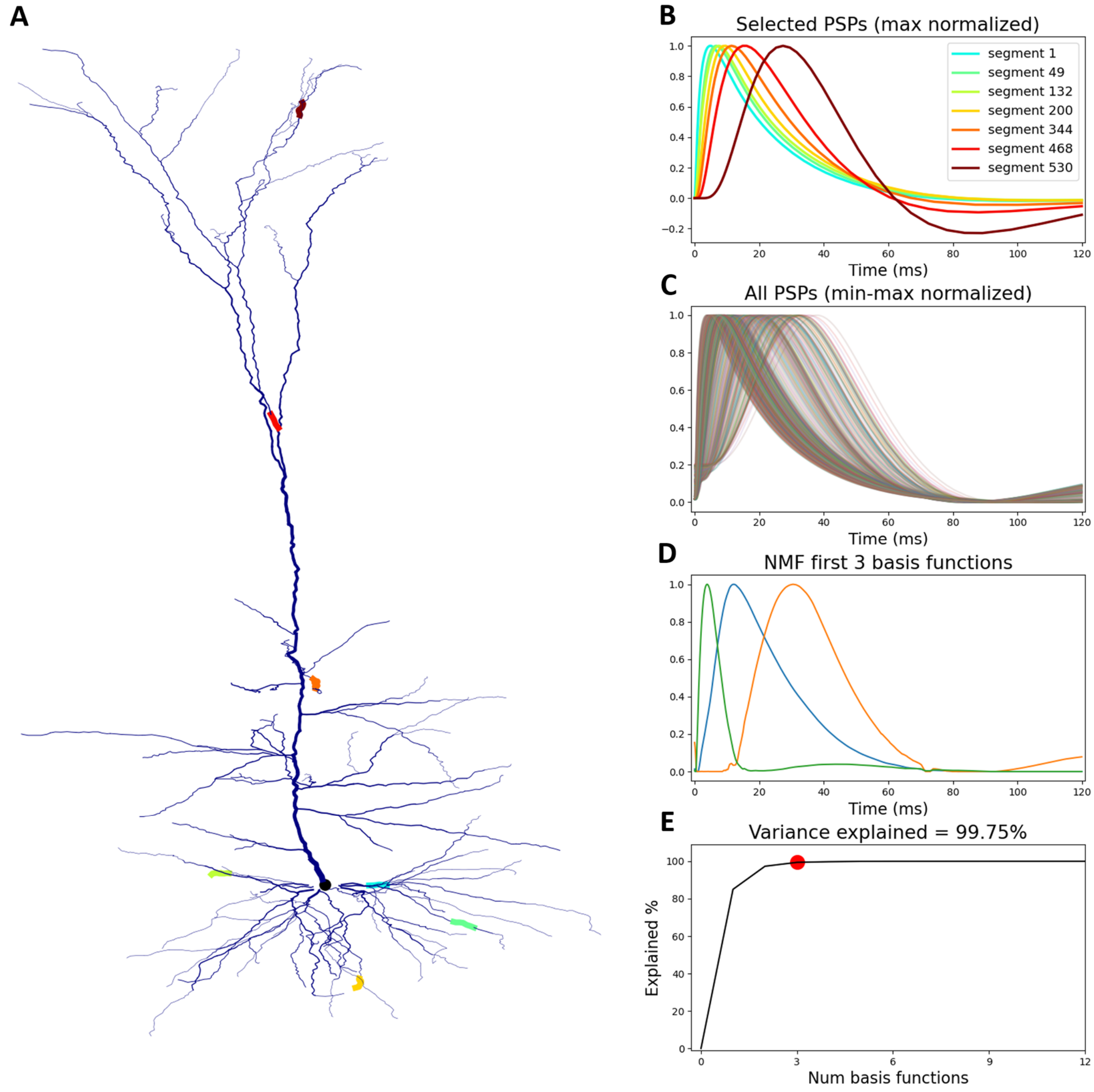

**Fig. S2. The dendritic filters of a reconstructed L5PC detailed biophysical model are also spanned by a three-dimensional basis set of PSPs as does our simplified F&F neuron throughout the paper** (**A**) The morphology of a L5PC that was biophysically modeled in detailed (Hay et al., 2011) with highlighted dendritic segments**.** (**B**) Normalized somatic PSPs in response to excitatory synaptic input at the highlighted dendritic segments in (A). Note that in some of the EPSPs there exists a small hyperpolarization that is due to nonlinear potassium channels at the soma. (**C**) Normalized somatic EPSPs from all dendritic segments in the modeled L5PC. The traces are normalized such that minimal value in the time window is 0 and maximal value is 1. (**D**) First three basis functions of the non-negative matrix decomposition of all EPSPs shown in (C). NMF instead of SVD was used for simplicity of interpretation and visualization purposes. These filters look very similar to those shown in Fig. 4**D**. (**E**) Cumulative variance explained by each basis component demonstrating that the three basis functions shown in (D) can span all EPSPs shown in (C).

## Acknowledgments

We thank all lab members of the Segev and London Labs for many fruitful discussions and valuable feedback.

## Funding

Patrick and Lina Drahi Foundation [https://plfa.info/]
Grant from the ETH domain for the Blue Brain Project
[https://www.epfl.ch/research/domains/bluebrain/]
The Gatsby Charitable Foundation [gatsby.org.uk]
NIH Grant Agreement U01MH114812 (I.S.).

## Author contributions

D.B: Conceptualization, Methodology, Investigation, Visualization, Software, Validation, Data curation, Writing - Original Draft. S.S: Visualization, Software. I.S. and M.L: Conceptualization, Methodology, Writing - Review & Editing, Supervision, Resources, Funding acquisition.

## Competing interests

The authors declare no competing interests.

## Data and materials availability

All code necessary to reproduce all results in this paper are available on GitHub via the link: https://github.com/SelfishGene/filter_and_fire_neuron

All data and live scripts to reproduce all figures are available on Kaggle at the following link: https://www.kaggle.com/selfishgene/fiter-and-fire-paper

# Chapter 3:

## Mapping any spatio-temporal function onto a single neuron, using the equivalent DNN for structural and functional plasticity with end-to-end supervised learning

### Introduction

In **chapter 1** we have established that one can approximate the full complexity of a single detailed biophysical neuron to a great degree of accuracy using a deep artificial neural network (DNN). An interesting property of a DNN is that it is differentiable. It is technically challenging and computationally intensive to calculate the precise derivative of the output of a neuron with respect to its input using strictly the partial differential equations that describe it. On the other hand, it is extremely straight forward to differentiate a deep neural network, as being differentiable is a corner stone of the deep learning revolution. In particular, the deep network that approximates a single cortical neuron from **chapter 1** is differentiable. One can therefore utilize this differentiability property of the DNN and use it to calculate an approximation of the gradient of the output of a realistic cortical neuron with respect to its inputs. The ability to quickly calculate to what extent increasing some synaptic input will increase or decrease the output voltage of the highly convoluted and complex cortical neuron could be used to learn synaptic weights to achieve a specific task.

Let us elaborate, imagine that we wish the neuron to fire for input pattern A, and keep silent for input pattern B. The patterns are fixed so the input spike trains cannot change, but by increasing or decreasing synaptic weights it's possible to increase or decrease the input current injected to the cell for each synapse. The task of making a neuron fire for pattern A and not for pattern B is therefore simply the task of finding which input synapse can be changed is such a way that the increase the output voltage for A is more than for B (e.g., an excitatory synapse located on a branch that is activated in A, but not activated in B will result in increased current injection for A relative to B due to NMDA synapse). Calculating the gradient of the output with respect to the input in one go is basically asking for all synapses simultaneously, be they excitatory or inhibitory, if increasing them a bit will result in increase or decrease of output voltage and by what amount. It is therefore easy to see how this capability allows us to iteratively learn the synaptic weights in order to achieve any desired output by the neuron, quickly and efficiently. Note that this process is analogous to functional plasticity. Now, consider a synaptic weight between an axon and a particular dendritic location is zero at some point in time during training, but the gradient for some pattern suddenly requires that it should be increased in order to improve performance on our task. Increasing this synapse is equivalent to "growing a connection" that wasn't there before. This is analogous to structural plasticity.

To conclude, if we have the gradient signal, it is possible to try and learn synaptic weights that map incoming input axons onto specific dendritic locations that will attempt to achieve any desired task. If there exists an axon-to-dendrite mapping that solves the task, the gradient signal helps us find it quickly and efficiently. All that is left is to try and find such mapping for several interesting tasks, and once successful, we have an existence proof that this task can be mapped onto a single neuron without any reservations.

As we've discussed in the introduction, the single neuron was historically typically abstracted as a simple computational unit that performs mostly perceptron like processing. As we've demonstrated in **chapter 1**, the single neuron is quite a bit more complex than that. But the question arises - can this complexity be utilized? As we demonstrated in **chapter 2**, just modeling a tiny fraction of the complexity of realistic neurons beyond that of a perceptron, i.e. dendritic filtering, can already be a useful computation. In this chapter we go one step further and attempt to ask - **can we utilize the full set of complex computations that take place inside a single neuron to good and targeted use?** Or what fraction of the complexity that we've observed in **chapter 1** is actually utilizable? If it is possible to do so in our simulations with a simple recipe, it might be that the nervous system does so as well during the normal day to day operation of the brain. Below we describe in detail how one can try and map any desired input/output relationship onto the neuron using our differentiable DNN analog of a single neuron.

## Mapping any I/O relationship to a neuron model using its DNN analogue

A single biological neuron receives multiple incoming input spike trains at its various dendritic locations. Formally, for mathematical precision, we denote the spike trains as $x_i(t)$ and their corresponding dendritic locations as $d_i$. When a presynaptic spike occurs, a synapse with conductance $g_i = g_0$ is activated in the neuron simulation. The neuron eventually emits a single output spike train $y(t)$. This is illustrated in **Figure 3.1**, top. In **chapter 1** we discussed creating a simulation that mimics the neuron's I/O function by training a deep temporally convolutional neural network (DNN). The equation that describes the DNN can be written as $\hat{y}(t) = f_{L5PC}(\boldsymbol{x}(t); \theta_{L5PC})$ (bold indicates that the variable is a vector). In **chapter 2** we discussed the phenomenon of multiple synaptic contacts with independent weights for each contact and we wish to update our model from Chapter 1 to accommodate this fact.

In **Figure 3.1,** middle, one can see how it is possible to apply simple rewiring of incoming input axons $a_j(t)$ each with identical synaptic weights on the different dendritic locations. Synaptic rewiring can be thought of as structural plasticity, and we wish to understand the potential of such a plasticity mechanism. If we wish to also support independent weight strengths for each location, we need to train a new model that supports weighted synapses. A potentially elegant way to do so is to keep the input to the network identical. But multiply each spike train by a weight. i.e., for each location $d_i$ the incoming spike train is multiplied by a factor $w_k$ for the DNN when the simulation activates a synapse with conductance $g_i = w_k \cdot g_0$. For this we could train a new DNN that also supports weighted inputs. In the full formalism that includes both multiple contact rewiring with synaptic weights the equation that describes the output of a DNN neuron is $\hat{y}(t) = f_{L5PC}(\boldsymbol{x}(t) = \boldsymbol{W}^{\boldsymbol{T}}\boldsymbol{a}(t); \theta_{L5PC})$. The matrix $\boldsymbol{W}$ is a rewiring matrix between each incoming input axon and each possible dendritic location. When this matrix is binary, this depicts pure rewiring. When this matrix is weighted, this depicts both rewiring and synaptic strengths of each contact. Note that now, $\boldsymbol{x}(t)$ can receive continuous variables instead of spike/no spike binary variables and this represents activation of a synapses with a different conductance at each location at each point in time. This is depicted in **Figure 3.1** bottom and in **Figure 3.2**.

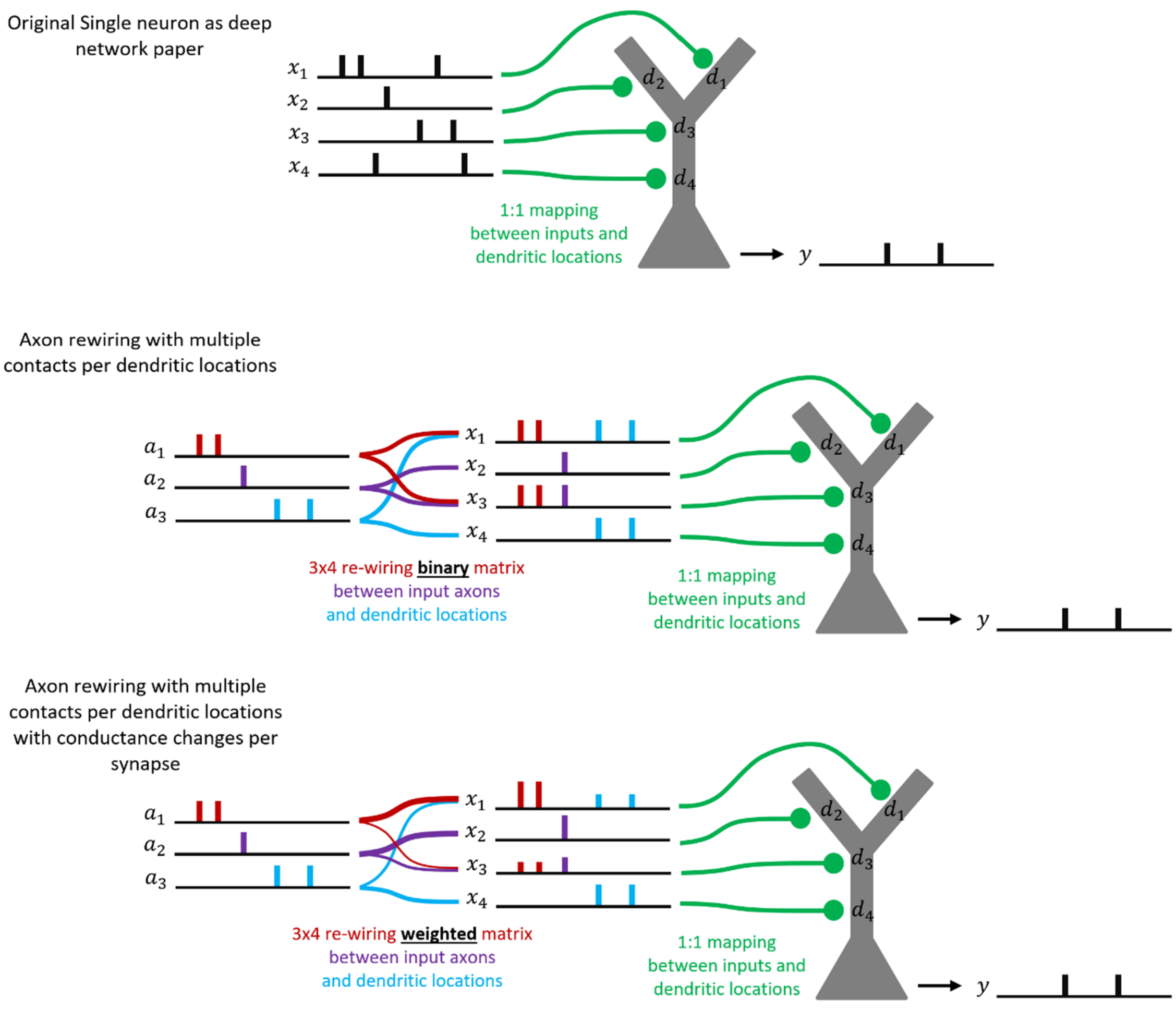

***Figure 3.1. Mapping input spike trains onto a single neuron**. (Top) A 1:1 mapping between incoming input spike trains (denoted by **x**) and dendritic locations (denoted by **d**) that are sent to a neuron simulation which generates a single output spike train (denoted by **y**). For visualization convenience, only excitation is shown, but each dendritic location has separate excitatory locations and inhibitory locations. (Middle) Mapping between incoming input axonal spike trains (denoted by **a**) and dendritic locations thought an intermediate layer of a spike trains (denoted by **x**), one for each input location (denoted by **d**). This non 1:1 mapping between axons and dendritic locations is achieved by a wiring layer that represents multiple synaptic contacts of incoming axons onto multiple different dendritic locations ("dendroplexing"). (Bottom) Similar to the middle figure, only this time the wiring is weighted, representing a weighted re-wiring matrix of synaptic strengths with independent weights assigned to each multiple contact.*

In **Figure 3.2** bottom we can see the result of the full scheme that contains multiple synaptic contacts re-wiring with independent weights for each contact flowing in a simple feedforward computational graph from incoming input axons $\boldsymbol{a}(t)$ to output spike train $y(t)$. The axons $\boldsymbol{a}(t)$ move thought a matrix multiply $\boldsymbol{W}^{\boldsymbol{T}}$ that mixes the incoming axons into an intermediate continuous weighted spike train representation of mixed inputs $\boldsymbol{x}(t)$ that feed the neuron. The neuron itself is replaced by a DNN $\hat{y}(t) = f_{L5PC}(\boldsymbol{x}(t); \theta_{L5PC})$ to create the output spike train $y(t)$. The DNN is pretrained and its weights $\theta_{L5PC}$ will be kept frozen. The only learnable parameters in the scheme we are about to describe below are those of $\boldsymbol{W}^{\boldsymbol{T}}$ that represent axonal rewiring and synaptic strengths.

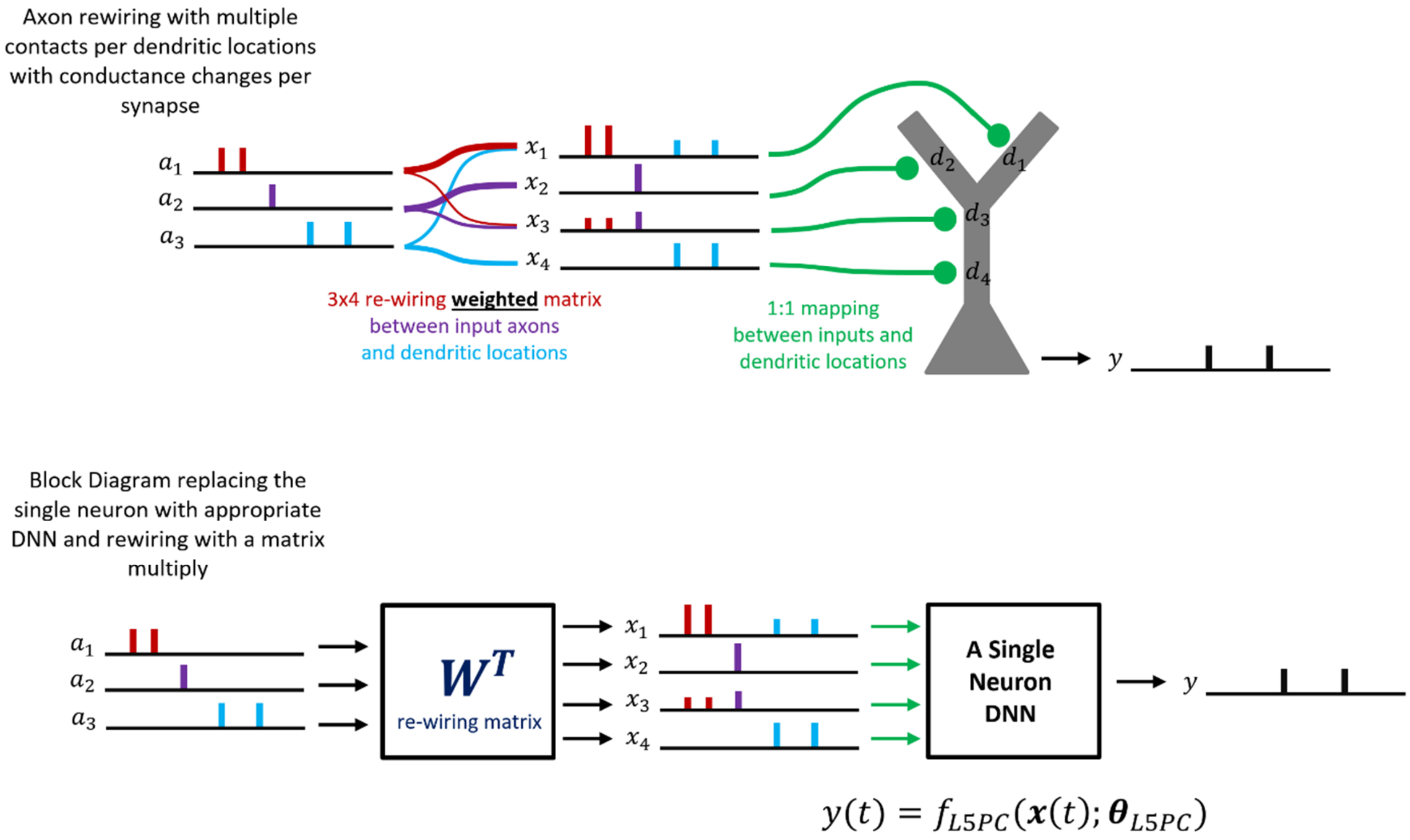

***Figure 3.2. The DNN equivalent of mapping input axons onto a single neuron resulting in a simple differentiable computational graph that can easily be used in any deep learning framework**. (Top) Mapping between incoming input axonal spike trains (denoted by **a**) and dendritic locations thought an intermediate layer of a weighted spike trains (denoted by **x**), one for each input location (denoted by **d**) that is sent to the neuron which generates a single output spike train (denoted by **y**). The non 1:1 mapping is achieved by a wiring layer that represents multiple synaptic contacts of incoming axons onto multiple different dendritic locations. This wiring is weighted, representing independent weights assigned to each multiple contact. (Bottom) A schematic illustration of the same circuit as at the top, this time replacing the neuron with an equivalent DNN that represents this neuron (the process of replacing a neuron with a DNN was discussed in Chapter 1), with and additional incoming matrix of a weighted re-wiring matrix of synaptic strengths that represent multiple synaptic contacts (See Chapter 2 for discussion on multiple synaptic contacts)*

Now, consider some dataset that is comprised of multiple input/output pairs, e.g., MNIST handwritten digit dataset or an auditory speech recognition dataset. Let us now represent the dataset in terms of multiple inputs and single output spatio-temporal spike train representation. Formally let's denote the spike train representation of all samples in the dataset as $\{\boldsymbol{a}_k(t), y_k(t)\}_{k=1}^{M}$, where $M$ is the number of total samples in the dataset. We wish to map this data onto the single neuron and ask various questions about the result, for example what level of performance we can achieve on this task using the specific spike train representation we have chosen and how does this compare to simple integrate and fire neuron models. We can do this using basic supervised learning and optimizing only the mixing weight matrix $\boldsymbol{W}^{\boldsymbol{T}}$. i.e.

$$W_{opt}^{T} = \underset{W^{T}}{\mathbf{argmin}} \sum \|y_k(t) - \hat{y}(t)\|^2 = \underset{W^{T}}{\mathbf{argmin}} \sum \left\| y_k(t) - f_{L5PC}\left(\boldsymbol{W^{T}} \boldsymbol{a}_k(t); \theta_{L5PC}\right) \right\|^2$$

Note that $\theta_{L5PC}$ remains frozen and only $\boldsymbol{W^T}$ is allowed to change.

In **Figure 3.3** one can see two examples of potential tasks that can be mapped onto input output spike train representation and evaluate whether they can be mapped onto the single neuron with what level of accuracy. Note that it is possible that a single neuron might achieve variable levels of performance on the same task, depending on the particular spike train representations scheme that we choose. This opens up the avenue of investigating what is the correct way to present information to a single neuron in order to unlock its pattern recognition capabilities. We discuss this further in **chapter 4**.

Overall, we find this formalism to be simple and extremely elegant. It's simplicity also allows it to be compositional and extendable to other neuron types and to much larger networks of highly complex detailed biological spiking neuron models. We describe this in detail in the chapters to come.

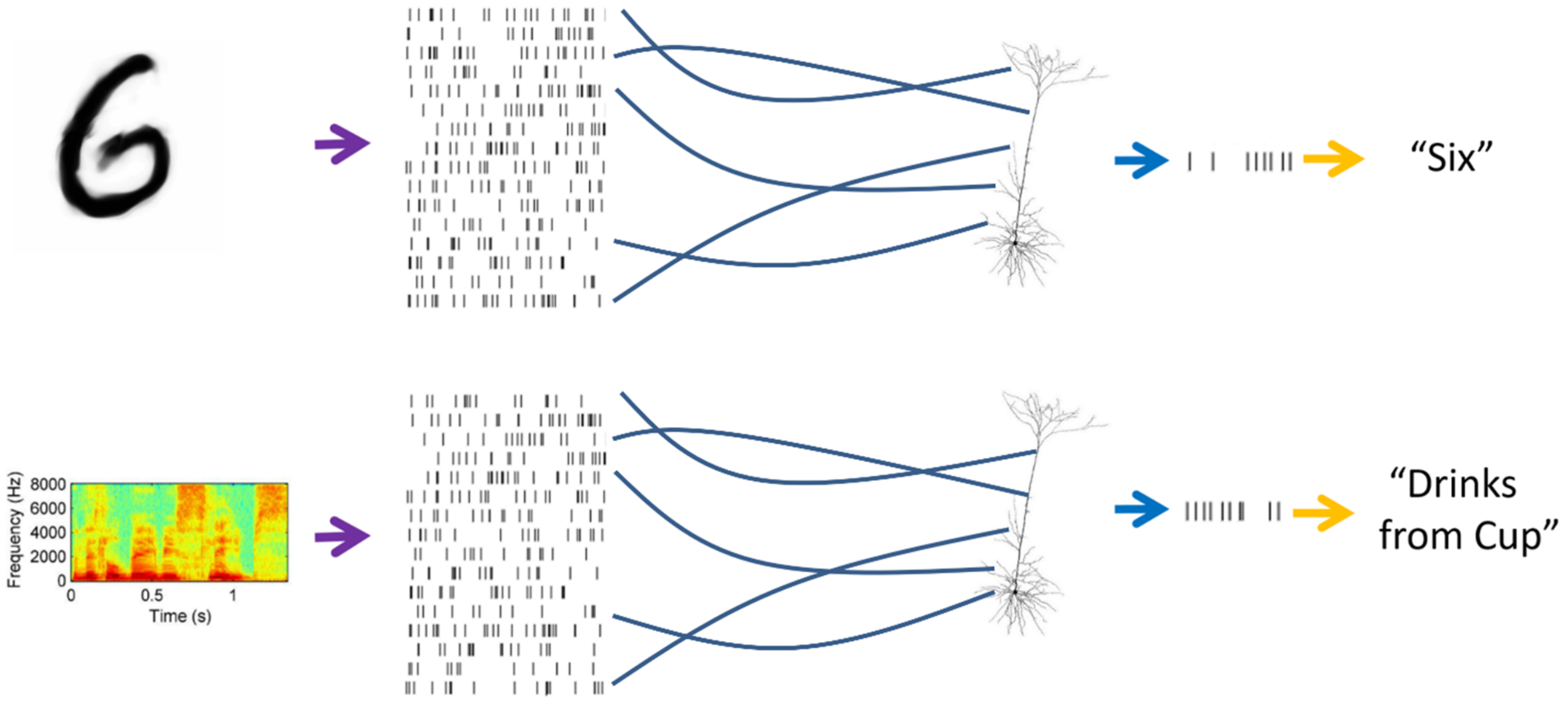

***Figure 3.3. Example of interesting tasks that can potentially be mapped onto a single neuron using our framework****. (Top) an image of the digits “6” is first converted into some spike train representation, this is then mixed and sent to a neuron simulation. The neuron emits an output spike train, and we seek to decode from the output spike train the category of the digit “six”. (Bottom) a similar example to the one above, only this time it is tasked to recognize speech e.g., “Drink from a cup”.*

## Mapping a temporally precise XOR function onto a single neuron – first attempt

In this section we describe our first attempt to map a temporally precise XOR function onto a single neuron and validating this transfer to a single neuron simulation. The XOR operation has a rich history in the field of neuroscience and machine learning, as it's demonstrably cannot be learned by a perceptron like neuron (Minsky and Papert, 1969), but more recent studies showed that a single biological neuron can, in principle, perform such a task (Gidon et al., 2020). This is a complete, concrete and simple end-to-end example of all stages of the process of mapping a simple function that allow us to get a feel for how results should look like and find the main technical challenges along the way. As we'll later discuss, the main challenge in this formalism appears to be related to the phenomenon of adversarial examples (Goodfellow et al., 2014) that typically occurs with models that are effectively being attacked. This phenomenon became quite famous with deep artificial neural networks in recent years.

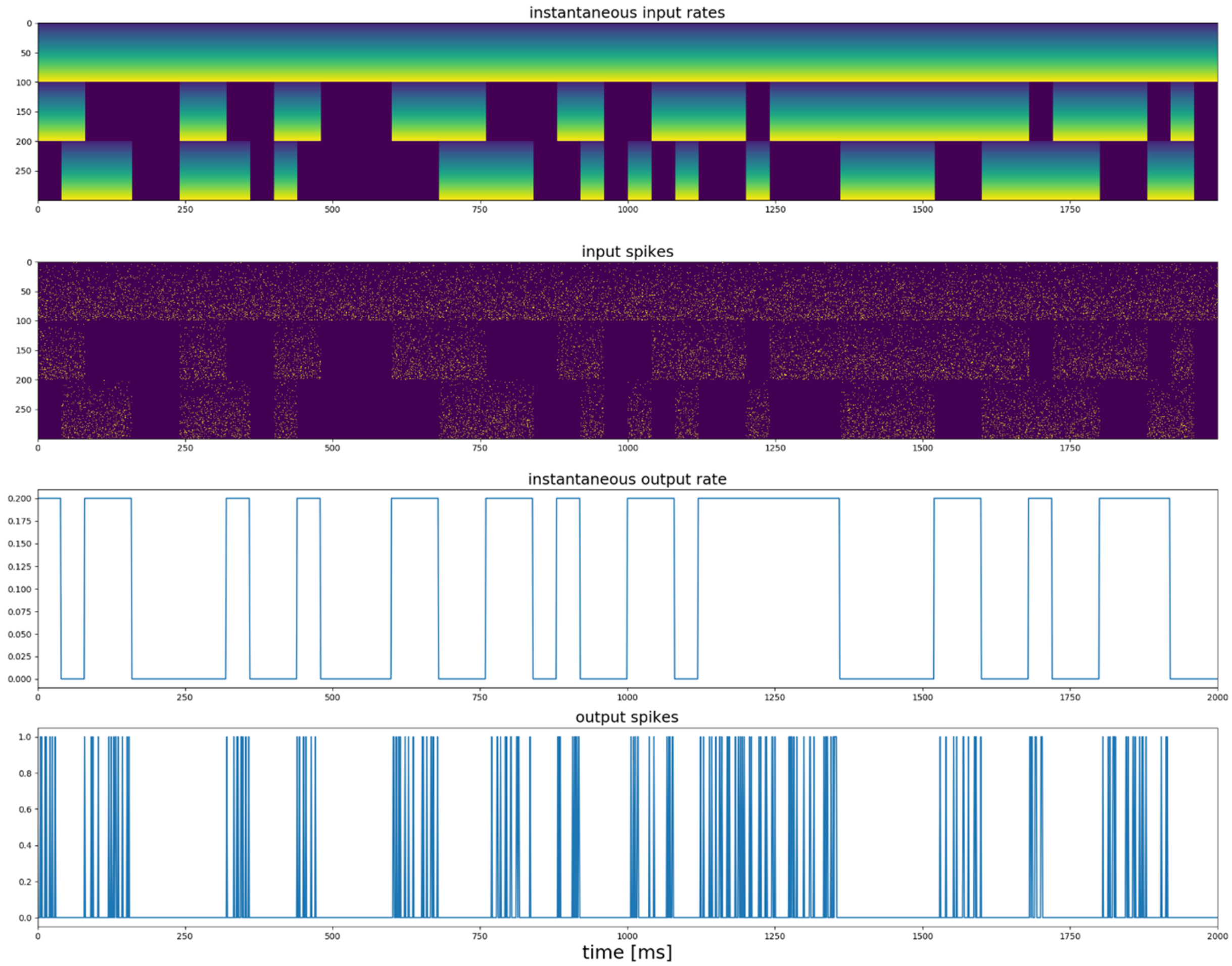

***Figure 3.4. Input and output spike train representation of the temporal XOR task**. (Top two plots) an example of instantaneous firing rates (as heatmaps) and sampled spike trains (as binary matrix) for 300*

*axons that are incoming and define two binary input and an additional constant input. (Bottom two plots) The output instantaneous firing rate of the XOR operation and sampled output spike train.*

In **Figure 3.4**, we can see how we chose to set up the XOR task in the temporal domain. For a period of 50 ms, the instantaneous firing rate of the input represents two binary inputs that vary as a function of time. To those we add a single constant input that is always positive in order to make the network learn also a bias/threshold term. These 3 inputs are converted to spike trains by sampling from piecewise constant Poisson distribution. For each point in time, the XOR operation is calculated on the two binary inputs and we calculate a desired output instantaneous firing rate for our neuron. From it we sample the desired output spike trains. Every 50ms the input firing rates are randomly resampled.

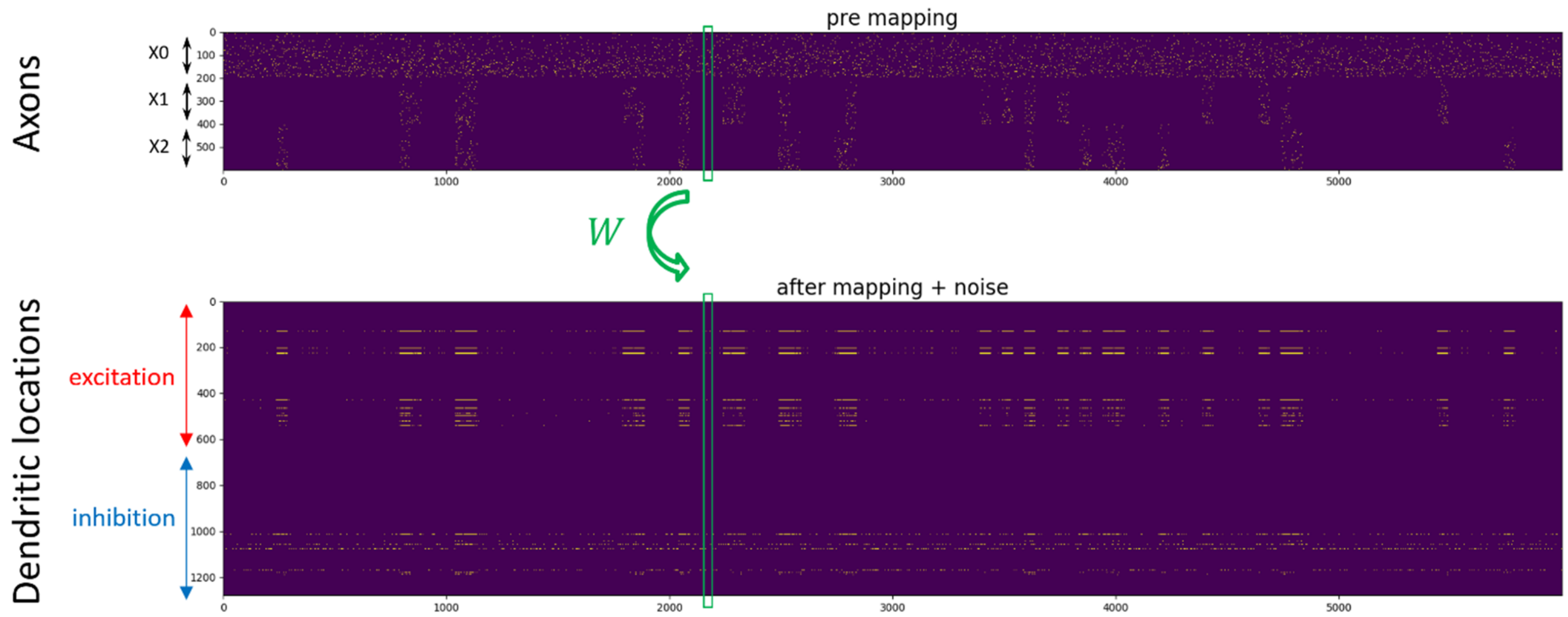

***Figure 3.5. Example of input spikes that are mapped to excitatory and inhibitory dendritic synaptic locations via binary mapping matrix W****. (Top) incoming input spike trains. (Bottom) the spike trains on the top mapped to all dendritic locations of a L5PC neuron using a mapping matrix W. In this example, we observe that only a handful of locations on the dendritic tree are used to map the input onto (vertical yellow lines), and those differ for excitation and inhibition.*

We can now attempt to map the input spike trains onto to the neuron at different locations on the dendritic tree in order to get the desired output spike train. We can see in **Figure 3.5**, the axons before and after the mapping using weight matrix $\boldsymbol{W}$. In **Figure 3.6** we can see the results of this experiment. After a brief optimization period, we manage to achieve a good fit where the output of the DNN matches our desired output spike probability (**Figure 3.6**B). But a problem appears when we wish to translate the learned mapping and test it on the real neuron simulation as can be seen in **Figure 3.6**D. Learning clearly occurs, but there is also spiking at times where we wish spikes not to occur. Note that the predicted spike probability of the neural network output indicates that there should not be any spiking at times when there are spikes in the neuron simulation (e.g., at around 2000ms timepoint). This indicates that we've reached the adversarial example regime in which our manipulation is able to alter the neural network according to our objective, but the underlying neuron is not similarly affected by this manipulation. We also note that, by examining **Figure 6**C, we can see that the neural network correctly predicts high somatic voltage at the times when the false positives occur. We can therefore add it to our optimization loss and see if performance improves.

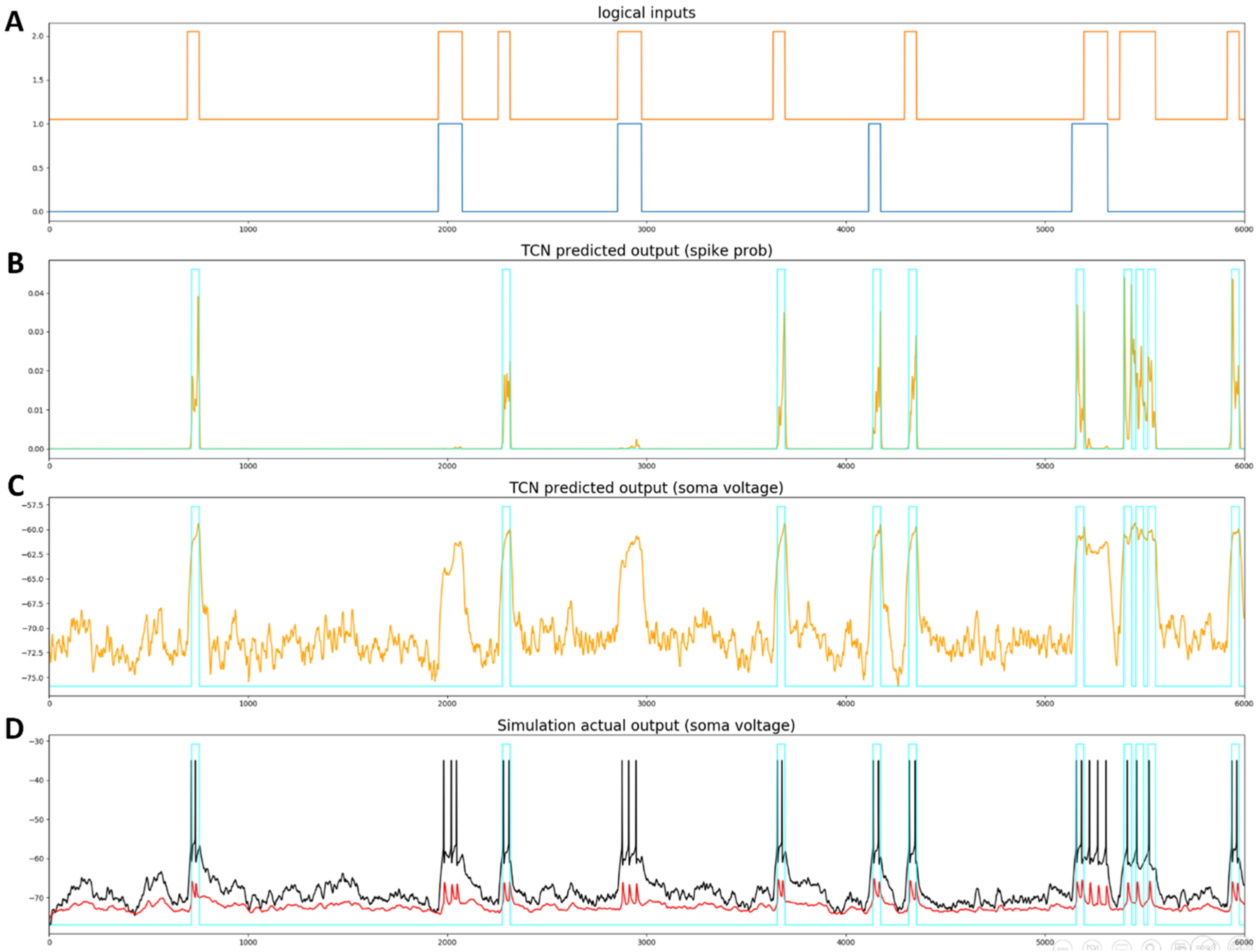

***Figure 3.6. Example of inputs and output for a learned mapping based on an output spikes loss, DNN succeeds, transfer to neuron simulation partially fails***. *(A, topmost). Two logical inputs for the XOR task. (B) In cyan is the desired logical output, and in orange is the output spike probability obtained by the neural network. We can see a good match (C) in cyan is the desired logical output and in orange is the predicted somatic voltage trace. We can potentially see a few false positives here (D, bottom most) in cyan is the desired logical output. In black is the result of neuron simulation after applying the learned mapping from the DNN. We can a non-perfect transfer as some firing occurring at non intended locations (e.g., second burst from left). In red is the apical nexus voltage trace.*

**Figure 3.7** shows a repeated attempt after adding a loss term that also tries to keep the somatic voltage output of the neuron low during times where there should be no spikes at the output. This is done in order to motivate the network to be more self-consistent with low voltages during periods of no spiking and have a large margin of safety regarding accidental crossing of spike threshold. Eventually, this is another failed attempt, and we still reach adversarial example regime here as well. Although if we squint, we might be able to see that the amount of spikes at false positives is much reduced compared to the number of spikes at true positive times and one could perhaps decode the XOR function by looking at firing rates. But, nevertheless, this is still a failed attempt since there is a discrepancy between what the DNN representation of the neuron output and the actual neuron simulation outputs.

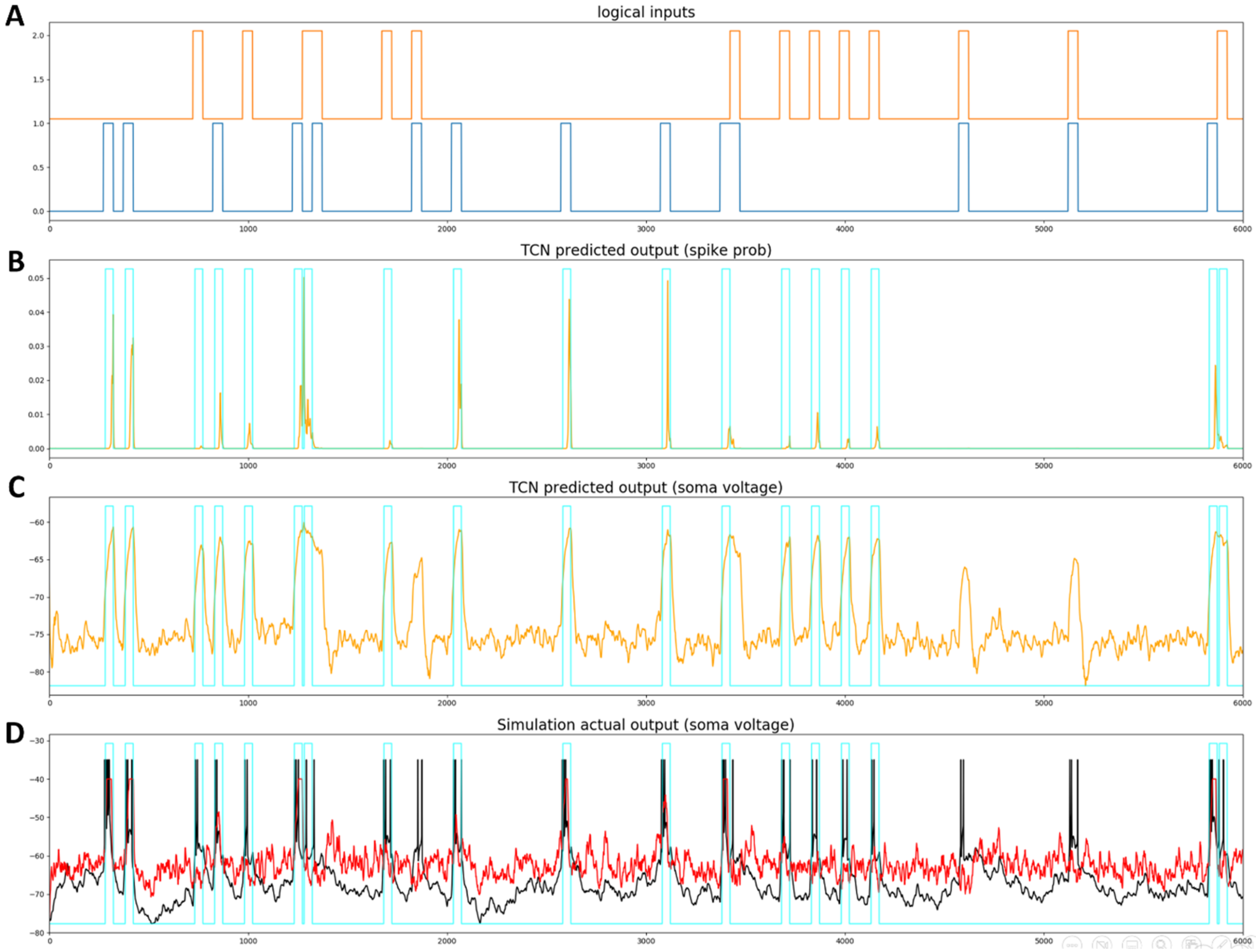

***Figure 3.7. Example of inputs and output for a learned mapping based on an output spikes loss and dendritic voltage loss, DNN succeeds, transfer to simulation still partially fails****. (A, topmost) two logical inputs for the XOR task. (B) in cyan is the desired logical output, and in orange is the output spike probability by the neural network. We can see a good match (C) in cyan is the desired logical output and in orange is the predicted somatic voltage trace. We can potentially see a few false positives here (D, bottom most) in cyan is the desired logical output. In black is the neuron simulation after applying the learned mapping from the DNN. We can a non-perfect transfer and some firing occurring at non intended locations. In red is the voltage trace in the apical nexus.*

We could speculate as to the source of this partial failure in mapping between the DNN and the NEURON simulation and suggest potential action to mitigate it. The literature on adversarial examples is full of various methods to mitigate it, but the most obvious and thus far best method is using adversarial training. i.e., during training of the DNN that mimics the single neuron, inject into the training set also some adversarial examples. We did not implement this here, but we hope that future work will do so. Instead, we continue below to describe possible future work. Before we do that, we also show that it is possible to visualize the mapping matrix and see which input is mapped onto which dendritic locations. This is depicted in **Figure 3.8** where heatmaps are shown of the neuron morphology for each of the 3 inputs of the XOR operation ($x_0$ is constant positive input). They depict dendritic locations that receive large weights from each input. This visualization method can be applied to any neuron and potentially provide hints about the specific strategy the neuron uses to complete this task (e.g., spatially clustering of inputs)

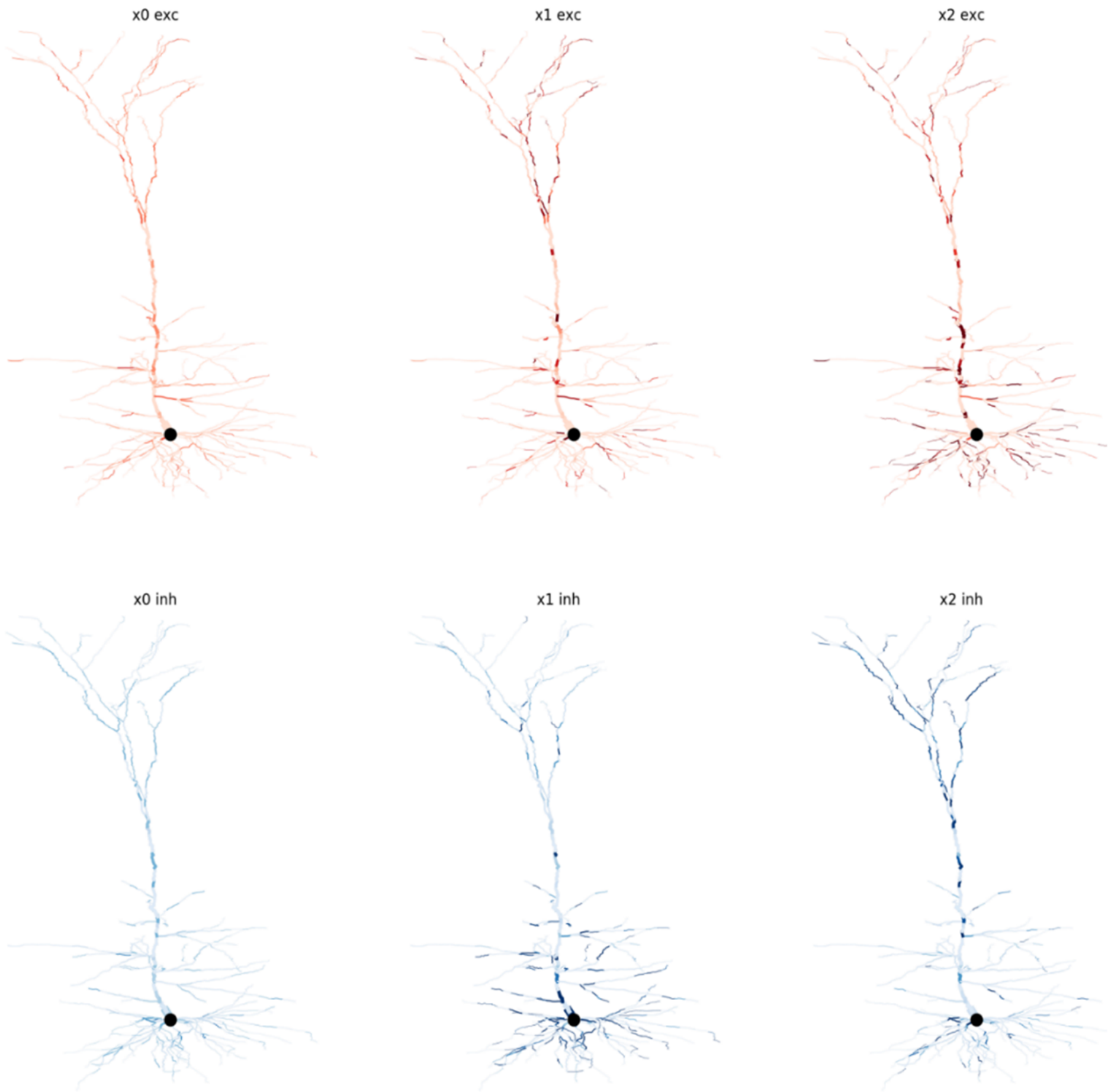

***Figure 3.8. Visualization of the learned mapping for the XOR problem showing which inputs in XOR task are mapped to which dendritic locations (for both excitation and inhibition).*** *(Top) heatmap of total synaptic weights coming from the specified logical input channel axon onto each dendritic segment for excitatory input. (Bottom) same as above but for axons that are mapped onto inhibitory inputs. (x0 is the constant input representing the bias term)*

From this attempt to map XOR onto a realistic L5 cortical pyramidal neuron we can see that the concept of our proposed approach works well, and that the failure of the process occurs only at the stage of transferring the input synapses from the DNN to the simulated neuron. Most importantly, gradients flow through the DNN neuron model very well and successfully manage to optimize the desired loss. The failure in the transfer to the simulation stage strongly suggests that there are two distinct possibilities as reasons for failure. (1) DNN approximation is not good enough for the specific regime of input spike train activations. (2) These are adversarial conditions that elucidate adversarial examples. There are two simple possible solutions that come to mind: (1) adversarial training (2) better DNN approximations to the simulated neuron, specifically under input conditions that better resemble the statistics of input activations after learning a mapping with a specific purpose. We hope that future work will be able to learn from this example.

### Constructing a morphology specific DNN architecture to describe the I/O of a single neuron type.

In this part we will describe a possible extension to our work in **chapter 1** with the goal to construct a more accurate, more compact, and much more interpretable DNN model that approximates a single realistic neuron model. These properties (accuracy, compactness, and interpretability) are all positive properties on their own right, but one additional reason to want them is that it might be the case that a better and more compact model will be more resilient against adversarial attacks and therefore useful in out attempt depicted in the chapter thus far.

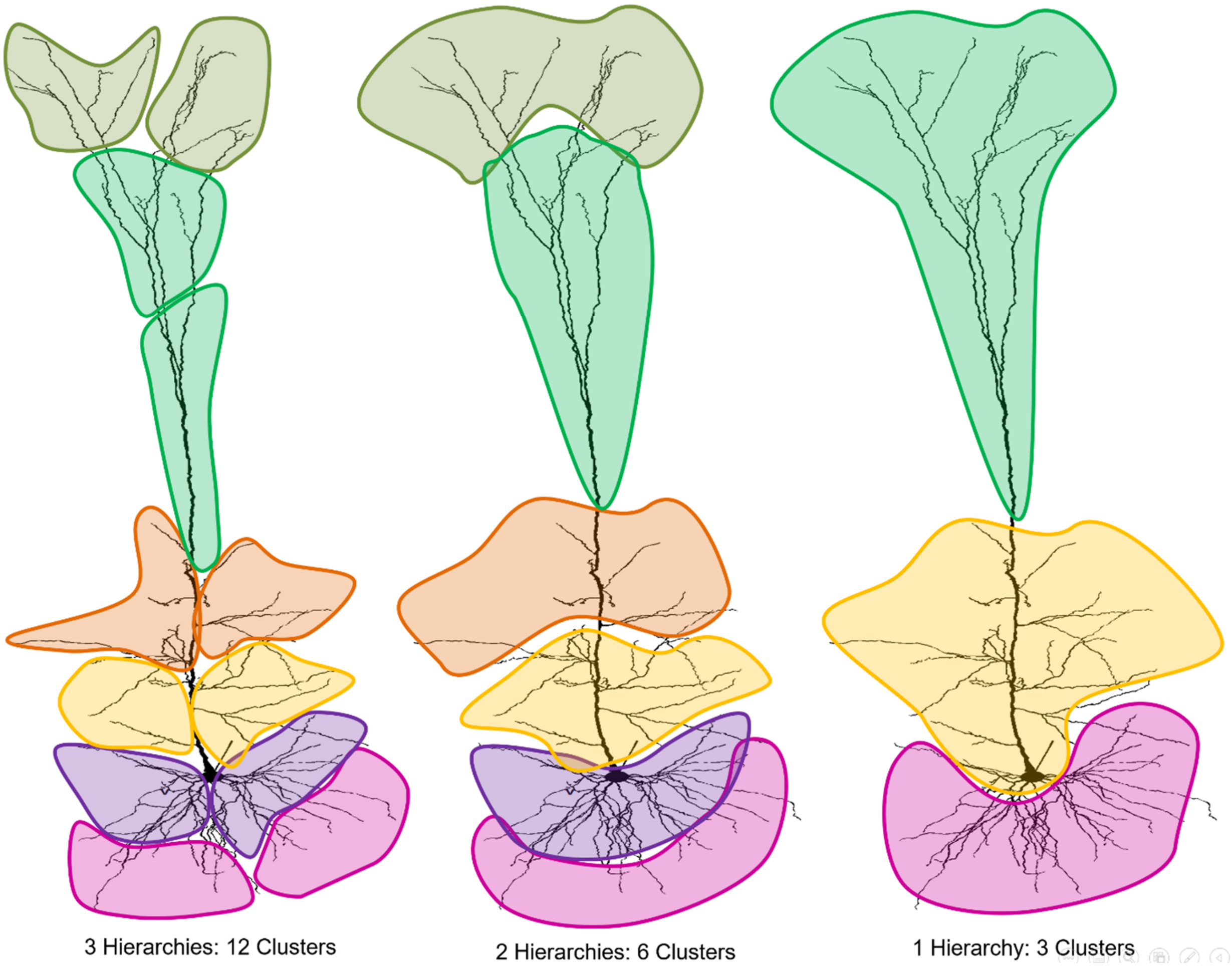

***Figure 3.9. Morphology of single neurons is inherently hierarchical and signal processing in a neuron inherently follows this hierarchy.*** *Depicted are 3 different example hierarchies by illustrating*

*nested clusters on the morphology of a single neuron. In this illustration, the nested hierarchical clustering of the morphology is divided into 12, 6 and 3 possible clusters.*

The dendritic tree of a single neuron is elaborate and vast, but it is also electrotonically distributed to a large extent (Rall and Rinzel, 1973). This means that some parts of the tree interact weakly with each other (e.g., two distal tips that have a large path length distance between them), and other parts are interchangeable (e.g., for two nearby inputs on the same branch, it doesn't matter which of them is active from the perspective of the rest of the neuron). The morphology of a dendritic tree therefore has a hierarchical structure and nearby locations usually are activated similarly. One can imagine that it is possible to coarse grain the morphology into its main regions/clusters. An example of several potential hierarchical clusters is illustrated in **Figure 3.9** (see also e.g., Fig. 6 in (Eyal et al., 2018))

The temporally convolutional neural network we trained in **chapter 1** does not account for these semi-independent dendritic subtrees. In the work described in **chapter 1**, along the spatial domain (dendritic location), the architecture is a fully connected all-to-all connectivity and special handling was performed only on the temporal axis (convolutions along the temporal axis). Based on what we know about neuron's biophysics and cable properties, this means that during training the fully connected structure was force learn these morphological constrains, instead of us constructing in advance a cleverer architecture that can automatically capture these features and make the training process simpler and more efficient.

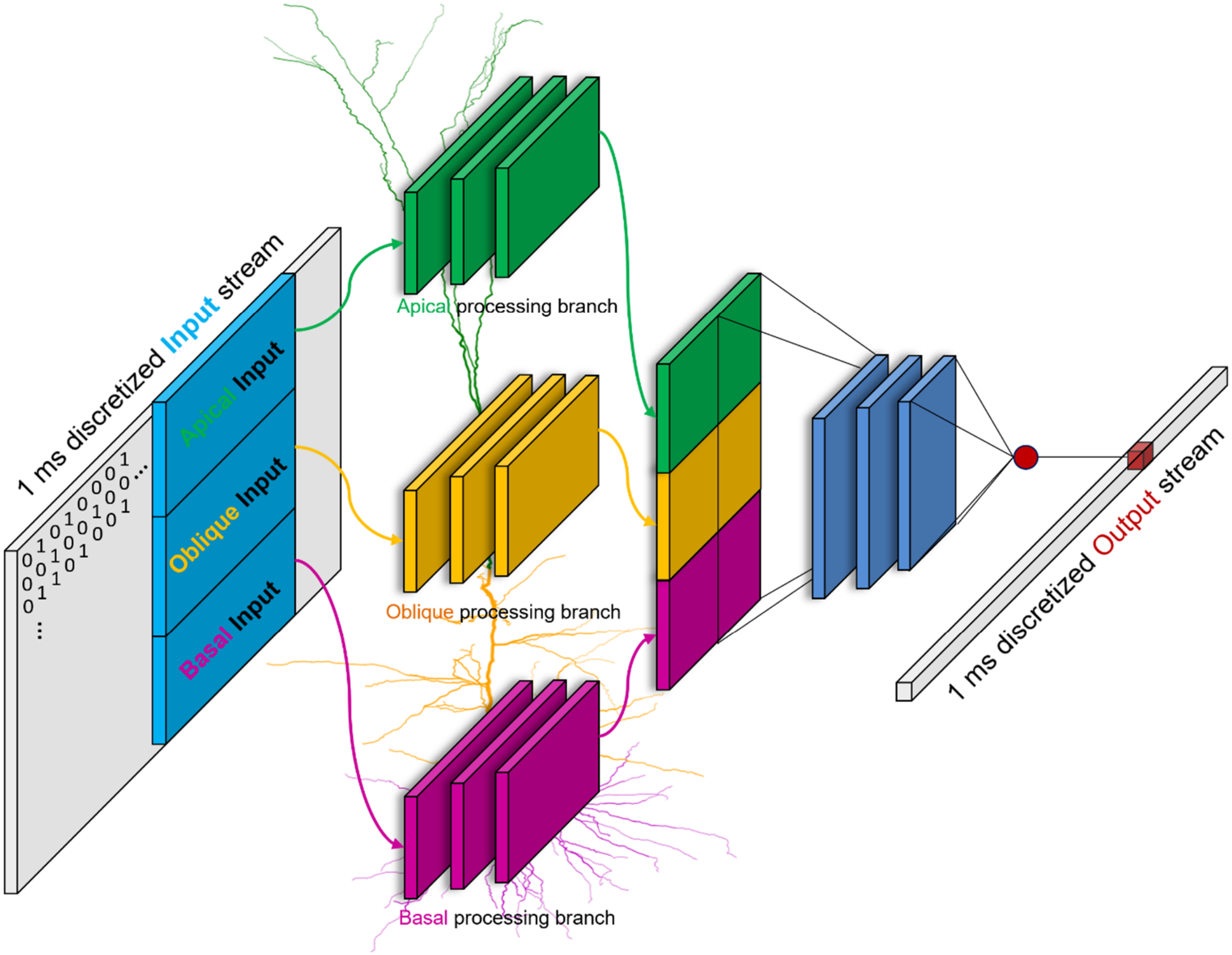

***Figure 3.10. Illustration of hierarchical and locally connected deep temporally convolutional network architecture that can mimic a single neuron.*** *Inputs are separated according to the putative dendritic clusters (in this case 3 separate clusters/subtrees are illustrated (apical, oblique, basal). First all the input associated to a single cluster are processed (in this case by 3 layers of temporally convolutional layers). Then the activations are concatenated and then processed by the higher-level clustering (in this case – the entire neuron). This is an illustration of a nested hierarchy with 2 hierarchies but, depending on the specific dendritic tree and input conditions, any number of such hierarchical structure is possible.*

In **Figure 3.10**, we illustrate a potentially new DNN architecture for a neuron, in this case separating the morphology into 3 main regions (subtrees) – the apical trunk, the oblique regions and the basal tree. For each of these regions, depicted are 3 layers of local processing, then the output of these 3 parallel branches are concatenated and processed together for 3 additional layers before giving the final output of the neuron. This architecture is much more parameter efficient compared to the networks we used in **chapter 1**. This is only an example of a 2-hierarchy architecture. It is possible to create several additional such hierarchies. For example, adding an additional hierarchy that separates the basal regions into two distinct regions (like in **Figure 3.9** middle), processes each independently before concatenating and processing together. And similarly for the oblique and apical tuft regions. The separation into clusters can be performed based on morphological path distances between dendritic compartments alone, or via empirical cross correlations between the voltage traces at each dendritic compartment and thus taking into account both the morphological features and electrical features in a joint fashion (Rabinowitch and Segev, 2006). In addition to being parameter efficient, it is also a much more interpretable representation of a neural network architecture. Suppose the best architecture of one neuron type will turn out to be to 3 hierarchies compared to a single hierarchy of a different neuron type. This will automatically give up some intuitive grasp about the computation going on in that neuron. **Figure 3.11**, left, depicts an example of potential results of such a comparison – plotting the ability of a specific neural network architecture to capture the output of a realistic neuron model similar to what we've done in **chapter 1**. An additional potential advantage is that it is possible that if we use the correct number of hierarchies for a particular neuron, then the network overall could turn out to be less deep, thus building a more compact DNN model for that neuron. This type of result is depicted in **Figure 3.11** on the right.

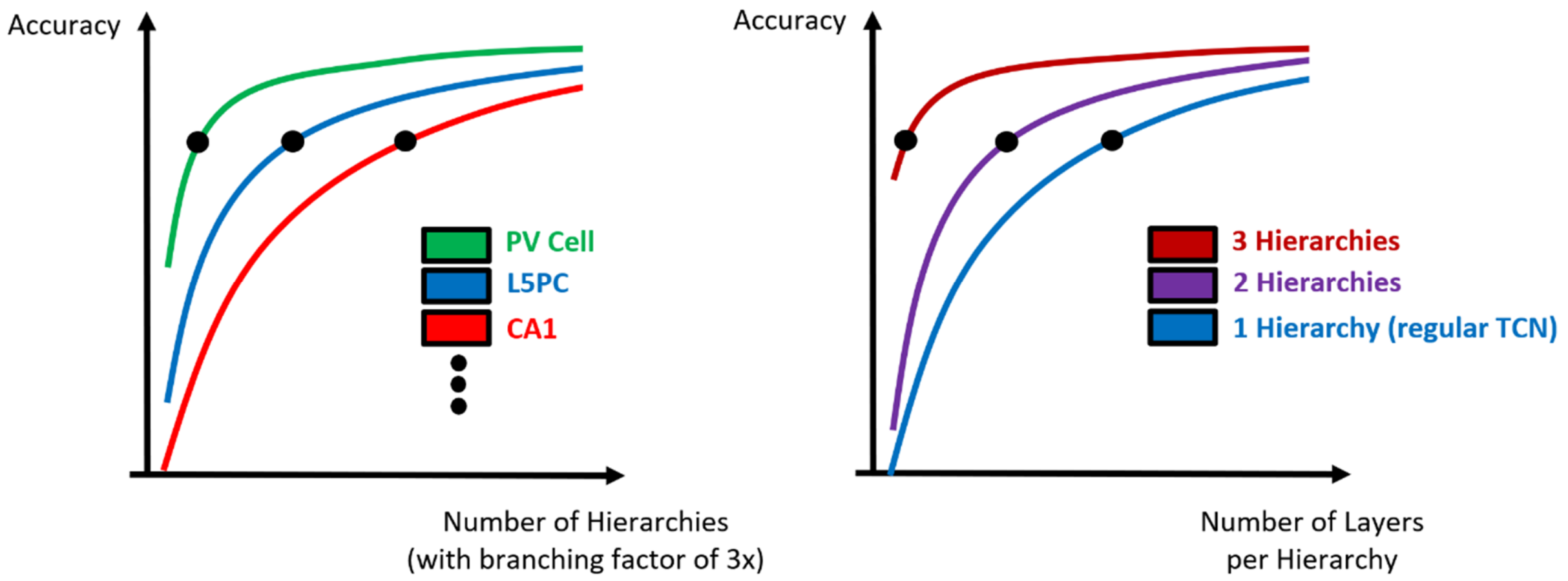

***Figure 3.11. Possible interesting questions to answer regarding different neuron types and their computationally equivalent DNN architectures.*** *(left). A possible result of the Accuracy in replicating the output of a realistic neuron model as a function of the number of hierarchies for various neuron types. In this case the green curve indicates that this (electrically compact PV inhibitory) neuron type is best thought of a shallow processing of information, whereas the red curve (electrically distributed hippocampal CA1 neuron) represents a neuron that has a large number of hierarchical computations within itself. (right). A potential result of Accuracy as function of number of layers per hierarchy. It is possible that due to reduced number of parameters, with more hierarchies one can reduce the total number of layers that achieve a similar level of accuracy.*

## Discussion

In this **chapter 3** we have attempted to embark on an ambitious synthesis project that joins both **chapters 1** and **2** and maps any input/output function on to a single neuron, similar to what we have done in **chapter 2**, but this time while taking into consideration all of the complexities of the single neuron that we have modeled in **chapter 1**. We precisely formalized the method, demonstrated that the method is indeed quite possible, although additional work is required to overcome the challenges we've encountered. We also outline a few possible ways in which to overcome these challenges including an architectural change that could yield merits on its own.

# Chapter 4 - Appendix:

## Beyond single neurons: Mapping any input/output task onto a network of realistic spiking neuron models

### Introduction

A single neuron is important, but computations in the brain are performed using a network of neurons. Thus far, mapping complex computational tasks to networks of spiking neurons has proved challenging, but see (Perez-Nieves et al., 2021). Moreover, training networks of realistic neuron models to perform tasks is basically nonexistent, but see (Chindemi et al., 2022) where the authors utilize a supercomputer to simulate the plasticity of a cortical column when presented with very simple inputs. In this chapter we wish to demonstrate how one can utilize the method presented in **chapters 1 & 3**, in order to train networks of realistic neurons to achieve specific tasks without using a supercomputer.

To date, although we have deep artificial neural networks that display very good, e.g., vision capabilities, we still don't have any spiking neuron network that is able to display even rudimentary vision capabilities (with physiological and realistic levels of firing rates). We therefore don't have any mechanistically plausible model of animal vision. Artificial neural networks, although loosely based on the operation of biological neurons, clearly display major differences. These differences are sufficiently large that there is not even an ad hoc realistic transfer to accommodate the most basic mechanism of operation in the biological nervous system – the spike. Note that nonrealistic transfers between artificial neural networks and spiking networks do exist, but they are possible but only if we allow firing rates to be unrealistically high. This modeling gap suggests that the way computations are utilized in artificial networks are potentially very different from how they are utilized in biological brains, and until this gap is addressed, there is no definitive proof otherwise.

Our hypothesis is indeed that there is a fundamental difference between the way biological neurons achieve their computational goals and the way artificial networks achieve their goals. Specifically, we hypothesize that biological computations are utilizing time as an auxiliary computing dimension. Namely that sensory neurons convert their input into spatio-temporal "packets" of coordinated neuronal activity that are then processed by dendrites in unique ways. In this chapter, we go into great depth about how one would go about proving such a hypothesis.

In the **introduction** we have discussed that there appear to be several conserved microcircuits in the brain with very distinct network connectivity structures between very distinct neuron types. A question arises, what are these highly specific and highly conserved structures used for? As we've previously seen, dendritic morphological structure and electrical properties directly related to how and which spatio-temporal patterns neurons are "attuned to". Our hypothesis is therefore that particular microcircuit structures are directly related to generating and processing spatio-temporal "packets" of highly coordinated activity, each with its own distinct characteristics (e.g. dense vs sparse spiking over shorter or longer periods of time). We address these questions in this chapter and suggest a very specific path to elucidating the answers. Namely, we describe in great detail a set of simulation experiments with increasing complexity that attempt to map which tasks can be achieved by various network circuits.

Finally, in this chapter we also suggest a surprisingly simple recipe about how one can, in principle, determine the precise overarching computation that is performed by any microcircuit or architecture by utilizing information about synaptic update rules from experimental work (In this context we refer to the word "computation" as in David Marr's highest level of analysis – i.e., determine the computational objective of the circuit). Namely, we describe how one can get at the computational level of analysis (in Marr's framework) of a neuronal circuit by incorporating knowledge from the implementational and algorithmic levels of analysis. Specifically, we describe in detail a way how this can be done by utilizing information about the biological learning rules implemented by the brain at the single synapse level.

## Mapping any input/output task onto a microcircuit of realistic spiking neurons

For simplicity's sake, we will first discuss here only a simplified case of Filter and Fire (F&F) neuron models (introduced in **chapter 2**), but eventually we will show how one can model networks of neurons of any set of realistic neuron types. Indeed, it is interesting to ask what kinds of tasks can be achieved using a network of spiking neurons and, in particular, whether a network consisting of say "Large Regular Spiking" (LRS) neurons can perform better (or differently) on specific computations than a network consisting of "Small Fast Spiking" (SFS) neurons.

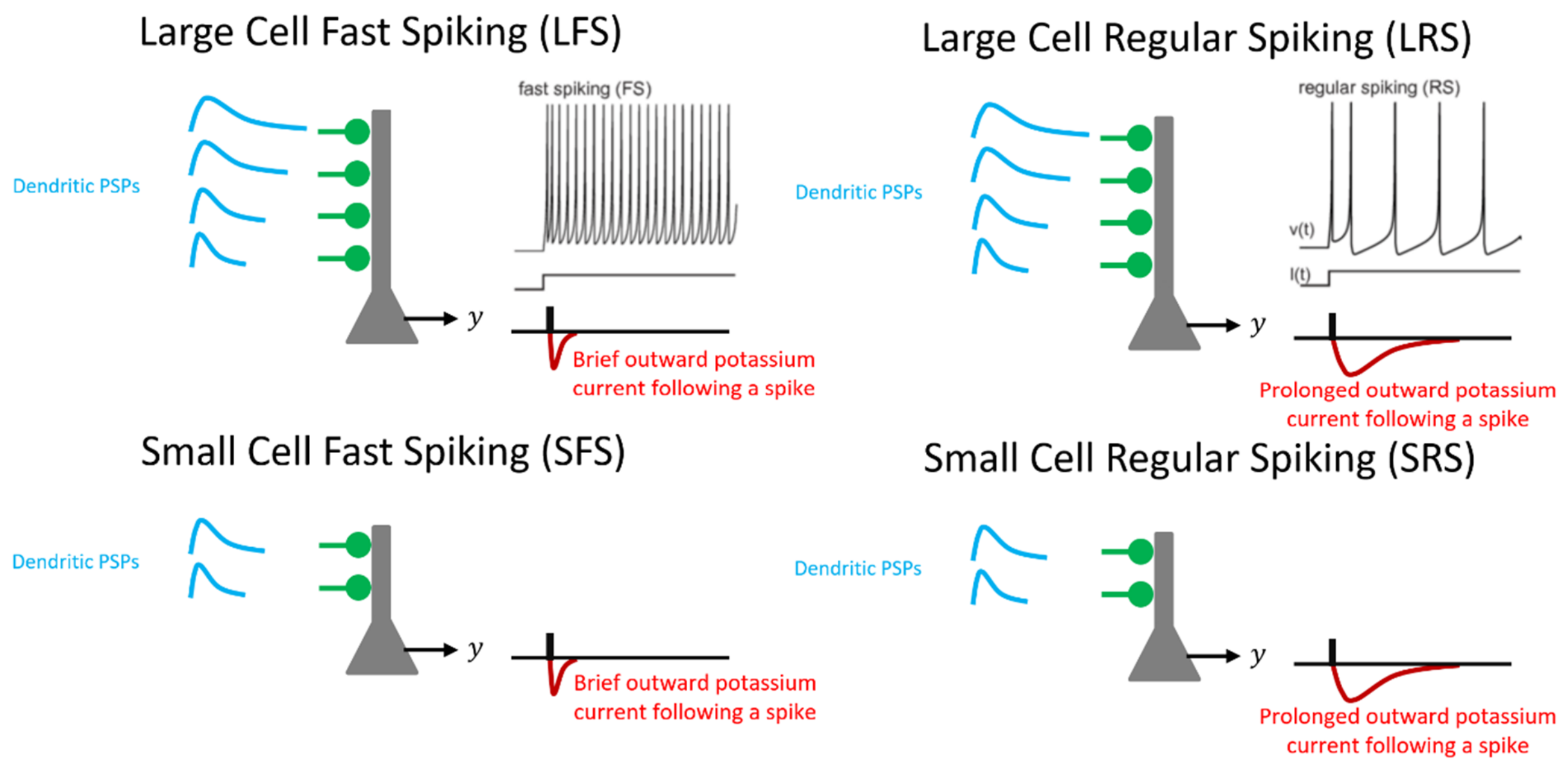

***Figure 4.1. Example of four simple F&F neuron types but encapsulate curious variability in the nervous system (fast/slow firing rate adaptation, large/small cell morphology).*** *(Top left) Illustration of a large F&F neuron model as indicated by an elongated dendritic branch and dendritic PSPs some are brief and some prolonged. This neuron is fast firing, i.e., has a brief outward potassium current after each spike, which allows it to fire at very high frequencies. (Top right) Illustration of a large F&F neuron model*

*as indicated by an elongated dendritic branch and dendritic PSPs that are both brief and prolonged. This neuron is regular firing, i.e., it displays firing rate adaptation since it has a prolonged outward potassium current after each spike which doesn't allow it to fire at very high frequencies but forces it to adapt. This type is akin to L5PC neuron for example. (Bottom left) Illustration of a small F&F neuron as indicated by a short dendritic branch and only brief PSPs. This type is most similar to most inhibitory interneurons in the cortex. (Bottom right) Small cell, regular spiking.*

In **Figure 4.1**, we depict four types of very simple F&F neurons that vary along two dimensions. Large vs small cells, that in our case will be related to the temporal extent of the variability of the dendrites, and Fast spiking vs Regular spiking neurons. Regular spiking neurons have a relatively long refractory period and they display firing rate adaptation. In order to investigate the effect of these two aspects (the morphological extent of dendritic tree and the excitable properties of somatic adaptation) on computation, we suggest creating DNN analogues for these four neuron types that replicate their behavior. Since they are simplified neuron models, and based on our experimentation in **chapter 1**, it's likely that we will be able to create a DNN analog of these neurons with near perfect replication accuracy. And since these are highly simplified neuron types and their DNN analogues are near perfect replicas, problems of adversarial examples are likely not to be encountered when attempting this. Regardless of the simplicity of these neurons, two potentially interesting computational aspects of neurons can be examined in depth in the context of an entire network of spiking neuron. One is the effect of large vs small dendritic delays, and the other is the effect of strong and prolonged firing rate adaptation vs brief firing rate adaptation.

First, we wish to build a framework of how we can incorporate DNN analogues of single neuron and place them in a network.

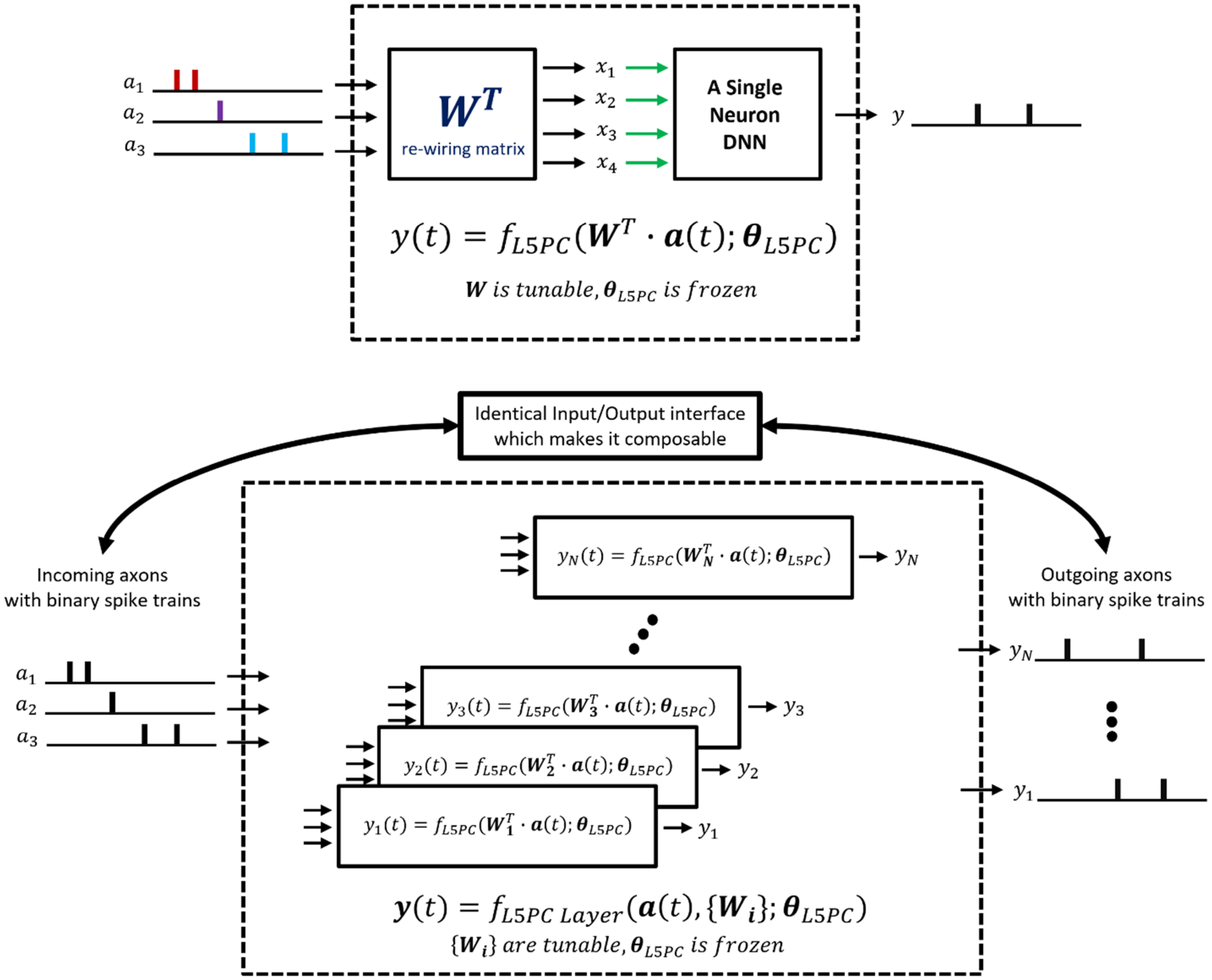

***Figure 4.2. Encapsulating multiple single realistic neuron DNNs into a DNN representing a whole layer of such neurons with a unified interface and learnable parameters that can be easily composed with additional layers.*** *(Top) A schematic illustration of a DNN circuit representing a single L5PC neuron with learned synaptic mapping onto it and with multiple synaptic contacts (the process of replacing a neuron with a DNN was discussed in* ***chapter 1****, multiple synaptic contacts were discussed in* ***chapter 2****). (Bottom). Extension of above plot to a layer of single neurons. Now some number of input spike trains is converted into a different number of output spike trains. Note the identical interface at the input and the output of this layer. Only mappings inputs to each neuron in a given layer are learned, the neurons themselves remain frozen.*

In **Figure 4.2**, depicted is a way to take the method we discussed thus far for mapping tasks onto single neurons using their DNN analogues, and extend this method to include a full network of neurons, in this case to an entire layer of parallelly connected neurons. The most important aspect of this formalism is the fact that the input and the output of a “layer” block (**Figure 4.2** bottom) are binary spike trains with variable number of inputs and outputs. i.e., they share a common interface, and thus these layers are composable.

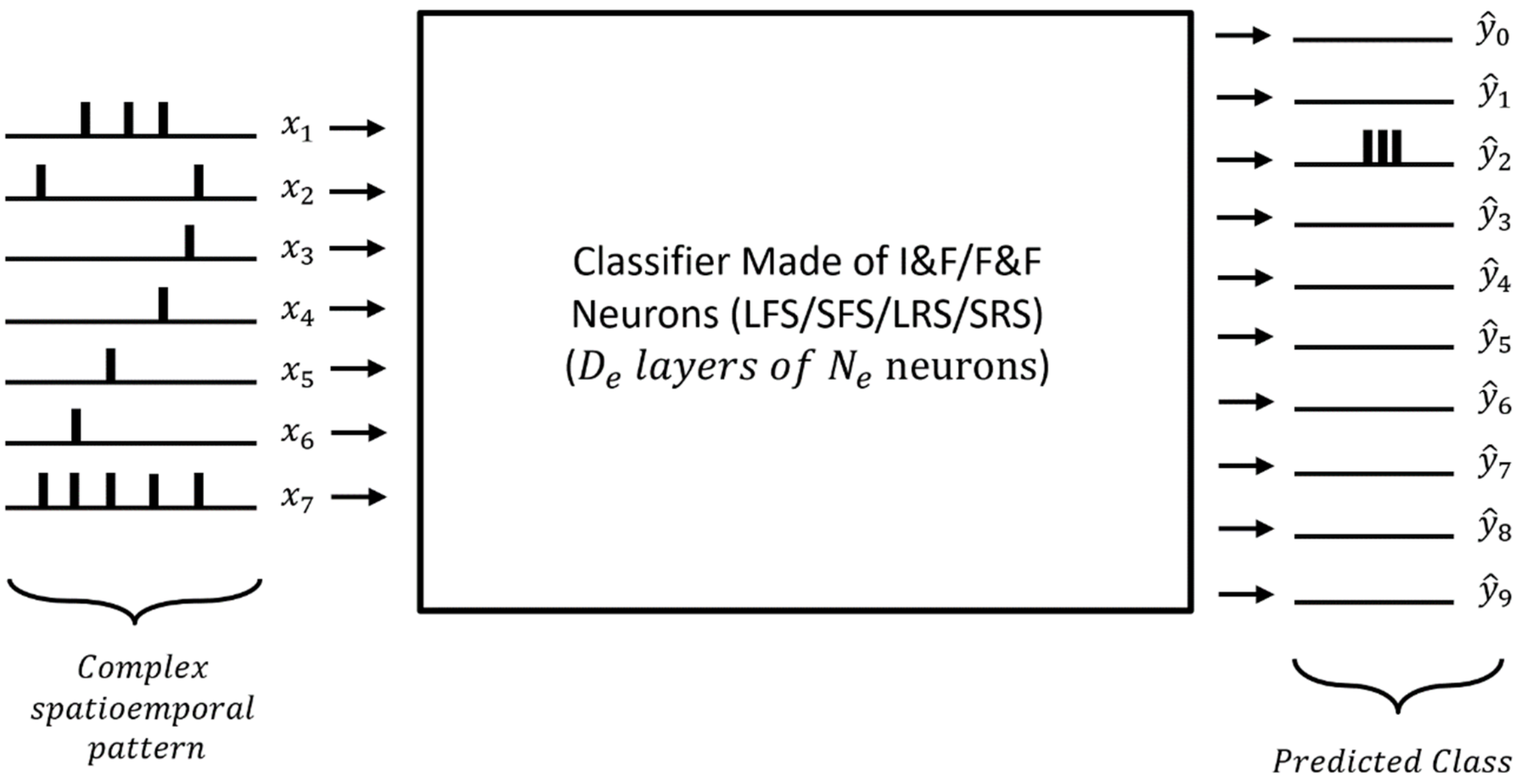

***Figure 4.3. An example task mapping onto a network of spiking neurons of various types for which it would be possible to examine task performance as a function of Neuron type and network parameters.*** *Left: Incoming spike trains with a complex spatiotemporal pattern encoded onto it, in this illustration representing the digit 2, can be learned to be classified as one of several categories. The mapping between inputs and output can consist of multiple layers of any number of neurons per layer, each neuron being of any type. The output* $\widehat{y2}$ *in this example successfully categorizes the input as the digit "2".*

It is interesting to ask what kinds of tasks can be achieved using a network of spiking neurons and whether a network of "Large Regular Spiking" (LRS) neurons can perform better or different computations than a network of "Small Fast Spiking" (SFS) neurons. For this reason, we suggest a large number of interesting tasks for these neurons and suggest training networks of variable depths and widths (number of neurons per layer). In **Figure 4.3** depicted is a basic multi class classification task that operates on spatiotemporal inputs, e.g., such as those we used in our study in **chapter 2**. Imagine plotting the classification accuracy for a network of SFS neuron as a function of the network size (depth and/or width) in one color, and in a different color plotting the accuracy of the other 3 neuron types discussed above. Will these curves appear identical? If the answer is not, then we are potentially starting a discussion about the inductive bias of large dendritic trees (and the long delays the incur) vs small dendritic trees. Also, in this way we are starting to get at the functional purpose of the amount of firing rate adaptation.

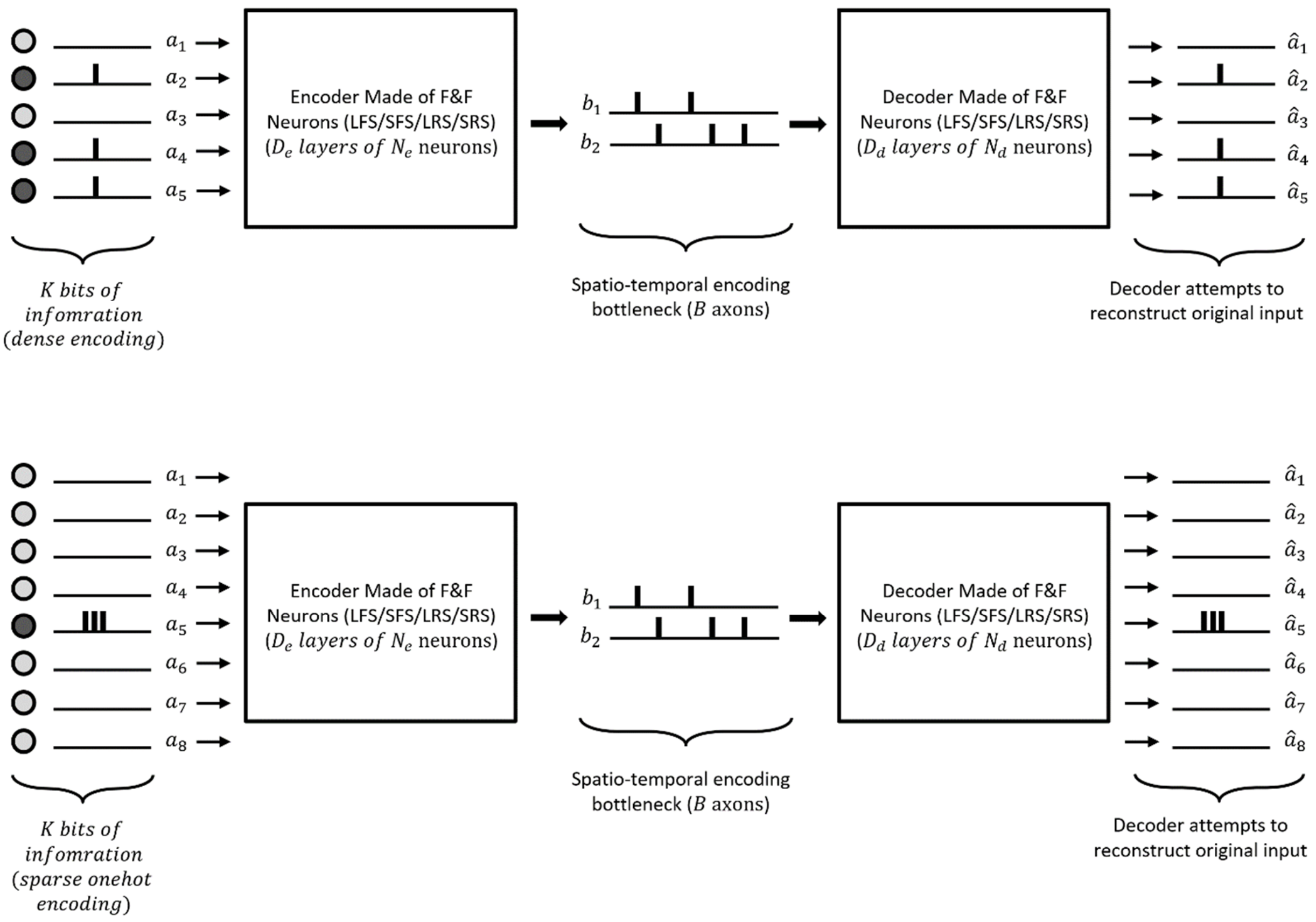

***Figure 4.4. Autoencoding spatial information with a network of spiking neurons, using a small number of axons to force temporal encoding.*** *This can help in understanding temporal coding properties of various network architectures, a task that specifically forces the neuron to represent spatial information using a small bottleneck of B number of axons, i.e., use temporal coding of that information. At top, one can represent the incoming information in a dense form to the encoder; on the bottom, the same number of bits of information can be encoded using a one "hot" encoding scheme.*

Of course, pattern recognition or spatiotemporal classification tasks are not the only type of task a network of neurons in the brain is required to solve. Another interesting avenue of inquiry that is related to spiking networks is the question of communication of information from one brain region into another, and this starts to get at the temporal coding of information. One can ask, how information that is encoded spatially can be processed by an encoder network and transmit information on a strict bottleneck using the temporal domain. E.g., how can one transmit five bits of information on a two-axon bottleneck transmission line? We can ask this type of question with several different variants depicted in **Figure 4.4**. it is particularly interesting if a network of Large Regular Firing cells could pack more information onto a fewer number of axons? Perhaps the encoder and the decoder could be comprised of different neuron types?

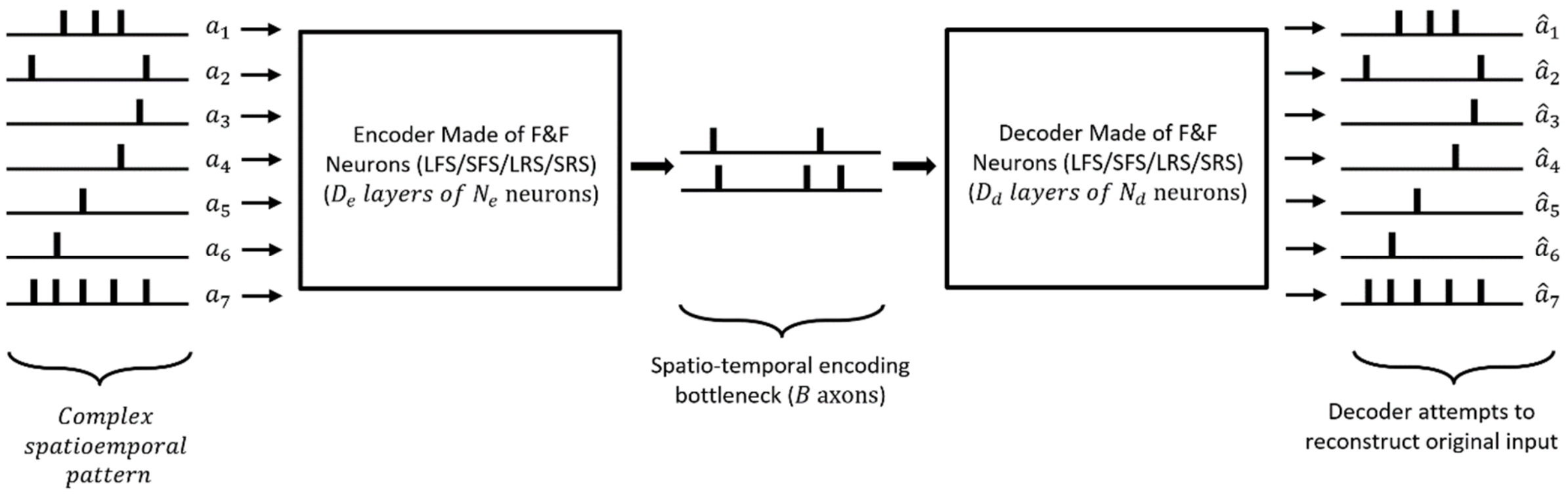

***Figure 4.5. Testing the ability of a network of spiking neurons to generate highly temporally precise spatiotemporal patterns of neuronal activations.*** *Full blown autoencoder architecture depicted here enables to test the ability of the neural network to generate highly coordinates spatiotemporal patterns of activations.*

Eventually, perhaps the most interesting aspect of spiking networks is how complex and highly coordinated multiple muscle activation sequences can be generated by a neural circuit with spiking neurons. This too can be probed with our method, by directly optimizing the connections between neurons to generate a complicated spatiotemporal output pattern. This can be coupled with the recognition and encoding of a highly complicated spatiotemporal input pattern using a spatiotemporal autoencoding task, depicted schematically in **Figure 4.5**. The questions here one again are numerous. How is information encoded in the bottleneck layer? What neuron types are required to create a highly coordinated output pattern of spikes. If the output spikes pattern is on the order of 100ms, will this be different than if the output spikes pattern is on the order of 1 second?

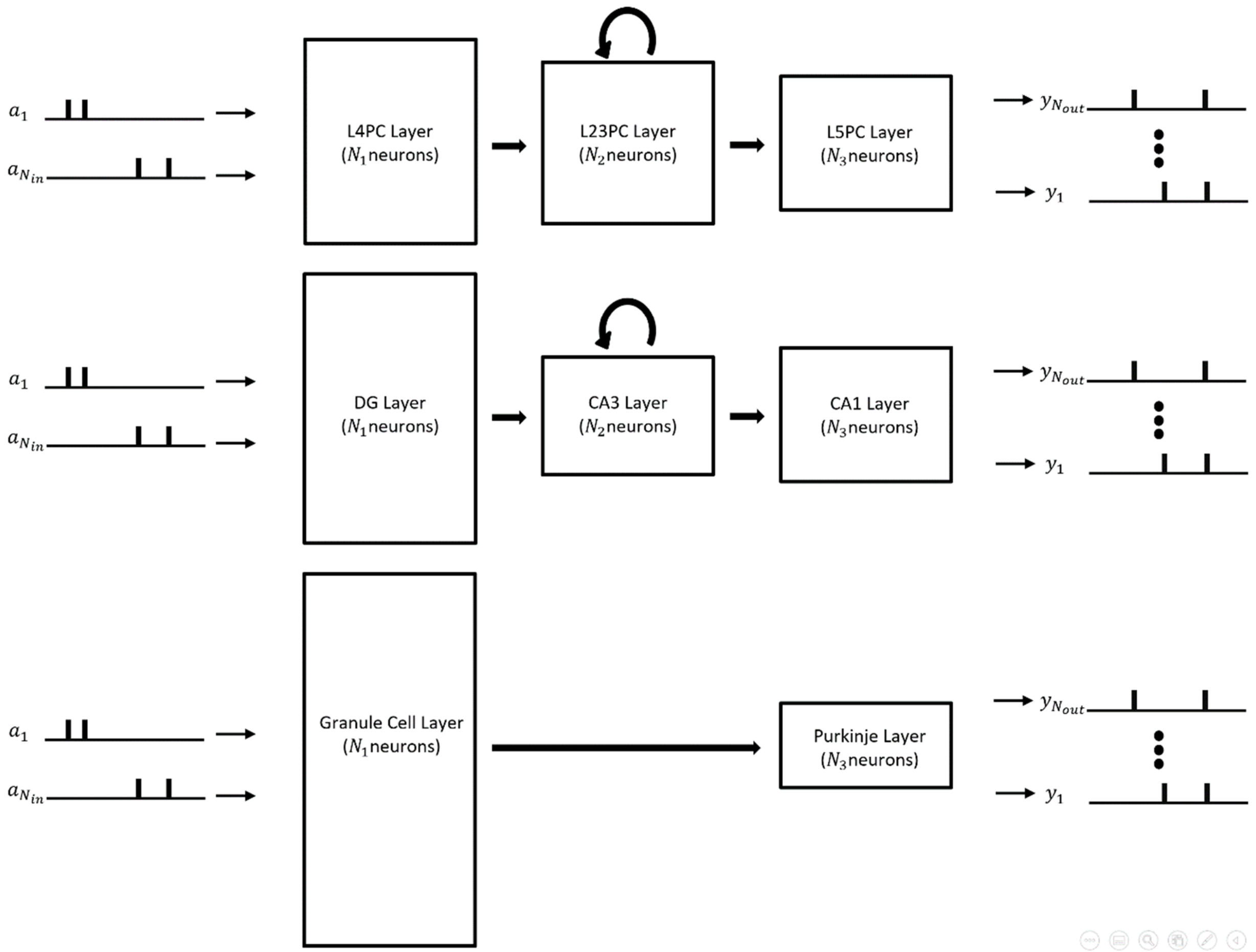

***Figure 4.6. Mapping the same task onto different known brain architectures using our DNN approach.*** *(top). Main circuit motifs of a typical cortical microcircuit. An expansive feedforward layer 4 sends inputs to a recurrent layer 2/3, which send information to layer 5 projection neurons to emit output. (middle). Main circuit motifs of a hippocampal microcircuit. An expansive feedforward dentate gyrus (DG) neurons send inputs to a recurrent CA3, which send information to layer CA1 projection neurons to emit an output. (bottom). Main circuit motifs of a cerebellar microcircuit. A vastly expansive feedforward granular cell (GC) layer sends inputs to a layer of Purkinje projection neurons to emit an output.*

Finally, once we have established that spiking neural networks can perform various tasks, one can now return to DNN that mimic detailed biophysical models of realistic neurons (instead of our simplified (LFS/SFS/LRS/SRS) neurons we've discussed thus far in this chapter) and ask questions that relate to various realistic microcircuit motifs and their computational properties. In **Figure 4.6** depicted are three examples of main microcircuit motifs, that of the cortex (**Figure 4.6** Top), hippocampus (**Figure 4.6** Middle) and cerebellum (**Figure 4.6** Bottom).

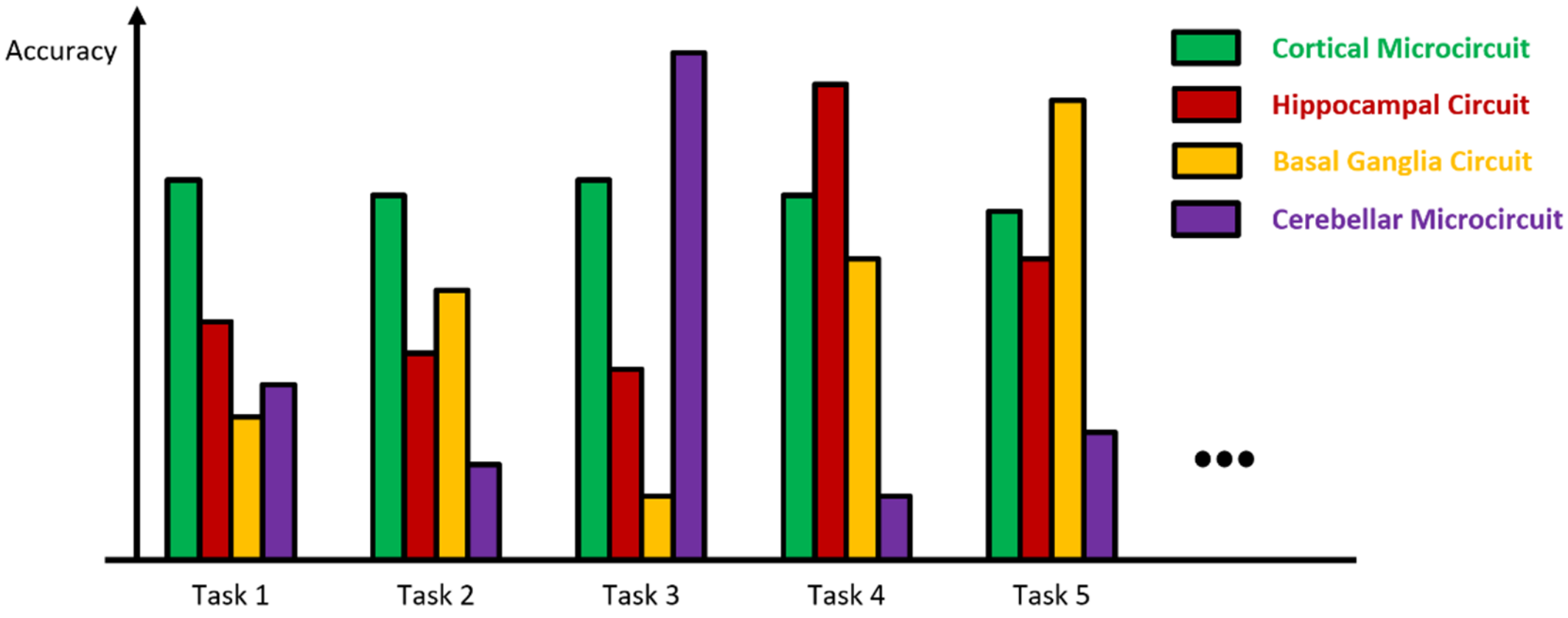

***Figure 4.7. Assessing the inductive bias of various brain macro and micro architectures.*** *One can try and present different tasks to all microcircuits in question and, by observing the pattern of their performance of the various task, asses the inductive bias of any neuronal microcircuit.*

It would be particularly interesting to compare various realistic circuit motifs of networks of realistic neurons against each other on an array of predefined set of tasks. This might reveal the inductive bias of each circuit micro-architecture, i.e., its strengths and weaknesses compared to the others. An illustrative example of how this could look like is depicted in **Figure 4.7**.

## Oja circuits are a concrete example of simple neurons and synaptic update rules which generate a clear computational statistical objective to a neural circuit

We discussed above how a neuron model, a microcircuit and several plasticity rules could potentially achieve specific computational goals. Plasticity rules, local connectivity and input/output relationships of neurons can be measured or readily deduced from biology, and we have seen in this document above that the neuroscientific community has made significant progress on these fronts. It is nevertheless a crucial question how one can convert these low-level details into higher level computational objectives of neuronal circuits. This question is also addressed in (Lillicrap and Kording, 2019). In order to answer this type of question, it is best to start with a simple yet compelling example – the Oja microcircuits and learning rules for achieving principle component analysis (PCA) and independent component analysis (ICA), which happens to be the most beautiful piece of computational neuroscience according to this author's humble opinion.

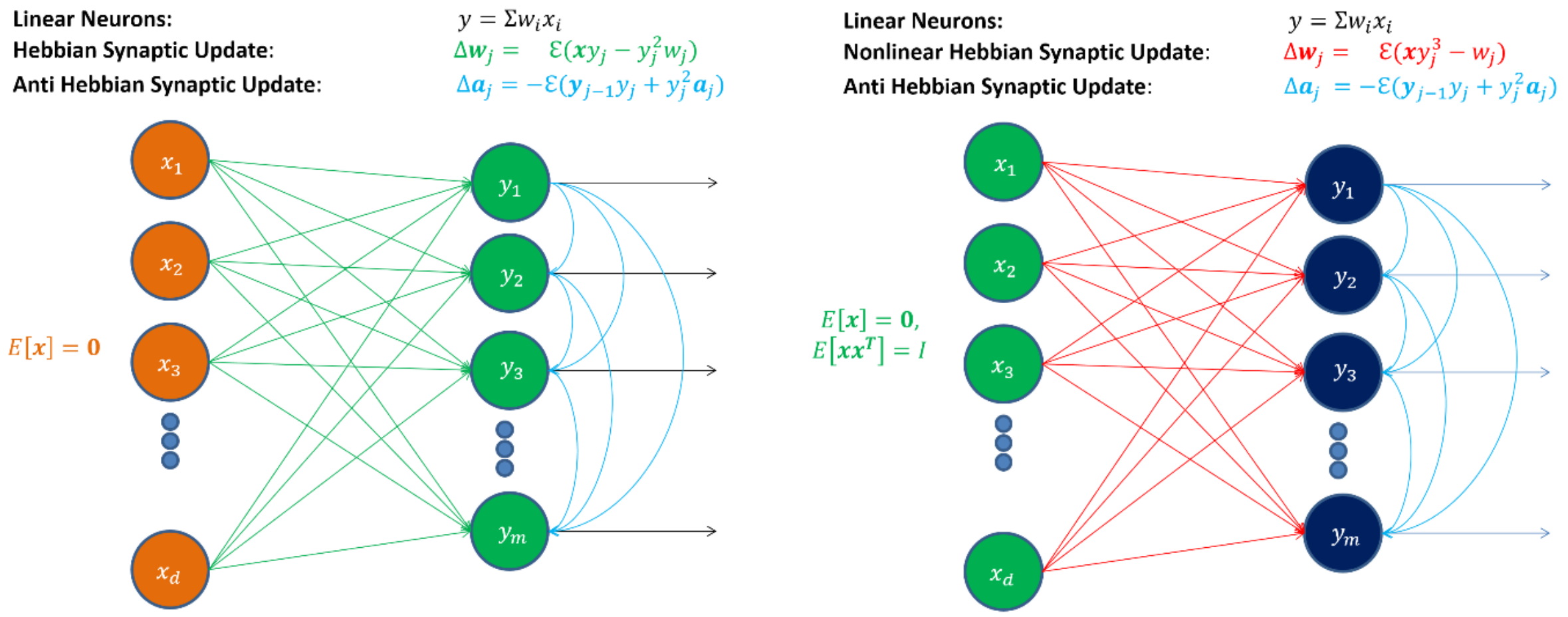

***Figure 4.8. Circuits of linear neurons with appropriate learning rules that achieve PCA and ICA*** *On the left, is a circuit diagram that achieves PCA. The update rules listed above correspond to weights at the same color. Feedforward in green, and lateral in blue. On the right, a circuit diagram that achieves ICA. Similar to illustration on the left, the feedforward weights are updated according to weight update in red, and lateral connections are updated according to weight rules colored in blue.*

In **Figure 4.8** depicted is a summary plot of a beautiful work done 3-4 decades ago (Hyvarinen and Oja, 1998; Karhunen et al., 1997; Oja, 1982; Sanger, 1989). Two circuit diagrams are shown that achieve PCA and ICA respectively when continuously presented with input taken from a constant probability distribution. Each of these circuits is comprised of simplified linear neurons, and different neurons and connections adhere to different learning rules. For the circuit on the left, the requirement on the input is that it will have zero mean, the network utilizes Hebbian learning for the feedforward weights (along with a normalization part to avoid explosion of weight magnitudes) and an anti-Hebbian learning rule for the lateral connections. After learning, this network finds the first M principal components of the data (PCA). i.e., it finds the subspace that spans the largest among of variance in the input. For the circuit on the right, the requirement on the input is that it is white. Meaning it has both zero mean and unit covariance. A diagonal covariance is achieved by the PCA circuit on the left so a simple normalization of each neuron to unit variance is all that is required to achieve such a distribution (that could be achieved by a homeostatic plasticity update). This circuit utilizes nonlinear Hebbian learning rule for the feedforward weights and an anti-Hebbian learning rule for the lateral connections as in the previous section and, after learning, it achieves independent component analysis (ICA) of the input.

Several points are worth noting: In the PCA example, we can see that Hebbian learning rules are used to learn the correlation structure in a feedforward manner, but also lateral anti Hebbian connections are used to de-correlate the outputs. The output of this network is therefore uncorrelated as expected from PCA calculation. In the ICA example, we can see that **nonlinear** Hebbian learning rule is used for the feedforward connection and the same anti Hebbian lateral connections are used to de-correlated the data. Another point to note is that the ICA circuit requires white data at its input in order to produce independent components at its output, which might pose a problem since it' is not an easy task to achieve generally. Conveniently, however, the PCA example produces de-correlated outputs, which is the major requirement for whitening. With proper homeostatic plasticity to normalize the variance of the output neurons to unity, we can imagine concatenating the output of the PCA circuit and the input of the ICA circuit to produce a slightly more complicated circuit that computes ICA directly on input that have no difficult constraint that it need to hold.

These Oja circuits that learn PCA and ICA using simplified linear neurons and specific connectivity structure using Hebbian, nonlinear Hebbian and anti-Hebbian biologically plausible synaptic learning rules are a clear, concrete, and compelling example of how simple microcircuits and several simple local learning rules give rise to precise mathematically well-defined objective functions. Here it is worth mentioning also the works from the lab of Dmitri Chklovskii that attempt to generalize Oja's work and create various additional circuit architectures to achieve various linear and properly defined computations (Golkar et al., 2020; Lipshutz et al., 2021; Pehlevan et al., 2020)

## An example of a spiking neural network circuit comprised of I&F neurons with feedback inhibition and spike timing dependent plasticity (STDP) learning

The Oja microcircuit example we have seen above might be a bit too simplistic and non-biologically plausible as it contains linear neurons that have both positive and negative activations, as well as synaptic weights that can be both positive or negative. Here we wish to highlight a microcircuit of slightly more biologically plausible neurons and an experimentally measured learning rule (STDP) that gives rise to interesting computation to illustrate that all principles discussed thus far still apply also when adding biological details.

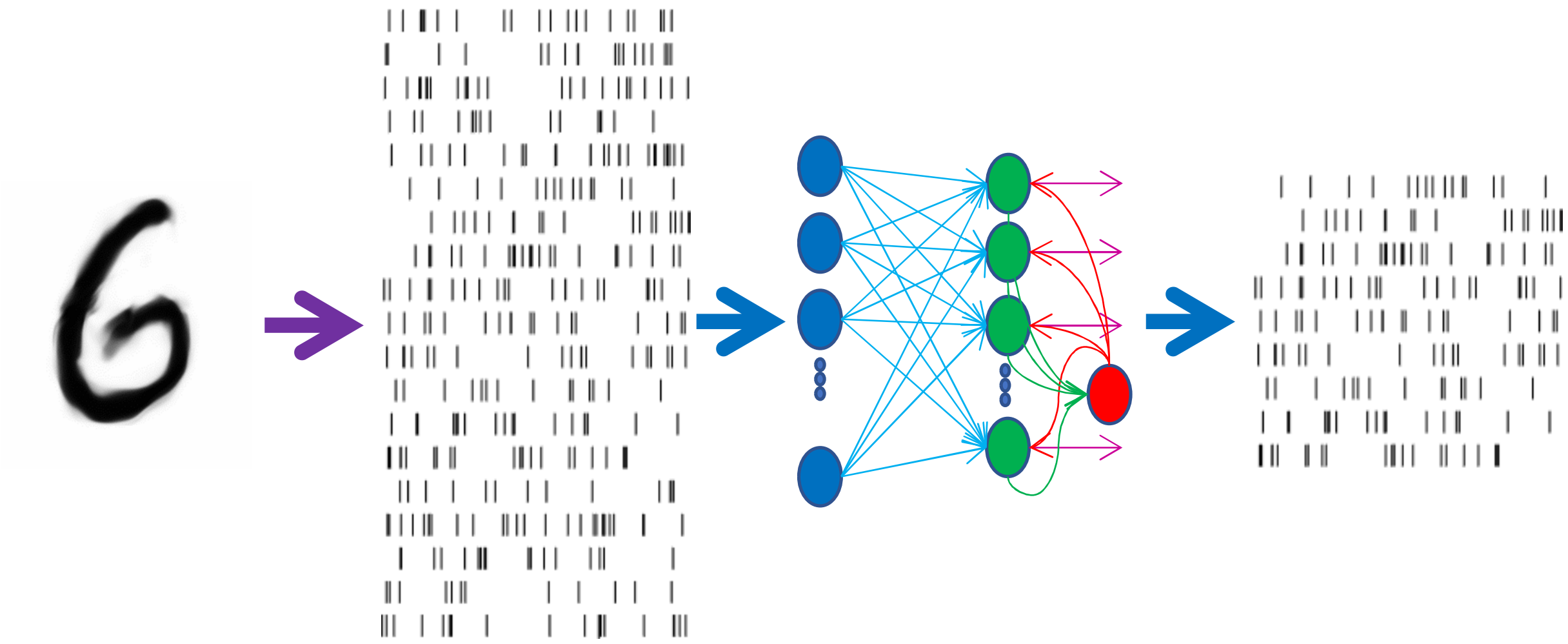

***Figure 4.9. A circuit comprised of spiking I&F neurons and STDP synaptic updates*** *Schematic of setup used in paper by Nessler et al.* (Nessler et al., 2013) *an input image is converted to spike trains, sent to a network where STDP learning rule is applied, local feedback inhibition microarchitecture is used and output spike trains are produced. this network can learn to idenify repeating patterns in the input in an unsupervised manner.*

In **Figure 4.9** we can see a work by Nessler et al.(Kappel et al., 2014; Nessler et al., 2013) that contains slightly more biologically plausible neuron models (Integrate and fire) that receive spike trains as input and produce spike trains as output, along with a more biologically plausible learning rule (standard STDP kernel). An interesting thing to note here is that the input is "interesting" in the sense that it is not a simply taken from a multivariate Gaussian distribution, nor itis a simple mix of statistically independent random variables but rather it is taken from an unknown probability distribution about which we can answer only partial questions (e.g., what is the digit it represents). Nonetheless, it is important to note that the input is "natural" in the sense that biological organisms need to face these kinds of visual inputs in their day-to-day experience. In the original paper, the authors used this network as an unsupervised learning scheme and tried to recognize the digit by decoding the output spike trains and did so successfully. This is an interesting example of much more biological components that are brought together and achieve some computational goal that has some interesting properties. But as opposed to the Oja example above, we do not know what the precise mathematical definition of this goal is. We therefore would like to find a systematic way to uncover the mathematical objective of a set of "low level" algorithmic and implementational details.

### Determine the optimization objective of a single neuron and a neural circuit

Neurons in the brain receive input, emit output, and change their intrinsic properties as a function of their activity. For example, if there is a correlation between a particular input axon activity and the resulting output of a neuron, typically the synaptic efficacy of the associated input will increase (fire together, wire together, Hebb). In the introduction we have seen that relatively simple local calcium-based plasticity rules can explain nearly all known synaptic plasticity phenomena that were discovered thus far for neurons in the brain. We have also seen in the section discussing the Oja circuits, that local synaptic plasticity rules can have a clear and well-defined statistical objective via the Hebbian Oja learning rule that achieves PCA. Similarly, it is quite reasonable to assume that any single neuron in the brain has some optimization objective that it tries to achieve and the intrinsic changes it undergoes (synaptic or other) are applied in order to achieve this objective. The question arises, how can one assess the objective of a local learning rule that was empirically measured by experiments without knowing the actual objective. Remember that the Oja Hebbian learning rule was derived as a gradient update of the optimization objective of PCA. Inspired by this fact, we will make a simple hypothesis: whatever the objective of a single neuron may be, the local synaptic learning rules are the gradient of that objective. Please note, this is not necessarily the truth, there are many different possible optimization algorithms that are not stochastic gradient descent (e.g. second order optimization updates, or evolutionary updates, etc.). Nevertheless, it is a very reasonable null hypothesis one can assume about the changes a neuron undergoes – they are iteratively moving the neuron's intrinsic parameters slowly towards its goal.

So how does one go about reverse engineering the optimization objective of a neuron given that we have good empirical approximations that simulate the synaptic updates for any input that is driven to the neuron? The answer could be that we simply need to find an objective whereby the updates that result from its gradient are similar to the simulated updates using biological learning rules (e.g., calcium-based plasticity). Due to our DNN formulation of a single neuron from the beginning of the current chapter we now can calculate the gradient with respect to the synaptic weights of any function of the output of a neuron.

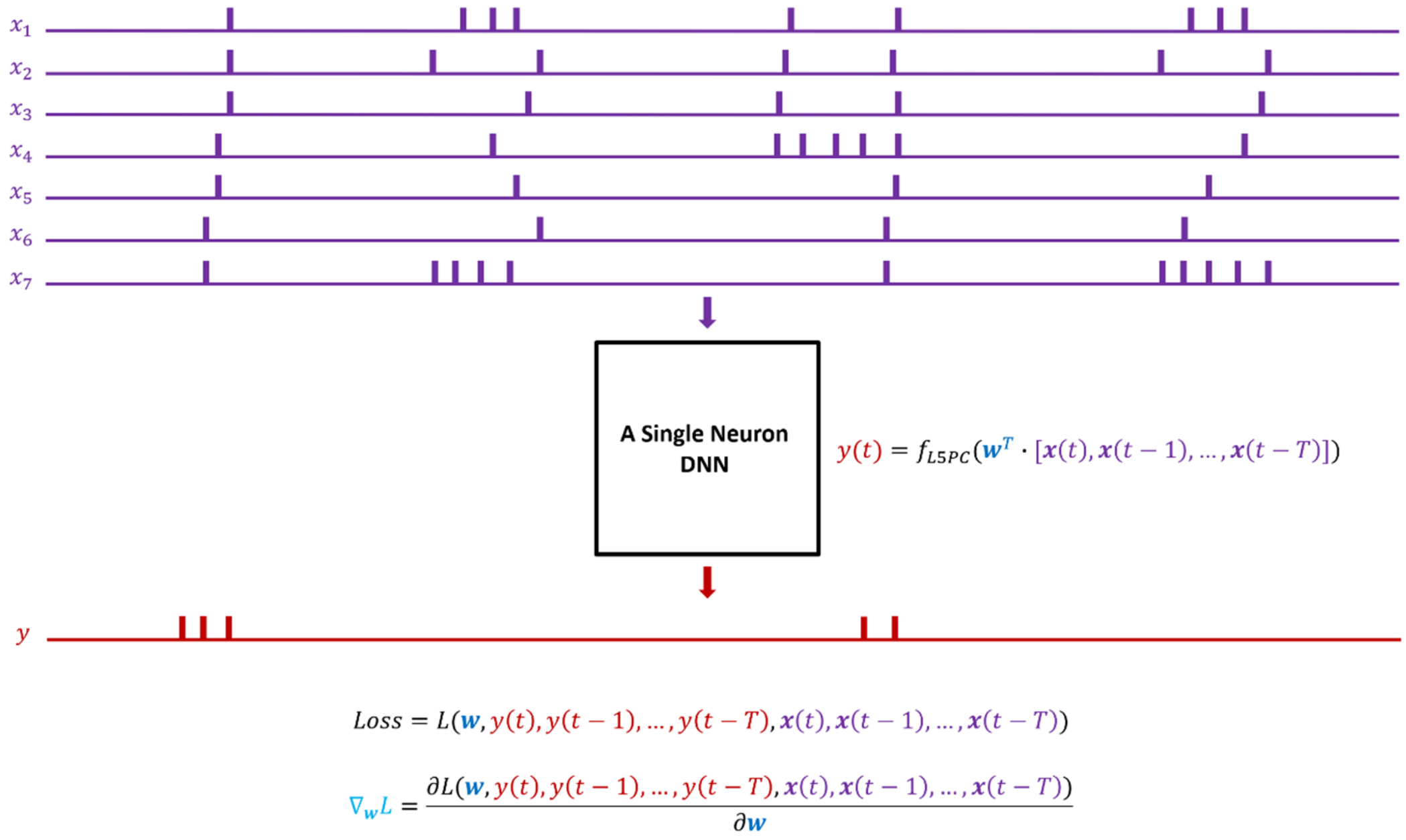

***Figure 4.10. For any input sent to the neuron, we can calculate the gradient with respect to any loss function we hypothesize.*** *(Top), incoming input spikes that are sent to a neuron. In this case, 7 axons are illustrated displaying complex spatiotemporal patterns representing the digit sequence "1","3","4","2" (middle), these input spikes are sent to a neuron-approximating DNN and emit an output spike train.*

*(bottom), for any loss function we might hypothesize that the neuron might attempt to achieve, we can easily derive what synaptic changes this loss function will imply. Note that this loss needs to be local, i.e. a function of only the current state of the neuron weights, and the inputs and the outputs history as the neuron doesn't have access to any other information when learning inside a living organism.*

In **Figure 4.10**, we can see an illustration of some input being driven to the neuron, some output is generated by the neuron using the DNN that approximates the neuron, and below are the equations that describe the loss function, this could be any loss we deem fit, and the equation that describes the derivative of this loss with respect to the synaptic weights. Because we have a DNN approximation of the neuron, this is very simple to calculate.

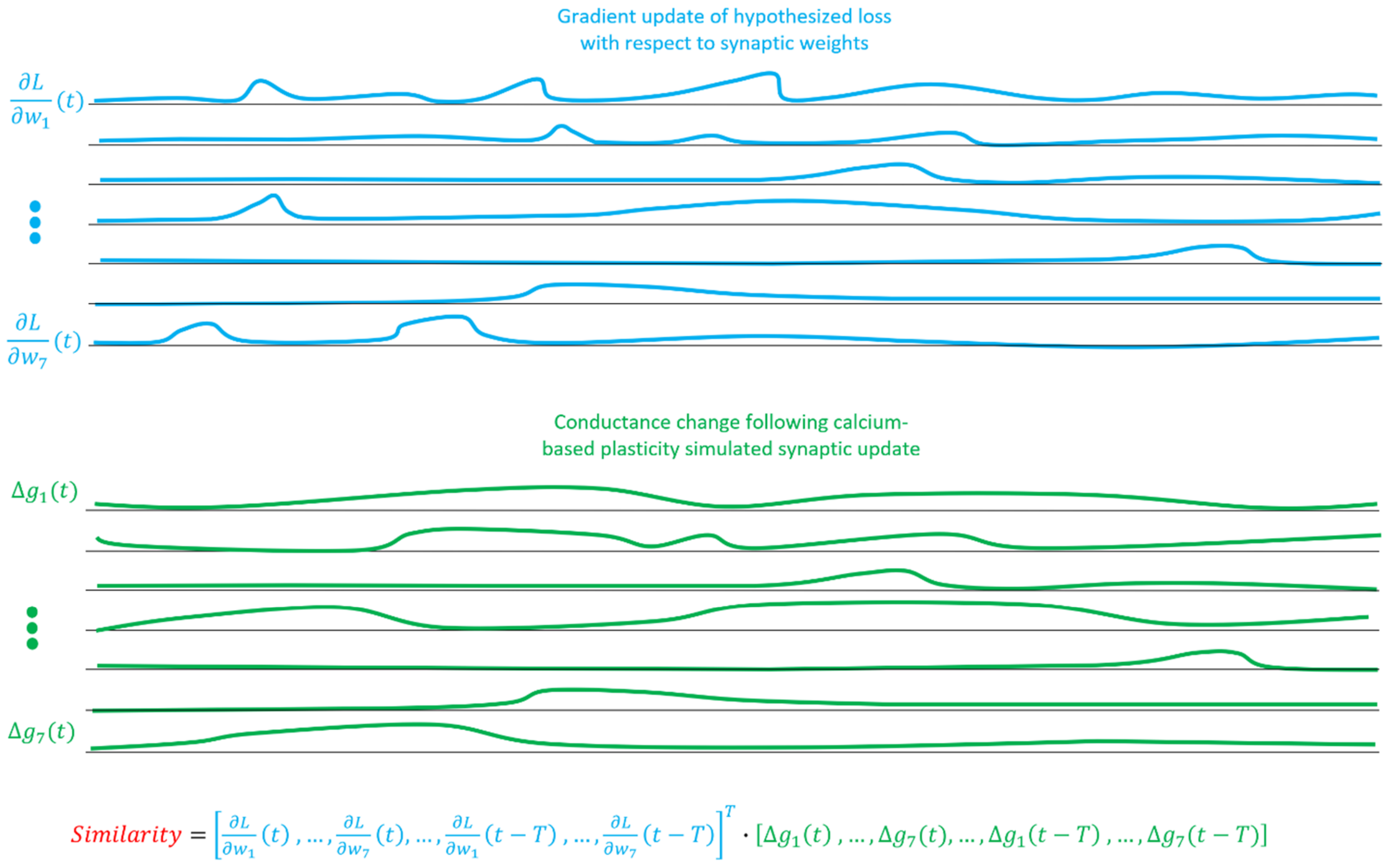

***Figure 4.11. Calculation of similarity between simulated synaptic updates and gradient of a particular optimization objective.*** *(top), the instantaneous weight updates that are the result of deriving the loss function according to each learnable parameter (middle) the instantaneous conductance changes that are the result of simulation a biological calcium-based plasticity learning rule. (bottom) the similarity measure is defined by the inner product between these instantaneous patterns. Note this is a highly dense similarity measure that creates a large number of datapoints to be compared so overfitting will most likely not be a problem.*

How can one use the ability to calculate any gradient to determine the optimization objective of the empirically determined synaptic learning rule (e.g., calcium-based plasticity)? We propose to simulate the plasticity of the synaptic update in detail (green curves in **Figure 4.11**) and compare the updates that are the result of a gradient of the hypothesized loss function (blue curves in **Figure 4.11**). The similarity measure can be as simple as inner product between these traces. If we find that the synaptic changes that are due to deriving our hypothesized loss agrees well with simulated synaptic conductance changes based on the realistic plasticity model, we know that it is likely that the loss is the optimization objective of the neuron. As we stated earlier, this assumes that true synaptic updates attempt to approximate the gradient with respect to some objective the neuron has. This is not necessarily true, but this nevertheless a good assumption to make.

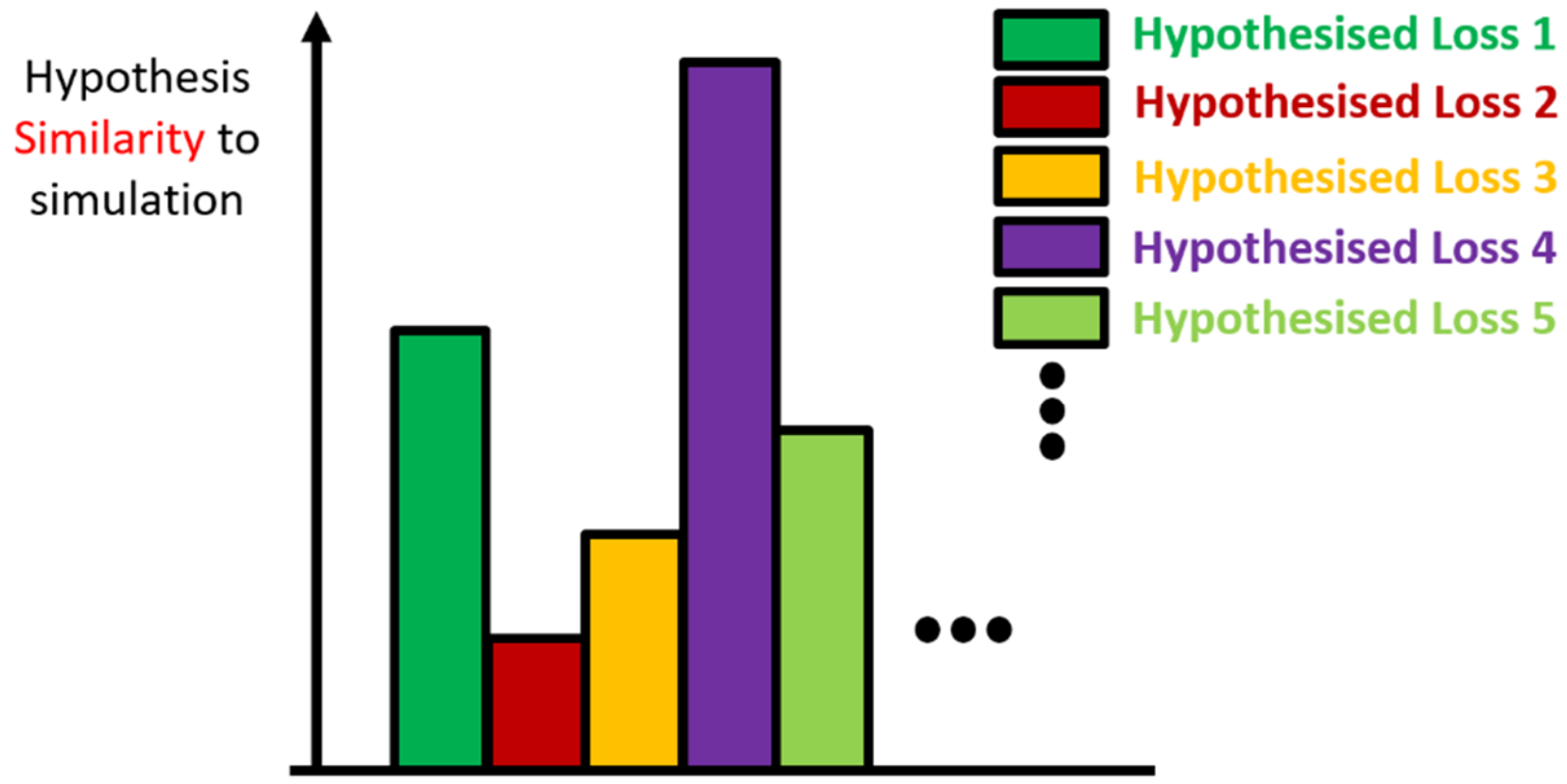

***Figure 4.12. Finding a suitable objective is essentially a process of guessing a potential objective and validating its compatibility against the simulated empirical synaptic updates.*** *An illustration of a bar plot depicting the similarity measure for various hypothesized computational objectives that the single neuron might seek to achieve.*

One can then come up with multiple hypothesized losses to serve as the potential objective of the circuits and searching which hypothesis fits best according to our similarity measure. We can stop hypothesizing once we have reached good agreement and declare that our hypothesized loss is indeed the computational objective of the single neuron (or very close to it). The potential results of this kind of process are depicted in **Figure 4.12**.

An interesting implication of what we are describing here is that we can also simulate the gradient updates for a simulated circuit composed of multiple neurons and, using the network formulation that we have described earlier in this chapter, we can try to assess the optimization objective of an entire microcircuit using exactly the same formulation described for a single neuron above. In this way, we can assess the optimization objective of a cortical microcircuit or a cerebellar microcircuit, etc. If we manage to find an objective that closely approximates the synaptic updates of an entire microcircuit, then it is hard to imagine a higher level of complete understanding of that circuit. We will call that complete *understanding of that circuit*.

In 1996, Marr & Poggio (Marr and Poggio, 1976) have put forward the idea that understanding of a neuronal system involves understanding at several different levels of analysis that are somewhat independent of each other. The main levels roughly map to computational, algorithmic and implementational levels of analysis. The optimization objective of a neuronal circuit that we've discussed above corresponds to Marr's 3$^{rd}$ and highest level of analysis, essentially describing the circuit goal. In order to achieve what we have discussed in this chapter; we need information about calcium-based plasticity and their synaptic updates (which is equivalent to Marr's 2$^{nd}$ algorithmic level of analysis) and we needed all the biophysical details about the operation of neurons (which is equivalent to Marr's 1$^{st}$ implementational level of analysis). The process described here is potentially a bridge that can translate detailed low-level knowledge, into full computational understanding of a neuronal circuit. Granted, we have a few assumptions here, but neither of them is preposterous. In this way, we can slowly but surely uncover the computational objective of neurons, microcircuits and eventually encompass the entire brain and understand it's objective. If there is even a slim chance of this to work, this author is under the impression this should be attempted.

# Discussion

Recent advances in the field of deep neural networks (DNNs) provide, for the first time, a powerful general-purpose tool that can learn complex mappings from examples. In **chapter 1** we used these tools to study the I/O mappings of single complex nonlinear neurons at millisecond temporal resolution. We constructed a large dataset of pairs of (synaptic) input - (axonal) output (I/O) examples by simulating a neuron model of layer 5 cortical pyramidal cell receiving a rich repertoire of synaptic inputs over its dendritic surface and recorded its spike output at millisecond precision, as well as its somatic subthreshold membrane potential. We then trained networks of various configurations on these input-output pairs until we obtained an analogous "deep" network with close performance to that of the neuron's detailed simulation. We applied this framework to a series of neuron models with various levels of morpho-electrical complexity and obtained new insights regarding the computational complexity of cortical neurons.

For simple I&F neuron models, our framework provides simple analogous DNNs with one hidden layer consisting of a single hidden unit that captures the full I/O relationship of the model, as expected. Surprisingly, even a model of layer 5 cortical pyramidal neuron with the full complexity of its dendritic trees and with a host of dendritic voltage-dependent currents and AMPA-based synapses is well-captured by a relatively simple network with a single hidden layer. However, in a full model of an L5 pyramidal neuron consisting of NMDA-based synapses, the complexity of the analogous DNN is significantly increased; we found a good fit to the I/O of this modelled cell when using a temporal-convolutional network that has 5-8 hidden layers.

The results in **chapter 1** suggest that the single cortical neuron with its nonlinear synaptic inputs is already, on its own, a sophisticated computational unit. Consequently, networks built from such units are "deeper" and computationally more powerful than they seem to be just based on their anatomical (pre-to-post) synaptic connections. Importantly, the implementation of the I/O function of single neurons using a DNN also provides practical advantages. It is computationally much more efficient than the traditional compartmental model, which required the solution of many thousands of partial differential equations (PDEs) per neuron. Indeed, for the full model of L5PC, we obtained a speedup of ~2000x when using the DNN instead of its compartmental-model counterpart. Furthermore, the size of the respective DNN for a given neuron could be used (under certain assumptions) as an index for its computational power; the larger it is the more sophisticated computations this neuron could perform. Such an index will enable a systematic comparison between different neuron types (e.g., CA1 pyramidal cell, cortical pyramidal cell, and Purkinje cell, or for the same type of cell in different species, e.g., mouse vs. human cortical pyramidal cells).

The physical properties of single neurons input/output transformation. E.g., temporal convolutions, nonlinear amplification/dampening, thresholding, etc. can be regarded as a bank of basic computational operations that can be utilized by the nervous system. Driving input to the neuron in a pattern that preferentially engages a particular mechanism of the above-mentioned basic operations can be regarded as utilizing that computation by the network. We therefore hypothesize that most patterns that can engage in a particular neuronal mechanism, can find a way to do so deliberately e.g., if the nervous systems “wants” to be able to perform a convolution operation, it must organize the inputs to a physical neuron in such a way that it performs a convolution using it's built in temporal convolution mechanism. For example, represent some spatial signal as a temporal signal, perform the operation required, and go back.

In order to highlight some of the computational advantages of dendrites in a more concrete way, we presented in **chapter 2** the Filter and Fire (F&F) neuron model that augments the commonly used Leaky Integrate and Fire (I&F) neuron model by incorporating to it multiple synaptic contacts per axon as well as the effect of dendritic filtering that transforms the incoming spike train to a set of multiple post synaptic potential (PSP) filters. Each filter corresponds to a particular synaptic contact, with varied time constants. This directly relates to the filtering of the dendritic synaptic inputs by the dendritic cable properties (Rall 1964; Rall 1967). In the F&F model, a single pre-synaptic spike results with PSP composed of multiple temporal profiles. We termed this phenomenon "dendro-plexing" of the presynaptic spike (**Figure 1** in **chapter 2**).

In **chapter 2** we have demonstrated that the capacity of the F&F neuron to memorize precise input-output relationships is increased by a factor of ~3 compared to that of the regular I&F. The capacity is measured as the ratio between the number of precisely timed output spikes and the number of incoming input axons. This ratio is ~0.15 spikes per axon for the I&F case as shown by Memmesheimer et al. (Memmesheimer et al., 2014), and ~0.45 spikes per axon for the F&F model. We show that the origin of this threefold increase in capacity is due to the fact that all possible PSPs are spanned by a 3-dimentional subspace which sets the effective upper limit on the capacity when using a large number of multiple synaptic contacts. i.e., even if using numerous synaptic contacts, they effectively serve as only 3 independent contacts in this simplified model of nonlinear dendrites.

Most importantly, we constructed a new spatiotemporal pattern discrimination task using the MNIST dataset and demonstrated that the F&F model can learn to detect single digits at well-above chance level performance on an unseen test set, whereas an I&F neuron model cannot learn the task at all. This is because of the fact that in the specific way we chose to represent each digit – the task does not contain enough spatial-only information suitable for I&F neuron discrimination. Our specific task design was deliberately chosen to highlight this temporal aspect of pattern discrimination that is possible when taking into account the temporal filtering due to cable properties of dendrites. We show that multiple synaptic connections with different PSP profiles allow the neuron to effectively parametrize the temporal profile of the PSP influence of each pre-synaptic axon on the somatic membrane potential. This is enabled by modifying the weight of the various (multiple) contacts made between the axon and the post synaptic cell.

This demonstrates that even when considering highly simplified (F&F) neuron model as used in **chapter 2**, one that only implements the passive temporal filtering aspect of dendrites, a computational role of dendrites is unraveled for the seeming redundancy of multiple synaptic contacts between pairs of neurons in the brain. Dendrites therefore allow us to salvage some of the redundant connectivity and put those synaptic weight parameters to a good use. This allows for increased memorization capacity, but perhaps more importantly to detect specific spatiotemporal patterns in the input axons. The spatiotemporal filtering properties of dendritic processing are prominently featured in **chapter 1** describing how single neurons produce output spikes in response to highly complex spatiotemporal patterns.

The question of how trains of spikes represent information in the nervous system has been a long-standing question in neuroscience, since its inception. A major debate revolves around whether information is largely carried by firing rates averaged over relatively long time periods, or rather that precisely timed spikes carry crucial bits of information. Evidence for both alternatives has been found for both sensory systems and motor systems and, consequently, much theoretical work on this key topic has been conducted (Abeles, 1982; Abeles et al., 1993; Castelo-Branco et al., 2000; Christopher Decharms and Merzenich, 1996; DeWeese et al., 2003; Florian, 2012; Gütig and Sompolinsky, 2006; Hopfield, 1995; Johansson and Birznieks, 2004; Kara et al., 2000; London et al., 2002, 2010; Maass and Schmitt, 1999; Meister et al., 1995; Memmesheimer et al., 2014; Naud et al., 2023; Neuenschwander and Singer, 1996; Schneidman et al., 1998; Thorpe et al., 2001; Wehr and Laurent, 1996). Since we have shown in **chapter 2** that dendrites and multiple synaptic connections per axon play a crucial role in decoding incoming spike trains and increase the neuronal repertoire in emitting precisely timed output spikes in response to spatiotemporal input patterns (and they might do this using a simple biologically plausible learning rule), we wish to suggest that dendritic hardware should be considered when discussing the question of the neural code and what information is transmitted via axons. Indeed, (Perez-Nieves et al., 2021) show that a diversity of time constants helps increasing the computational repertoire of spiking networks.

In **chapter 2** we suggest that diversity of time constants that helps in increasing the computational repertoire already happens at the neuronal level. The fact that a single neuron can decode complex spatiotemporal patterns on its own and does not require a highly coordinated decoding network of neurons to extract temporal information from incoming spike trains, not only allows for potential "hardware" savings, but also suggest that information might be ubiquitously encoded by precise spike times throughout the central nervous system. A single neuron can emit precisely timed output spikes in response to spatiotemporal inputs, as we showed here and was previously shown in simpler I&F models by (Memmesheimer et al., 2014). It is therefore not required to have a large and highly coordinated network of neurons to encode temporally precise patterns transmitted via axons. We believe that if already a single neuron can learn to generate such precise spike timing without relying on network mechanisms, this might increase the likelihood that neuronal information is encoded by precise spike timing as opposed to average firing rates (over relatively long periods of time) throughout the CNS. In **chapters 3 & 4** we further suggest investigating how these temporal properties of the F&F neuron model can aid a network of such F&F neurons represent signals using temporal coding in a systematic way. Network level investigation of F&F neurons, that are elaborated upon in **chapter 2,** is only possible due to the application of the method we develop in **chapter 1.**

In **chapters 3 & 4** we describe a few potential future direction that **chapters 1 & 2** build towards. Importantly, a deep neural network (DNN) is a differentiable function. Although it might be difficult to differentiate a large and complex simulation of a detailed biophysical neuron model, after approximating that neuron with a DNN, taking the derivative of the artificial neural network is straightforward. We seek to take advantage of this property and describe in detail how one can map any task onto to any neuron model. We discuss this in depth in **chapter 3**. Although there are several technical challenges yet to overcome, this approach appears extremely promising based on the initial attempts that we describe. The basic approach is that we keep the neuron DNN frozen and learn only how to map inputs onto its dendritic arbor. Inputs mapped onto dendritic locations can take advantage of various dendritic mechanisms distributed on the neuron. In **chapter 2** we discussed and demonstrated the benefits of the temporal aspect of dendrites, but **chapter 1** strongly suggests that the nonlinear cooperative amplification of neurons mediated by the NMDA ion channels is a great candidate to bind features and perform useful computations as well.

Beyond mapping tasks onto single complex neuron models in order to better understand their operation, since the input/output interface of the approach we describe in **chapter 3** is essentially comprised of binary space-time matrices that represent input or output spike trains, it is very simple to use the exact same framework in a network of neurons (each represented by an artificial network itself). The expansion to a network of neurons is expanded upon in **chapter 4**. Once inside our network-in-network framework, we can start to better tackle the questions of temporal coding of information by spiking networks of neurons of various types, ask what are the computations that can be performed by various microcircuits, etc.

Specifically, we wish to highlight the ability of potentially training very large spiking networks using our framework. Much more than previously possible, due to the nature of our formalism taking advantage of state-of-the-art software packages of training deep neural networks and utilizing efficiently specialized hardware such as GPUs. Also, an additional interesting aspect is that we can potentially construct a spiking network and test its ability to create a large number of highly coordinated and temporally precise patterns of activation, e.g., to drive complex motor actions. This of course is in addition to demonstrating the ability to perceive complex spatio-temporal stimuli such as video or audio. These two properties, recognition of complex spatio-temporal spike trains and generation of complex spatio-temporal patterns, are the bedrock of all processing and communication in the brain.

Overall, this network-in-network framework that we greatly elaborate on in **chapter 4** will allow seamless incorporation of information about any neuron type of any complexity level with information about the connectivity of any brain circuit with any myriad of feedforward or feedback connections or network motifs between neurons of different types into a single model that can be efficiently run-on specialized hardware. Therefore, we, hereby propose a concrete, biologically-inspired network architecture for brain networks that seamlessly incorporates single neuron complexity in them. Focusing on architecture as a key element of neuroscience research was recently advocated by (Richards et al., 2019). Indeed, the search of the appropriate architecture of artificial neural networks is one of the most rewarding avenues of machine learning today (He et al., 2015; Lin et al., 2014; Vaswani et al., 2017), and studying the specific architecture suggested in this thesis may unravel some of the inductive biases hidden within the various neuronal brain microcircuits and harness them for future AI applications.

In addition to seamless incorporation of all architectural elements that the neuroscience community empirically gathers about the brain (any microcircuit connectivity, all single neuron biophysical and morphological complexities) inside a single network-in-network framework, we can also use the information gathered by neuroscience community about synaptic updates that relate to any learning that goes on inside that architecture, and use it to determine the computational objective of any neural circuit. This can be achieved by first hypothesizing a computational objective to a microcircuit, then simulating the gradient updates for the circuit composed of multiple neurons using the network-in-network formulation that we have described earlier, and finally compare those updates to a simulated biologically inspired plasticity update based on grounded neuroscientific knowledge. We can iterate and hypothesize multiple computational objectives until we reach a good match between the resultant gradient updates and the ground truth biological plasticity. In this way, once we found a good match, we know our hypothesized objective is potentially correct or at least quite close to being correct. We can assess the optimization objective of a cortical microcircuit or a cerebellar microcircuit, or any other circuit. If we manage to find a simple objective that closely approximates the function and synaptic updates of an entire microcircuit, then it's hard to imagine any higher level of complete understanding of that circuit. We will call that understanding and go home happy with a good day's work.

The optimization objective of a neuronal circuit is Marr's 3$^{rd}$ and highest level of analysis, essentially describing the circuit goal. In order to achieve what we have discussed and get to the optimization objective of a neuronal circuit, we need information about calcium-based plasticity rules (synaptic updates are part of Marr's 2$^{nd}$ algorithmic level of analysis), we needed all the biophysical details about the operation of neurons (which fall into the category of Marr's 1$^{st}$ implementational level of analysis) and we need circuit connectivity information (which also falls into the category of Marr's 1$^{st}$ level of analysis). In this way, we can essentially encompass the experimentally accessible information from lower levels of implementational details about the brain and use them to get to the highest level of desired description.

To summarize, this thesis describes in great detail a systematic method that under certain (yet reasonable) assumptions has the potential to eventually explain the computational objective of all neuronal circuits in the brain. If there is even a slim chance for this to work, then this author is under the impression that it might be worth taking the risk and attempting to follow the path outlined here. This thesis summarizes the first several few steps taken on this path. The scenery was breathtaking thus far, and I am hopeful that the neuroscientific community will continue walking along the path described in this thesis. I am certain what lies ahead is many orders of magnitude more beautiful.

# נוירונים ביולוגיים בודדים כמזהי תבניות מרחביות-זמניות מדויקות

חיבור לשם קבלת תואר דוקטור לפילוסופיה

מאת

ולדיסלב (דוד) בניאגוייב

בהנחיית עידן שגב ומיכאל לונדון

מרכז אדמונד ולילי ספרא למדעי המוח, האוניברסיטה העברית בירושלים

# תקציר

"הכול צריך להיות פשוט ככל האפשר, אך לא יותר פשוט מזה"

אלברט איינשטיין

התזה הזו מתמקדת ברעיון המרכזי שעל נוירונים בודדים במוח יש להתייחס כמזהים תבניות זמני-מרחביות מדויקות ומורכבות. זאת בניגוד לדעה הרווחת של רוב הנוירוביולוגים היום, שמתייחסים לנוירונים ביולוגיים כמזהי תבניות פשוטות ומרחביות בלבד. בתזה זו אני אנסה להראות שזוהי הבדלה חשובה, בעיקר משום שהתכונות המוזכרות של נוירונים בודדים יש תוקף על כלל החישובים שיכולים להתבצע על ידי נוירונים, על כלל המעגלים השונים במוח שהם מרכיבים, ועל איך שמידע מקודד על ידי פעילות נוירונית במוח. כלומר, שלפרטים "נמוכים" אלו ברמת הנוירון הבודד יש השלכות משמעותיות על כלל המערכת. במבוא נדגיש את הרכיבים העיקריים שמרכיבים מעגל נוירוני מיקרוסקופי המבצע חישובים שימושיים ונציג את התלות הדדית של רכיבים אלו מנקודת מבט של המערכת.

בפרק 1 נדון במורכבות הגדולה של קשר הקלט/פלט בין התבניות הזמני-מרחביות של נוירונים קורטיקליים שנובעת מהמבנה המורפולוגי ומאפיינים ביופיזיים של תעלות היונים השונות בנוירון, במיוחד תעלת הNMDA שהיא תעלה מבוססת קולטן המעורבת בפעולה של סינפסות קורטיקליים אקסיטטוריות. אנו מראים שמיפוי הקלט/פלט של נוירון קורטיקלי שכבה 5 (L5) שקול לרשת נוירונים קונבולטיבית עמוקה מלאכותית (DNN) בעלת 5-8 שכבות. מחקר זה פורסם לאחרונה בכתב העת Neuron תחת הכותרת "נוירונים קורטיקליים בודדים כרשתות נוירוניות מלאכותיות עמוקות" (2021).

בפרק 2 נדגיש כי נוירונים בודדים יכולים ליצור תבניות פלט מדויקות זמנית בתגובה לתבניות קלט זמני-מרחביות מסוימות באמצעות כלל למידה ביולוגי פשוט והגיוני. זה אפשרי בזכות הוספת שתי תכונות ביולוגיות לדגם הנוירון המקוטב ומפעיל (I&F) המפושט – יציאות סינפטיות מרובות היוצאות מצינור פרה-סינפטי בודד וסינון זמני הנובע מהמאפיינים הכבל של הדנדריטים. עבודה זו פורסמה ב bioRxiv תחת הכותרת "סינפסות מרובות חיבורים בשילוב עם סינון דנדריטי מגדילים את יכולות זיהוי התבניות הזמני-מרחביות של נוירונים בודדים" (2022) וכיום מוגשת לכתב עת עם ביקורת עמיתים.

בפרק 3, באמצעות הכלים שפיתחנו בפרקים 1 ו-2, אנו מדגישים דרכים חדשות לחקירה עתידית (עם כמה התחלות בכיוונים אלו). הפרק מתמקד בכיצד ניתן לטפל בשאלה הגדולה של: איזה סוג של חישובים יכולים להתבצע על ידי נוירון בודד? אנו מתארים זאת בפרטים רבים עם תקווה שזה ישמש כמדריך לעבודה עתידית.

בפרק 4 אנו מרחיבים את פרק 3 ומתארים הרחבה של רעיונותינו לרשתות נוירוניות המורכבות ממספר רב של נוירונים ביולוגיים ריאליסטיים המייצגים מעגלים מיקרוסקופיים קטנים או אף אזורי מוח שלמים. בפרק זה אנו מתארים בפרטי פרטים סדרה של ניסויי הדגמה בעלי מורכבות גוברת שמנסים להשתמש ברשתות של נוירונים ריאליסטיים כדי לבצע סדרה מסוימת של משימות חישוביות עם המטרה לקבוע אילו תכונות ביולוגיות חשובות עבור אילו היבטים חישוביים של מערכת העצבים. בנוסף, בפרק זה, אנו גורסים מתכון מפתיע בפשטותו של כיצד ניתן, באופן עקרוני, לקבוע את החישוב המובהק שמבוצע על ידי כל מעגל מיקרוסקופי או ארכיטקטורה על ידי שימוש במידע אודות כללים לעדכון סינפטי מעבודה ניסויית (במסגרת זו אנו מתייחסים למילה "חישוב" כפי שמגדיר זאת דיוויד מאר ברמה הגבוהה ביותר של ניתוח – כלומר, לקבוע את היעד החישובי של המעגל). כלומר, אנו מתארים כיצד ניתן להשיג את רמת הניתוח החישובית (במסגרת מאר) של מעגל נוירוני על ידי שילוב ידע מרמות הניתוח המימושית והאלגוריתמית. בפרט, אנו מתארים בפרטים רבים דרך כיצד ניתן לעשות זאת על ידי שימוש במידע על כללי הלמידה הביולוגיים שמיושמים במוח ברמה של סינפס בודד ובאמצעות שימוש בייצוג של רשת נוירונית עמוקה של נוירונים בודדיםריאליסטיים (שפורמליזם מדויק שלהם בוצע בפרק 1) במעגלים נוירוניים המורכבים רשתות מלאכותיות שכאלה.

בדיון, אנו מפרטים על איך העקרונות שמסופקים באמצעות התזה זו יכולים להיות מוכנים ללמידה של מגוון רחב של מעגלים נוירוניים ובניה הדרגתית לקראת רמת ההבנה החישובית של כל המעגלים הנוירוניים המיקרוסקופיים במוח, ומשם בניה לקראת קביעת היעד החישובי של אזורי המוח השולמים שבסופו של דבר כוללים את היעד החישובי של כלל המוח.